\documentclass{article}

\usepackage[scr=boondoxo,scrscaled=1.05]{mathalfa}
\usepackage[pdftex]{graphicx}
\usepackage{amsmath,amssymb,hhline,tikz,mathtools}
\usepackage{enumerate}
\usetikzlibrary{shapes.geometric,decorations.markings}
\usetikzlibrary{calc}
\usepackage[title]{appendix}
\usepackage{etoolbox}
\patchcmd{\appendices}{\quad}{: }{}{}
\usepackage{algorithm}
\usepackage{algorithmicx}
\usepackage{algpseudocode}
\usepackage{rotating}
\usepackage{afterpage}
\usepackage{xstring}%%% needed for \rowvec macros
\usepackage{xcolor}
\usepackage{cmap}
\usepackage{mathdots}
\usepackage{colortbl}
\usepackage{booktabs}
\usepackage{textcomp}     % access \textquotesingle
\usetikzlibrary{backgrounds}
\usepackage{pgfplots}
\pgfplotsset{compat=1.12}
\pgfmathsetseed{5112}
\usepackage{multirow}
\usepackage{pgfplotstable}
\usepackage{adjustbox}
\usepackage{diagbox}

\PassOptionsToPackage{hyphens}{url}
\usepackage{hyperref}
\hypersetup{
    pdftitle={Word Embedding Techniques for Malware Classification},
    pdfauthor={Mark Stamp},
    bookmarksnumbered=true,     
    bookmarksopen=true,         
    bookmarksopenlevel=3,       
    colorlinks=true,linkcolor=black,citecolor=black,urlcolor=blue,filecolor=blue,
    pdfstartview=Fit,           
    pdfpagemode=UseOutlines,
    pdfpagelayout=SinglePage
}

\definecolor{darkgreen}{rgb}{0.125,0.5,0.169}

\algrenewcommand{\algorithmiccomment}[1]{{\color{red}{\tt //}\ #1}}
\algnewcommand{\Initialize}[1]{%
  \State \textbf{Initialize:}
  \Statex \hspace*{\algorithmicindent}\parbox[t]{.8\linewidth}{\raggedright #1}
}
\algnewcommand{\Given}[1]{%
  \State \textbf{Given:}
  \Statex \hspace*{\algorithmicindent}\parbox[t]{.8\linewidth}{\raggedright #1}
}

\long\def\symbolfootnotetext[#1]#2{\begingroup%
\def\thefootnote{\fnsymbol{footnote}}\footnotetext[#1]{#2}\endgroup}

\let\oldsqrt\sqrt
\def\sqrt{\mathpalette\DHLhksqrt}
\def\DHLhksqrt#1#2{%
\setbox0=\hbox{$#1\oldsqrt{#2\,}$}\dimen0=\ht0
\advance\dimen0-0.2\ht0
\setbox2=\hbox{\vrule height\ht0 depth -\dimen0}%
{\box0\lower0.4pt\box2}}
\def\clap#1{\hbox to 0pt{\hss#1\hss}}

\allowdisplaybreaks

\def\figureFastSpeed{s}\def\figureSpeed{f}
\let\figureFastSpeed=\figureSpeed
\def\selectFigureSpeed#1#2{
\if\figureSpeed\figureFastSpeed #1\else #2\fi}

\def\srowvecc#1#2{(\!\begin{array}{cc} 
      \noexpandarg\IfBeginWith{#1}{-}{\! #1}{#1}
    & #2\kern-0.5pt\end{array}\!)}
\def\rowvecc#1#2{\left(\!\begin{array}{cc} 
      \noexpandarg\IfBeginWith{#1}{-}{\! #1}{#1}
    & #2\kern-0.5pt\end{array}\!\right)}
\def\rowveccc#1#2#3{\left(\!\begin{array}{ccc} 
      \noexpandarg\IfBeginWith{#1}{-}{\! #1}{#1}
    & #2 
    & #3\kern-0.5pt\end{array}\!\right)}
\def\rowvecccc#1#2#3#4{\left(\!\begin{array}{cccc}
      \noexpandarg\IfBeginWith{#1}{-}{\! #1}{#1}
    & #2 
    & #3 
    & #4\kern-0.5pt\end{array}\!\right)}
\def\srowvecccc#1#2#3#4{\bigl(\!\begin{array}{cccc}
      \noexpandarg\IfBeginWith{#1}{-}{\! #1}{#1}
    & #2 
    & #3 
    & #4\kern-0.5pt\end{array}\!\bigr)}
\def\rowveccccc#1#2#3#4#5{\left(\!\begin{array}{ccccc} 
      \noexpandarg\IfBeginWith{#1}{-}{\! #1}{#1}
    & #2
    & #3
    & #4
    & #5\kern-0.5pt\end{array}\!\right)}
\def\srowvecccccc#1#2#3#4#5#6{(\!\begin{array}{cccccc} 
      \noexpandarg\IfBeginWith{#1}{-}{\! #1}{#1}
    & #2
    & #3
    & #4
    & #5
    & #6\kern-0.5pt\end{array}\!)}
\def\rowvecccccc#1#2#3#4#5#6{\left(\!\begin{array}{cccccc} 
      \noexpandarg\IfBeginWith{#1}{-}{\! #1}{#1}
    & #2
    & #3
    & #4
    & #5
    & #6\kern-0.5pt\end{array}\!\right)}
\def\figureType{*}\def\figureSlowType{slowType}
\def\selectFigureType#1#2{
\if\figureType\figureSlowType #1\else #2\fi}
\makeatletter
\newcommand{\lowsub}[1]{\mathpalette{\raisem@th{#1}}}
\newcommand{\raisem@th}[3]{\raisebox{-#1}{$#2#3$}}
\makeatother
\def\halfthin{\kern 0.083em}

\def\un{\underline{\hspace*{0.2cm}}}

\def\Csim{\cos_{\theta}}

\DeclareMathOperator{\ma}{a}
\DeclareMathOperator{\me}{e}
\DeclareMathOperator{\ms}{s}
\DeclareMathOperator{\mt}{t}

\def\eref#1{{\color{black}(\ref{#1})}}

\pgfkeys{
    /pgf/number format/precision=2, 
    /pgf/number format/fixed zerofill=true }
    
\pgfplotstableset{
    /color cells/min/.initial=0,
    /color cells/max/.initial=1000,
    /color cells/textcolor/.initial=,
    color cells/.code={%
        \pgfqkeys{/color cells}{#1}%
        \pgfkeysalso{%
            postproc cell content/.code={%
                \begingroup
                \pgfkeysgetvalue{/pgfplots/table/@preprocessed cell content}\value
\ifx\value\empty
\endgroup
\else
                \pgfmathfloatparsenumber{\value}%
                \pgfmathfloattofixed{\pgfmathresult}%
                \let\value=\pgfmathresult
                \pgfplotscolormapaccess
                    [\pgfkeysvalueof{/color cells/min}:\pgfkeysvalueof{/color cells/max}]%
                    {\value}%
                    {\pgfkeysvalueof{/pgfplots/colormap name}}%
                \pgfkeysgetvalue{/pgfplots/table/@cell content}\typesetvalue
                \pgfkeysgetvalue{/color cells/textcolor}\textcolorvalue
                \toks0=\expandafter{\typesetvalue}%
                \xdef\temp{%
                    \noexpand\pgfkeysalso{%
                        @cell content={%
                            \noexpand\cellcolor[rgb]{\pgfmathresult}%
                            \noexpand\definecolor{mapped color}{rgb}{\pgfmathresult}%
                            \ifx\textcolorvalue\empty
                            \else
                                \noexpand\color{\textcolorvalue}%
                            \fi
                            \the\toks0 %
                        }%
                    }%
                }%
                \endgroup
                \temp
\fi
            }%
        }%
    }
}
\DeclareMathOperator{\zzNN}{-NN}
\def\kNN#1{{#1}\!\zzNN}
\makeatletter
\newcommand*\bigcdot{\mathpalette\bigcdot@{.5}}
\newcommand*\bigcdot@[2]{\mathbin{\vcenter{\hbox{\scalebox{#2}{$\m@th#1\bullet$}}}}}
\makeatother

\def\O{{\cal O}}

\def\k{\kern 2.75pt}

\def\p{\underline{\phantom{0}}}

\newlength{\xxxxx}
\def\log{\mbox{log}}

\def\n{\phantom{0}}

\def\sscoin{%
  \leavevmode
  \vtop{\offinterlineskip %\bfseries
    \setbox0=\hbox{\scriptsize S}%
    \setbox2=\hbox to\wd0{\hfil\hskip-.03em
    \vrule height .3ex width .08ex\hskip .08em
    \vrule height .3ex width .08ex\hfil}
    \vbox{\copy2\box0}\box2}}
\newcommand\affil[2]{%
  \begingroup
  \renewcommand\thefootnote{}\footnote{\llap{$\hbox{}^{#1}\hbox{}$}#2}%
  \addtocounter{footnote}{-1}%
  \endgroup
}
\newcommand\markonly[1]{%
$\hbox{}^{\mbox{\kern4.5pt,\kern0.75pt #1}}$
}

\advance\oddsidemargin by -0.525in
\advance\textwidth by 1.05in
\advance\topmargin by -0.625in
\advance\textheight by 1.25in

\title{\vspace{-0.5in}A Comparison of Word2Vec, HMM2Vec, 
	and PCA2Vec for Malware Classification}

\author{Aniket Chandak\footnote{chandakaniket537@gmail.com}\ \ \ \ 
Wendy Lee\footnote{wendy.lee@sjsu.edu}\markonly{\sscoin}\ \ \ \ 
\setcounter{footnote}{3}
Mark Stamp\footnote{mark.stamp@sjsu.edu}\markonly{\sscoin}
}
\date{}

\begin{document}

\maketitle

\vglue-0.35in

\affil{\sscoin}{Department 
of Computer Science,
San Jose State University,
San Jose, California}

\abstract
Word embeddings are often used in natural language processing
as a means to quantify relationships between words. More generally,
these same word embedding techniques can be used to quantify relationships
between features. In this paper, we first consider multiple different word embedding
techniques within the context of malware classification.
We use hidden Markov models to obtain embedding vectors in an
approach that we refer to as HMM2Vec, and we generate vector
embeddings based on principal component analysis. We also
consider the popular neural network based word embedding 
technique known as Word2Vec.
In each case, we derive feature embeddings based on opcode sequences
for malware samples from a variety of different families.
We show that we can obtain better classification accuracy
based on these feature embeddings, as compared to
HMM experiments that directly use the opcode sequences, and serve
to establish a baseline. These
results show that word embeddings can be a useful feature 
engineering step in the field of malware analysis.

\section{Introduction}

Malware detection and analysis are critical 
aspects of information security. 
The~2019 Internet Threat Security Report~\cite{isrt}
claims an increase of~25\%\ in one 
year in the number of attack groups using malware to 
disrupt businesses and organizations. According to the~2016
California Data Breach Report~\cite{cdb}, malware 
contributed to~54\%\ of all breaches and~90\%\ of total 
records breached, with a staggering~44 million records breached
due to malware in the years 2012--2016. Statistics 
such as these imply that malware is an increasing threat.

In this paper,
we apply machine learning classification techniques to engineered
features that are derived from malware samples. This feature engineering
involves machine learning techniques. %---specifically, word embeddings. 
In effect, we apply machine
learning to higher-level features, where these features are themselves
obtained using machine learning models.
The motivation is that machine learning can serve to distill useful information
from training samples, and hence the classification techniques may perform
better on such data. In this research, we consider the effectiveness of
using these derived features in the context of malware classification.

Specifically, we use word embeddings based on opcodes to derive features for
subsequent classification. We consider three distinct word embedding techniques.
First, we derive word embeddings from trained hidden Markov models (HMM).
We refer to this technique as HMM2Vec. We then consider and analogous
technique based on principal component analysis (PCA), which we refer to 
as PCA2Vec. And, as a third approach, we experiment with the popular
neural network based word embedding technique known as Word2Vec.
In each case, we generate word embeddings for a significant number of samples
from a variety of malware families. We then use several classification
techniques to determine how well we can classify these samples
using word embeddings as features.

The remainder of this paper is organized as follows. We provide a selective
survey of relevant related work in Section~\ref{sect:RW}.
Section~\ref{sect:back} contains an extensive and wide-ranging 
discussion of machine learning
topics that play a role in this research. In Section~\ref{sect:wordEmbed},
we provide details on the word embedding techniques that
form the basis of our experiments.
Section~\ref{sect:exp} gives our experiments and results,
while Section~\ref{sect:conc} provides our conclusion and
some paths for future work.

\section{Related Work}\label{sect:RW}

Malware analysis and detection are challenging problems due to a
variety of factors, including the large volume of mlaware and
obfuscation techniques~\cite{dhammibhehav}. 
Every day, thousands of 
new malware are generated---manual analysis techniques cannot
keep pace. Obfuscation is widely used by malware developers to make 
it difficult to analyze their malicious code.

Signature-based malware detection methods rely on pattern 
matching with known signatures~\cite{vinod2009survey}. 
Signature detection is relatively fast,
and it is effective against ``traditional'' malware. 
However, extracting signatures is a labor intensive process, 
and obfuscation techniques can
defeat signature scanning.

Anomaly-based techniques are based on ``unusual'' or
``virus-like'' behavior or characteristics. 
An example of anomaly detection is behavior based analysis, 
which can be used to analyze
a sample when executed or under emulation~\cite{vinod2009survey}. 
When an executable file performs any action that does not 
fit its expected behavior, an alarm can be triggered. Such a method 
can detect obfuscated and zero-day malware, but it is slow, and 
generally subject to excessive false positives.

Recently, machine learning techniques have proven 
extremely useful for malware detection.
The effectiveness of machine learning algorithms 
depends on the characteristics of the features used by such models. 
In malware detection and classification, a sample can be represented 
by a wide variety of features, including mnemonic opcodes, raw bytes, 
API calls, permissions, header information, etc. Opcodes are a popular
feature that form the basis of the analysis considered in this paper.

In~\cite{bilar}, the author experiments with opcodes 
and determines that such
features can be successfully used to detect malware. The 
paper~\cite{malwaredetapi} achieves good results using API calls 
as a feature. Such features can be somewhat more difficult for 
malware writers to obfuscate, since API calls relate to 
the essential activity of software. However, extracting API calls from an 
executable is more costly than extracting opcodes.

Another example of malware research involving opcodes
can be found in~\cite{markovblanket}. This paper features 
opcode $n$-grams, with a Markov blanket used to
select from the large set of available $n$-gram.
Classification is based on hidden Markov models, and
experiments are based on five malware families.

In~\cite{8422083}, malware opcodes are treated as a language,
with Word2Vec used to quantify contextual information. 
Classification relies on $k$-nearest neighbors ($\kNN{k}$). 
The research in~\cite{w2vmalware} also uses Word2Vec to generate 
feature vectors based on opcode sequences, with a deep neural network 
employed for malware classification. In this latter research,
the number of opcodes is in the range of~50 to~200,
and the length of the Word2Vec embeddings 
range from~250 to~750.

Word2Vec embeddings are used as features
to train bi-directional LSTMs in~\cite{w2vfeature2}. 
The experiments achieve good accuracy for malware detection,
but training is costly. In~\cite{Hashemi2016GraphEA}, the author 
proposed a word embedding method based on opcode graphs---the 
graph is projected into vector space, which yields 
word embeddings. This technique is also computationally expensive.

%%%%% Efficiency experiments????? Word2Vec timings?????
In comparison to previous research,
we consider additional vector embedding
techniques, we experiment with a variety of classification algorithms, 
we use a smaller number of opcodes, and we generate
short embedding vectors. Since we use a relatively
small number of opcodes and short embedding
vectors, our techniques are all highly efficient
and practical. In addition, our experiments are based 
on a recently collected and challenging malware dataset.

\section{Background}\label{sect:back}

In this section, we present background information on the various
learning techniques that are used in the experiments
discussed in Section~\ref{sect:exp}.
Specifically, we introduce 
neural networks, beginning with some historical background and moving
on to a modern context. 
We also introduce HMMs and PCA, which form the basis for the word embedding
techniques that we refer to as HMM2Vec and PCS2Vec, respectively. 
Finally, we introduce four classification techniques,
which are used in our experiments.

In Section~\ref{sect:wordEmbed}, we discuss HMM2Vec, PCA2Vec,
and the neural network based word embedding technique, Word2Vec,
in detail.
For our experiments in Section~\ref{sect:exp}, 
we use these three word embedding techniques
to generate features to classify malware samples. 
%In each
%case, we compare multiple classifiers.

\subsection{Neural Networks}\label{sect:NNintro}

The concept of an artificial neuron~\cite{lhardesty,cwallis} is not new, as the 
idea was first proposed by McCulloch and Pitts in the 1940s~\cite{MP}. However, 
modern computational neural networks begins with the perceptron,
as introduced by Rosenblatt in the late 1950s~\cite{frosenblatt}.

\subsubsection{McCulloch-Pitts Artificial Neuron}

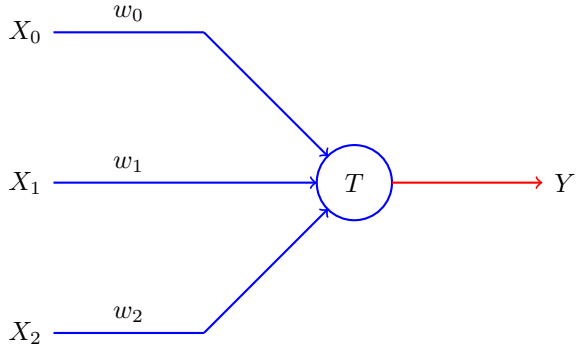
\begin{figure}[!htbp]
  \centering
    \begin{tikzpicture}[thick]
%    \node[draw,circle,xshift=2.2cm,minimum size=10mm,outer sep=0] (big) {};
%    \node[draw,circle,minimum size=10mm,outer sep=0] (big) {$T$};
    \draw[color=blue,thick] (2,2) circle (0.5cm);
    \node at (2, 2)   {$T$};
    \node at (-2.375, 0)   {$X_2$};
    \draw[blue] (-2,0) -- (0,0);
    \draw[->,blue] (0,0) -- (1.65,1.65); 
    \node at (-1, 0.25)   {$w_2$};
    \node at (-2.375, 4)   {$X_0$};
    \draw[blue] (-2,4) -- (0,4);
    \draw[->,blue] (0,4) -- (1.65,2.35);
    \node at (-1, 4.25)   {$w_0$};
    \node at (-2.375, 2)   {$X_1$};
    \draw[->,blue] (-2,2) -- (1.5,2.00);
    \node at (-1, 2.25)   {$w_1$};
    \draw[->,red,thick] (2.5,2) -- (4.5,2);
    \node at (4.8, 2)   {$Y$};
%    \node(2,2) {$T$};
%    \node[draw,circle,minimum size=2mm,outer sep=0] (small) {};
%    \draw (tangent cs:node=small,point={(big.south)},solution=2) -- (tangent cs:node=big,point={(small.south)});
%    \draw (tangent cs:node=small,point={(big.north)},solution=1) -- (tangent cs:node=big,point={(small.north)},solution=2);
\end{tikzpicture}
%  \vglue-0.1in
  \caption{Artificial neuron}\label{fig:AN}	
\end{figure}

An artificial neuron with three inputs is illustrated in Figure~\ref{fig:AN}. 
In the original McCulloch-Pitts formulation, the inputs~$X_i\in\{0,1\}$, 
the weights~$w_i\in\{+1,-1\}$,
and the output~$Y\in\{0,1\}$. The output~$Y$ is~0 (inactive) 
or~1 (active), based on whether or not the linear 
function~$\sum w_i X_i$ exceeds the specified
threshold~$T$.
This form of an artificial neuron
was modeled on neurons in the brain, 
which either fire or it do not
(thus, $Y\in\{0,1\}$), and have input that comes from other
neurons (thus, each~$X_i\in\{0,1\}$). The weights~$w_i$
specify whether an input is excitatory (increasing the chance of
the neuron firing) or inhibitory (decreasing the chance of the neuron
firing). Whenever~$\sum w_i X_i > T$, the excitatory response wins, and
the neuron fires---otherwise the inhibitory response wins and the
neuron does not fire.

\subsubsection{Perceptron}\label{sect:percept}

A \textit{perceptron} is less restrictive than a McCulloch-Pitts artificial neuron.
With a perceptron, both
the inputs~$X_i$ and the weights~$w_i$ can be real valued, as opposed
to the binary restrictions of McCulloch-Pitts. As wth the
McCulloch-Pitts formulation, the output~$Y$ of a perceptron
is generally taken to be binary.
%McCulloch and Pitts chose their more restrictive formulation because
%they were trying to model logic functions. At the time, it was felt that
%encoding elementary logic into artificial neurons would be a
%necessary step towards constructing systems with artificial intelligence. However, that
%point of view has certainly not panned out, while the additional generality offered by
%perceptrons (and further generalizations) has proven extremely useful in practice.

Given a real-valued input vector~$X=(X_0,X_1,\ldots,X_{n-1})$, a perceptron 
can be viewed as an instantiation of a function of the form
$$
  f(X) = \sum_{i=0}^{n-1} w_i X_i + b ,
$$
that is, a perceptron computes a weighted sum of the input components.
Based on a threshold, a single perceptron can defines a binary classifier. That is,
we can classify a sample~$X$ as ``type~1'' provided that~$f(X) > T$, for some specified 
threshold~$T$, and otherwise we classify~$X$ as ``type~0.''  

%Note that
In the case of two dimensional input, the decision boundary of a 
is of the form
\begin{equation}\label{eq:perceptron_2}
  f(x,y) = w_0 x + w_1 y + b 
\end{equation}
which is the equation of a line.
In general, the decision boundary of a perceptron is a hyperplane.
Hence, a perceptron can only provide ideal separation in cases
where the data itself is linearly separable.

%There was considerable research into artificial neural networks (ANN) 
%in the 1950s and 1960s, and that era is often considered  
%the first ``golden age'' of artificial intelligence (AI).
%But the gold turned to lead in~1969
%when an influential work by Minsky and Papert~\cite{Minsky} emphasized
%the limitations of perceptrons. Specifically, they pointed out that
%since the XOR function is not linearly separable, it follows 
%that a single perceptron cannot model something as elementary as XOR.
%The OR, AND, and XOR functions are illustrated in Figure~\ref{fig:XOR},
%where we see that OR and AND are linearly separable, while XOR is not.

%\begin{figure}[!htbp]
%\centering
%\advance\tabcolsep by -1pt
%\begin{tabular}{cccccc}
%    \input figures/figXOR_a.tex
%&\ \ \ \ \ &
%    \input figures/figXOR_b.tex
%&\ \ \ \ \ &
%    \input figures/figXOR_c.tex
%\\[2ex]
%(a) OR & & (b) AND & & (c) XOR 
%\end{tabular}
%\caption{OR and AND are linearly separable but XOR is not}\label{fig:XOR}	
%\end{figure}

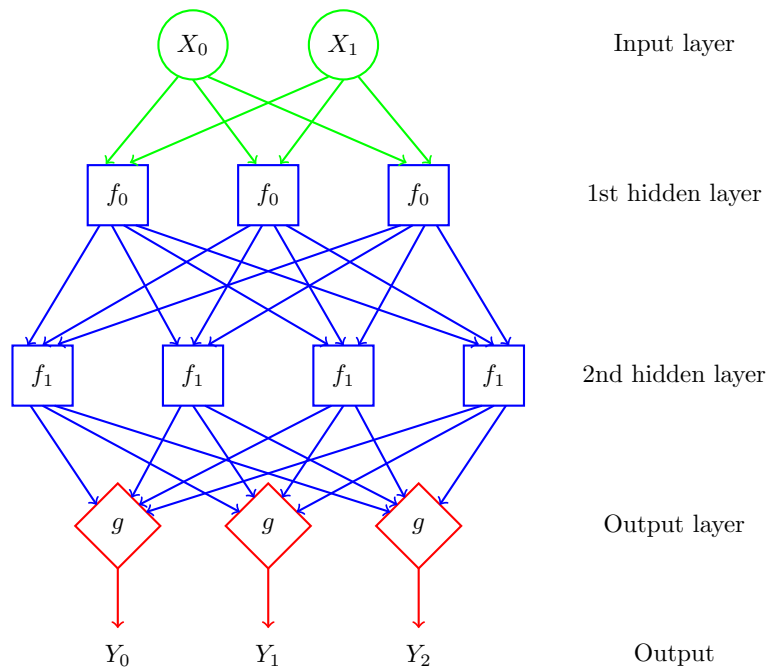
\begin{figure}[!htb]
\centering
    \begin{tikzpicture}[scale=0.8,every node/.style={scale=0.95}]
    
    % circles
    \draw[thick,color=green] (3.0,8.5) circle (0.575);
    \draw[thick,color=green] (5.5,8.5) circle (0.575);

    % squares (top)
    \draw[thick,color=blue] (1.25,5.5) rectangle (2.25,6.5);
    \draw[thick,color=blue] (3.75,5.5) rectangle (4.75,6.5);
    \draw[thick,color=blue] (6.25,5.5) rectangle (7.25,6.5);
    
    % squares (bottom)
    \draw[thick,color=blue] (0.0,2.5) rectangle (1.0,3.5);
    \draw[thick,color=blue] (2.5,2.5) rectangle (3.5,3.5);
    \draw[thick,color=blue] (5.0,2.5) rectangle (6.0,3.5);
    \draw[thick,color=blue] (7.5,2.5) rectangle (8.5,3.5);
    
    % diamonds
    \draw[thick,color=red,rotate around={45:(1.75,0.5)}] (1.25,0.0) rectangle (2.25,1.0);
    \draw[thick,color=red,rotate around={45:(4.25,0.5)}] (3.75,0.0) rectangle (4.75,1.0);
    \draw[thick,color=red,rotate around={45:(6.75,0.5)}] (6.25,0.0) rectangle (7.25,1.0);

    % circle to square
    \draw[thick,color=green,->] (2.75,7.97) -- (1.55,6.52);
    \draw[thick,color=green,->] (3,7.92) -- (4.05,6.52);
    \draw[thick,color=green,->] (3.25,7.97) -- (6.55,6.54);
    
    \draw[thick,color=green,->] (5.25,7.97) -- (1.95,6.54);
    \draw[thick,color=green,->] (5.5,7.92) -- (4.45,6.52);
    \draw[thick,color=green,->] (5.75,7.97) -- (6.95,6.52);

    % square to diamond
    \draw[thick,color=blue,->] (0.3,2.5) -- (1.4,0.85);
    \draw[thick,color=blue,->] (0.5,2.5) -- (3.78,0.73);
    \draw[thick,color=blue,->] (0.7,2.5) -- (6.28,0.73);
    
    \draw[thick,color=blue,->] (2.8,2.5) -- (1.98,0.97);
    \draw[thick,color=blue,->] (3.0,2.5) -- (4.02,0.97);
    \draw[thick,color=blue,->] (3.2,2.5) -- (6.4,0.85);
    
    \draw[thick,color=blue,->] (5.3,2.5) -- (2.1,0.85);
    \draw[thick,color=blue,->] (5.5,2.5) -- (4.48,0.97);
    \draw[thick,color=blue,->] (5.7,2.5) -- (6.52,0.97);
    
    \draw[thick,color=blue,->] (7.8,2.5) -- (2.22,0.73);
    \draw[thick,color=blue,->] (8.0,2.5) -- (4.72,0.73);
    \draw[thick,color=blue,->] (8.2,2.5) -- (7.1,0.85);

    % square to square
    \draw[thick,color=blue,->] (1.45,5.5) -- (0.25,3.5);
    \draw[thick,color=blue,->] (1.65,5.5) -- (2.75,3.5);
    \draw[thick,color=blue,->] (1.85,5.5) -- (5.25,3.51);
    \draw[thick,color=blue,->] (2.05,5.5) -- (7.75,3.52);

    \draw[thick,color=blue,->] (3.95,5.5) -- (0.5,3.51);
    \draw[thick,color=blue,->] (4.15,5.5) -- (3.0,3.5);
    \draw[thick,color=blue,->] (4.35,5.5) -- (5.5,3.5);
    \draw[thick,color=blue,->] (4.55,5.5) -- (8.0,3.51);

    \draw[thick,color=blue,->] (6.45,5.5) -- (0.75,3.52);
    \draw[thick,color=blue,->] (6.65,5.5) -- (3.25,3.51);
    \draw[thick,color=blue,->] (6.85,5.5) -- (5.75,3.5);
    \draw[thick,color=blue,->] (7.05,5.5) -- (8.25,3.5);

    % output
    \draw[thick,color=red,->] (1.75,-0.2) -- (1.75,-1.2);
    \draw[thick,color=red,->] (4.25,-0.2) -- (4.25,-1.2);
    \draw[thick,color=red,->] (6.75,-0.2) -- (6.75,-1.2);

    % labels for circles
    \node at (3.0,8.5) {$X_0$};
    \node at (5.5,8.5) {$X_1$};

    % labels for squares (top)
    \node at (1.75,6.0) {$f_0$};
    \node at (4.25,6.0) {$f_0$};
    \node at (6.75,6.0) {$f_0$};

    % labels for squares (bottom)
    \node at (0.5,3.0) {$f_1$};
    \node at (3.0,3.0) {$f_1$};
    \node at (5.5,3.0) {$f_1$};
    \node at (8.0,3.0) {$f_1$};

    % labels for diamonds
    \node at (1.75,0.5) {$g$};
    \node at (4.25,0.5) {$g$};
    \node at (6.75,0.5) {$g$};

    % labels for output
    \node at (1.75,-1.65) {$Y_0$};
    \node at (4.25,-1.65) {$Y_1$};
    \node at (6.75,-1.65) {$Y_2$};
    
    % labels
    \node at (11.0,8.5) {Input layer};
    \node at (11.0,6.0) {1st hidden layer};
    \node at (11.0,3.0) {2nd hidden layer};
    \node at (11.0,0.5) {Output layer};
    \node at (11.0,-1.65) {Output};

\end{tikzpicture}
%  \vglue-0.1in
\caption{MLP with two hidden layers}\label{fig:mini_MLP}
\end{figure}

As the name suggests, a multilayer perceptron (MLP) is an ANN that includes
multiple (hidden) layers in the form of perceptrons. An example of an MLP with two 
hidden layers is given in Figure~\ref{fig:mini_MLP}, where each edge represent a weight
that is to be determined via training. Unlike a single layer perceptron, MLPs are not 
restricted to linear decision boundaries, and hence an MLP can accurately
model more complex functions. For example,
the XOR function---which cannot be modeled by a single layer 
perceptron---can be modeled by an MLP.
%However, the perceptron training method originally
%proposed by Rosenblatt~\cite{frosenblatt} 
%cannot be used to effectively traing an MLP~\cite{akurenkov}.

To train a single layer perceptron, %such as that illustrated in Figure~\ref{fig:AN}, 
simple heuristics will suffice, assuming that the data is actually linearly separable. 
From a high level perspective, training a single layer perceptron is
somewhat analogous to training a linear support vector machine (SVM),
except that for a perceptron, we do not require that the 
margin (i.e., minimum separation between the classes) be maximized.
But training an MLP is clearly far more challenging, 
since we have hidden layers between the
input and output, and it is not obvious how changes to the
weights in these hidden layers will affect each other or the output.

As an aside, it is interesting to note that for SVMs, we
deal with data that is not linearly separable
%by employing a soft margin (i.e., we allow for training errors) and
by use of the ``kernel trick,'' where
the input data is mapped to a higher dimensional ``feature space'' via
a (nonlinear) kernel function. In contrast, perceptrons (in the form of MLPs)
overcome the limitation of linear separability by the use of
multiple layers. With an MLP, it is 
as if a nonlinear kernel function has been embedded directly
into the model itself through the use of hidden layers, as opposed
to a user-specified explicit kernel function, which is the case for an SVM. 
We can view the relationship between ANNs and deep learning as being somewhat akin
to that of Markov chains and hidden Markov models (HMM). That is, ANNs serve as a basic 
technology that can be used to build powerful machine learning techniques,
analogous to the way that an HMM is built on the foundation of an elementary 
Markov chain. 

\subsection{Hidden Markov Models}\label{sect:HMMintro}

\begin{figure}[!htb]
\centering
        \begin{tikzpicture}[scale=0.9,every node/.style={scale=0.9}]
    
    % squares
    \draw[thick,color=blue] (0,0) rectangle (1,1);
    \draw[thick,color=blue] (2.5,0) rectangle (3.5,1);
    \draw[thick,color=blue] (5,0) rectangle (6,1);
    \draw[thick,color=blue] (10,0) rectangle (11,1);

    % circles
    \draw[thick,color=green] (0.5,4.5) circle (0.575);
    \draw[thick,color=green] (3,4.5) circle (0.575);
    \draw[thick,color=green] (5.5,4.5) circle (0.575);
    \draw[thick,color=green] (10.5,4.5) circle (0.575);
    
    % observations
%    \node at (-1.5,0.5){Observations:};
    \node at (0.5,0.5){$\O_0$};
    \node at (3,0.5){$\O_1$};
    \node at (5.5,0.5){$\O_2$};
    \node at (8,0.5){$\cdots$};
    \node at (10.5,0.5){$\O_{T-1}$};

    % States
%    \node at (-1.75,4.5){Markov process:};
%    \node at (0.5,4.5){$X_0$};
%    \node at (3,4.5){$X_1$};
%    \node at (5.5,4.5){$X_2$};
%    \node at (8,4.5){$\cdots$};
%    \node at (10.5,4.5){$X_{T-1}$};
    \node at (0.5,4.5){$X_0$};
    \node at (3,4.5){$X_1$};
    \node at (5.5,4.5){$X_2$};
    \node at (8,4.5){$\cdots$};
    \node at (10.5,4.5){$X_{T-1}$};
       
    % A's
    \node at (1.7,4.8){$A$};
    \node at (4.2,4.8){$A$};
    \node at (6.7,4.8){$A$};
    \node at (9.2,4.8){$A$};
    
    % B's
    \node at (0.2,2.1){$B$};
    \node at (2.7,2.1){$B$};
    \node at (5.2,2.1){$B$};
    \node at (10.2,2.1){$B$};
    
    % circle to circle
     \draw[thick,color=black,->] (1.075,4.5) -- (2.425,4.5);
     \draw[thick,color=black,->] (3.575,4.5) -- (4.925,4.5);
     \draw[thick,color=black,->] (6.075,4.5) -- (7.425,4.5);
     \draw[thick,color=black,->] (8.575,4.5) -- (9.925,4.5);

    % circle to square
     \draw[thick,color=black,->] (0.5,3.925) -- (0.5,1);
     \draw[thick,color=black,->] (3.0,3.925) -- (3.0,1);
     \draw[thick,color=black,->] (5.5,3.925) -- (5.5,1);
     \draw[thick,color=black,->] (10.5,3.925) -- (10.5,1);

    % curtain
    \draw[thick,dashed,color=red] (-0.3,3) -- (11.2,3);
   
    \end{tikzpicture}
%  \vglue-0.1in
  \caption{Hidden Markov model}\label{fig:hidden}
\end{figure}
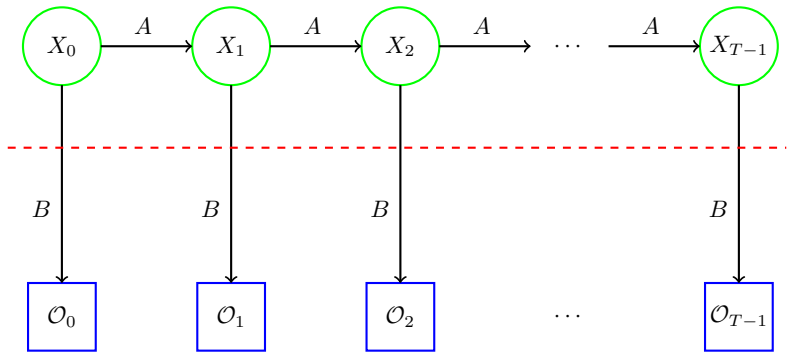

A generic hidden Markov model is illustrated in
Figure~\ref{fig:hidden}, where the~$X_i$ represent the
hidden states and all other notation is as in Table~\ref{tab:HMMnotation}.
The state of the Markov process, which we can view as being hidden 
behind a ``curtain'' (the dashed line in Figure~\ref{fig:hidden}), 
is determined by the current state and the~$A$
matrix. We are only able to observe the observations~$\O_i$, 
which are related to the (hidden)
states of the Markov process by the matrix~$B$.

\subsubsection{Notation and Basics}

The notation used in an HMM is summarized in Table~\ref{tab:HMMnotation}.
Note that the observations are assumed to come from 
the set~$\{0,1,\ldots,M-1\}$, 
which simplifies the notation with no loss of
generality. That is, we simply associate each of the~$M$ 
distinct observations with one of the elements~$0,1,\ldots,M-1$, so that
we have $\O_i\in V=\{0,1,\ldots,M-1\}$ for~$i=0,1,\ldots,T-1$.

\begin{table}[!htb]
  \caption{HMM notation}\label{tab:HMMnotation}
  \vglue 0.125in
  \centering
  \begin{tabular}{cl} \midrule\midrule
    \textbf{Notation} & \hspace*{0.5in}\textbf{Explanation}\\ \midrule
    $T$ & Length of the observation sequence\\
    $N$ & Number of states in the model\\
    $M$ & Number of observation symbols\\
    $Q$ & Distinct states of the Markov process, $q_0,q_1,\ldots,q_{N-1}$\\
    $V$ & Possible observations, assumed to be $0,1,\ldots,M-1$\\
    $A$ & State transition probabilities\\
    $B$ & Observation probability matrix\\
    $\pi$ & Initial state distribution\\
    $\O$ & Observation sequence, $\O_0,\O_1,\ldots,\O_{T-1}$ \\ \midrule\midrule
  \end{tabular}
\end{table}

%For the temperature example in the previous section, the
%observations sequence is given in~\eref{eq:obs}, and we have
%$T=4$, $N=2$, $M=3$, $Q=\{H,C\}$, and~$V=\{0,1,2\}$. Note that we 
%let~$0,1,2$ represent small, medium, and large
%tree rings, respectively. For this example, the matrices~$A$, $B$, 
%and~$\pi$ are given by~\eref{eq:A}, 
%\eref{eq:B}, and~\eref{eq:pi}, respectively.

The matrix~$A=\{a_{ij}\}$ is~$N\times N$ with
$$
  a_{ij} = P(\mbox{state } q_j \mbox{ at } t+1\,|\,
           \mbox{state }q_i \mbox{ at } t) .
$$
The matrix~$A$ is row stochastic, that is, each row
satisfies the properties of a discrete probability distribution. Also,
the probabilities~$a_{ij}$ are independent of~$t$,
and hence the~$A$ matrix does not vary with~$t$.
The matrix $B=\{b_j(k)\}$ is of size~$N\times M$, with
$$
  b_j(k) = P(\mbox{observation } k \mbox{ at } t\,|\,
           \mbox{state } q_j \mbox{ at } t) .
$$
As with the~$A$ matrix, $B$ is row stochastic,
and the probabilities~$b_j(k)$ are independent of~$t$.
The somewhat unusual notation~$b_j(k)$ is convenient
when specifying the HMM algorithms.

An HMM is defined by~$A$, $B$, and~$\pi$
(and, implicitly, by the dimensions~$N$ and~$M$). 
Thus, we denote an HMM as~$\lambda = (A,B,\pi)$.

Suppose that we are given an observation sequence
of length four, that is,
$$
  \O = \bigl( \O_0,\O_1,\O_2,\O_3 \bigr) .
$$
Then the corresponding (hidden) state sequence is denoted as
$$
  X = \bigl( X_0,X_1,X_2,X_3 \bigr) .
$$
We let~$\pi_{\!\lowsub{1.5pt}{X_{0}}}$ denote the probability of starting in 
state~$X_0$, and~$b_{\lowsub{1.5pt}{X_0}}(\O_0)$ denotes the probability of
initially observing~$\O_0$, 
while~$a_{\lowsub{1.5pt}{X_0,X_1}}$ is the probability of transiting
from state~$X_0$ to state~$X_1$.
Continuing, we see that the probability of a
given state sequence~$X$ of length four is
\begin{equation}\label{eq:Px}
  P(X,\O) = \pi_{\!\lowsub{1.5pt}{X_{0}}}b_{\lowsub{1.5pt}{X_{0}}}(\O_0)
         a_{\lowsub{1.5pt}{X_{0},X_{1}}}b_{\lowsub{1.5pt}{X_{1}}}(\O_1)
         a_{\lowsub{1.5pt}{X_{1},X_{2}}}b_{\lowsub{1.5pt}{X_{2}}}(\O_2)
         a_{\lowsub{1.5pt}{X_{2},X_{3}}}b_{\lowsub{1.5pt}{X_{3}}}(\O_3) .
\end{equation}
Note that in this expression, the~$X_i$ represent indices
in the~$A$ and~$B$ matrices, not the names of the corresponding 
states.%\footnote{Your kindly author sincerely regrets this slight abuse of notation.}

To find the optimal state sequence in the
dynamic programming (DP)
sense, we simply choose the sequence (of length four, in this case)
with the highest probability. In contrast,
to find the optimal state sequence in the HMM sense,
we choose the most probable symbol at each position.
The optimal DP sequence and
the optimal HMM sequence can differ.
% and all state transitions are valid.

%Next, we discuss the three problems that can be solved using HMMs.
%Then we turn our attention to a detailed discussion of the the algorithms
%used to solve each of these three problems.

\subsubsection{The Three Problems}\label{sect:HMM3probs}

There are three fundamental problems\index{hidden Markov model!three problems}
that we can solve using HMMs.
Here, we briefly describe each of these problems.
%then in the next section we
%discuss efficient algorithms for their solution.

\begin{description}

\item[{\bf Problem~1}]
Given the model~$\lambda = (A,B,\pi)$ and 
a sequence of observations~$\O$, determine~$P(\O\,|\,\lambda)$.
That is, we want to compute a score for the observed 
sequence~$\O$ with respect to the given model~$\lambda$. 

\item[{\bf Problem~2}]
Given $\lambda = (A,B,\pi)$ and an observation sequence~$\O$,
find an optimal state sequence for the underlying Markov process.
In other words, we want to uncover the  
hidden part of the hidden Markov model. 

\item[{\bf Problem~3}]
Given an observation sequence~$\O$ and the 
parameter~$N$, determine a model~$\lambda = (A,B,\pi)$ 
that maximizes the probability of~$\O$. This can be 
viewed as training a model to best fit the observed data. 
This problem is generally solved using Baum-Welch 
re-estimation~\cite{LRR,Stamp04arevealing}, which is a discrete hill climb 
on the parameter space
represented by~$A$, $B$, and~$\pi$.
There is also an alternative gradient ascent
technique for HMM training~\cite{BaldiChauvin,Stamp19deep}.

\end{description}

Since the technique we use to train an HMM (problem~3) is a hill climb,
in general, we obtain a local maximum. Training with
different initial conditions can result in different 
local maxima, and hence it is often beneficial to train 
multiple HMMs with different initial conditions,
and select the highest scoring model.

\subsubsection{Example}

Consider, for example, the problem of speech recognition which,
not coincidentally, is one of the earliest and best-known
successes of HMMs. In speech problems, the hidden states can be 
viewed as corresponding to movements of the vocal cords, which 
are not directly observed. Instead, we observe the sounds that are
produced, and extract training features from these sounds.
In this scenario, we can use the solution to HMM Problem~3 to
train an HMM~$\lambda$ to, for example, recognize the spoken word ``yes.''
Then, given an unknown spoken word, we can use the 
solution to Problem~1 to score the word 
against the trained model~$\lambda$ and determine 
the likelihood that the word is~``yes.''
In this case, we do not need to solve Problem~2, but it
is possible that such a solution (i.e., uncovering the hidden
states) might provide additional insight into the underlying
speech model. 

English text analysis is another classic application of 
HMMs, which appears to have been first considered
by Cave and Neuwirth~\cite{CaveNeuwirth}.
This application nicely illustrates the strength of HMMs
and it requires no background in any specialized 
field, such as speech processing or information security.

Given a length of English text, we
remove all punctuation, numbers, etc., and converts
all letters to lower case. This leaves~26 distinct
letters and word-space, for a total of~27
symbols. We assume that
there is an underlying Markov process (of order one)
with two hidden states. For each of these two hidden states, we
assume that the~27 symbols are observed according 
to fixed probability distributions.

This defines an HMM with~$N=2$ and~$M=27$, where 
the state transition probabilities of the~$A$ matrix
and the observation probabilities of the~$B$ 
matrix are unknown, while the observations~$\O_t$
consist of the series of characters we have extracted from the given text. 
To determine the~$A$ and~$B$ matrices, we
must solve HMM Problem~3, as discussed above.

We have trained such an HMM,
using the first~$T=50{,}000$ observations 
from the Brown Corpus,\footnote{Officially,
it is the Brown University Standard Corpus of Present-Day American English,
which includes various texts totaling about~1,000,000 words. Here, ``Present-Day'' 
means~1961.}
which is available at~\cite{BrownCorpus}.
We initialized each element of~$\pi$ and~$A$ randomly 
to approximately~$1/2$, taking care to sure that the matrices
are row stochastic. For one specific iteration of this
experiment, the precise values used were
$$
  \pi = \rowvecc{0.51316}{0.48684}
$$
and
$$
  A = \left(
  \begin{array}{cc}
    0.47468 & 0.52532\\
    0.51656 & 0.48344
  \end{array}\right) \! .
$$
Each element of~$B$ was initialized to
approximately~$1/27$, again, under the constraint
that~$B$ must be row stochastic. The values in 
the initial~$B$ matrix (more precisely, the transpose
of~$B$) appear in the second and third 
columns of Table~\ref{tab:bt}.

\begin{table}[!htb]
  \caption{Initial and final $B^T$}\label{tab:bt}
  \vglue 0.125in
  \centering
  {\small
  \begin{tabular}{c|cc|cc} \midrule\midrule
    \textbf{Observation} & \multicolumn{2}{c|}{\textbf{Initial}}
    	 & \multicolumn{2}{c}{\textbf{Final}}\\ \midrule
    {\tt a}  & 0.03735  & 0.03909 & 0.13845 & 0.00075\\
    {\tt b}  & 0.03408  & 0.03537 & 0.00000 & 0.02311\\
    {\tt c}  & 0.03455  & 0.03537 & 0.00062 & 0.05614\\
    {\tt d}  & 0.03828  & 0.03909 & 0.00000 & 0.06937\\
    {\tt e}  & 0.03782  & 0.03583 & 0.21404 & 0.00000\\
    {\tt f}  & 0.03922  & 0.03630 & 0.00000 & 0.03559\\
    {\tt g}  & 0.03688  & 0.04048 & 0.00081 & 0.02724\\
    {\tt h}  & 0.03408  & 0.03537 & 0.00066 & 0.07278\\
    {\tt i}  & 0.03875  & 0.03816 & 0.12275 & 0.00000\\
    {\tt j}  & 0.04062  & 0.03909 & 0.00000 & 0.00365\\
    {\tt k}  & 0.03735  & 0.03490 & 0.00182 & 0.00703\\
    {\tt l}  & 0.03968  & 0.03723 & 0.00049 & 0.07231\\
    {\tt m}  & 0.03548  & 0.03537 & 0.00000 & 0.03889\\
    {\tt n}  & 0.03735  & 0.03909 & 0.00000 & 0.11461\\
    {\tt o}  & 0.04062  & 0.03397 & 0.13156 & 0.00000\\
    {\tt p}  & 0.03595  & 0.03397 & 0.00040 & 0.03674\\
    {\tt q}  & 0.03641  & 0.03816 & 0.00000 & 0.00153\\
    {\tt r}  & 0.03408  & 0.03676 & 0.00000 & 0.10225\\
    {\tt s}  & 0.04062  & 0.04048 & 0.00000 & 0.11042\\
    {\tt t}  & 0.03548  & 0.03443 & 0.01102 & 0.14392\\
    {\tt u}  & 0.03922  & 0.03537 & 0.04508 & 0.00000\\
    {\tt v}  & 0.04062  & 0.03955 & 0.00000 & 0.01621\\
    {\tt w}  & 0.03455  & 0.03816 & 0.00000 & 0.02303\\
    {\tt x}  & 0.03595  & 0.03723 & 0.00000 & 0.00447\\
    {\tt y}  & 0.03408  & 0.03769 & 0.00019 & 0.02587\\
    {\tt z}  & 0.03408  & 0.03955 & 0.00000 & 0.00110\\
   space    & 0.03688  & 0.03397 & 0.33211 & 0.01298\\ \midrule\midrule 
  \end{tabular}
  }
\end{table}

After the initial iteration, we find
$
  \log\bigl(P(\O\,|\lambda)\bigr) = -165097.29
$
and after~100 iterations, we have
$
  \log\bigl(P(\O\,|\,\lambda)\bigr) = -137305.28 .
$
These model scores indicate that training has improved the model significantly 
over the~100 iterations.
%\footnote{Provided a model converges, 
%we typically see~$\log\bigl(P(\O\,|\lambda)\bigr)$ improve slowly 
%over an initial series of iterations, followed by
%a period of rapid convergence, followed again by negligible improvement. 
%Once we reach this second plateau, the hill climb is complete.}

In this particular experiment, after~100 iterations, 
the model~$\lambda=(A,B,\pi)$ has converged to
$$
  \pi = \rowvecc{0.00000}{1.00000} \mbox{\ \ and\ \ }
  A =
  \left(
  \begin{array}{cc}
    0.25596 & 0.74404\\
    0.71571 & 0.28429
  \end{array}
  \right)
$$
with the converged~$B^T$ appearing in the last two
columns of Table~\ref{tab:bt}. 

The most interesting part of an HMM is generally the~$B$ 
matrix. Without having made any assumption about the
two hidden states, the~$B$ matrix in Table~\ref{tab:bt} 
shows us that one hidden state consists of vowels
while the other hidden state
consists of consonants. Curiously, from this perspective,
word-space acts more like a vowel,
while~{\tt y} is not even sometimes a vowel.

Of course, anyone familiar with English
would not be surprised that there 
is a significant distinction between
vowels and consonants. But, the crucial point here
is that the HMM has
automatically extracted this statistically important
distinction for us---it has ``learned'' to distinguish
between consonants and vowels. And, thanks
to HMMs, this feature of English text could be 
easily discovered by someone who 
previously had no knowledge whatsoever of the language.

Cave and Neuwirth~\cite{CaveNeuwirth} obtain additional
results when considering HMMs with more than
two hidden states. In fact, they are able to 
sensibly interpret the results for models
with up to~$N=12$ hidden states.

For more information on HMMs,
see~\cite{Stamp04arevealing}, which includes detailed algorithms
including scaling, or Rabiner's classic paper~\cite{LRR}.

\subsection{Principal Component Analysis}\label{sect:PCA}

Principal component analysis (PCA) is a linear algebraic technique that provides
a powerful tool for dimensionality reduction. Here, we provide a very brief
introduction to the topic; for more details, Shlens' tutorial
is highly recommended~\cite{Shlens05}, while a good sources for the
math behind PCA is~\cite{Shalizi}. The discussion at~\cite{StackExchange}
provides a brief, intuitive, and fun introduction to the subject.

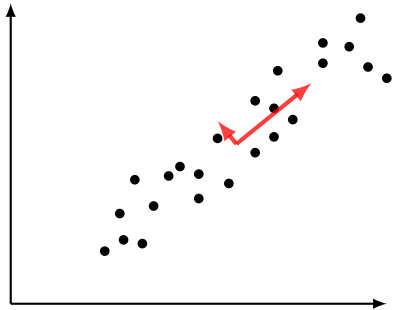
\begin{figure}[!htb]
    \begin{center}
    \begin{tikzpicture}[scale=1,>=latex]
 
%    \draw[ultra thick,color=red] (0,0) -- (0,0.5);

    % circles
%    \draw[thick,color=black,fill=black] (0.5,0.5) circle (0.05);    
%    \draw[thick,color=black,fill=black] (0.75,0.6) circle (0.05);    
    \draw[thick,color=black,fill=black] (1.25,0.7) circle (0.05);    
    \draw[thick,color=black,fill=black] (1.5,0.85) circle (0.05);    
    \draw[thick,color=black,fill=black] (1.65,1.65) circle (0.05);    
    \draw[thick,color=black,fill=black] (2.5,1.725) circle (0.05);    
    \draw[thick,color=black,fill=black] (2.25,1.825) circle (0.05);    
    \draw[thick,color=black,fill=black] (2.75,2.2) circle (0.05);    
    \draw[thick,color=black,fill=black] (3.25,2.7) circle (0.05);    
    \draw[thick,color=black,fill=black] (3.5,2.6) circle (0.05);    
    \draw[thick,color=black,fill=black] (3.55,3.1) circle (0.05);    
    \draw[thick,color=black,fill=black] (4.15,3.47) circle (0.05);    
    \draw[thick,color=black,fill=black] (4.65,3.8) circle (0.05);    

    \draw[thick,color=black,fill=black] (1.75,0.8) circle (0.05);    
    \draw[thick,color=black,fill=black] (1.45,1.2) circle (0.05);    
    \draw[thick,color=black,fill=black] (1.9,1.3) circle (0.05);    
    \draw[thick,color=black,fill=black] (2.5,1.4) circle (0.05);    
    \draw[thick,color=black,fill=black] (2.9,1.6) circle (0.05);    
    \draw[thick,color=black,fill=black] (2.1,1.7) circle (0.05);    
    \draw[thick,color=black,fill=black] (3.25,2.01) circle (0.05);    
    \draw[thick,color=black,fill=black] (3.5,2.22) circle (0.05);    
    \draw[thick,color=black,fill=black] (3.75,2.45) circle (0.05);    
    \draw[thick,color=black,fill=black] (4.15,3.2) circle (0.05);    
    \draw[thick,color=black,fill=black] (4.5,3.42) circle (0.05);    
    \draw[thick,color=black,fill=black] (4.75,3.15) circle (0.05);    

    \draw[thick,color=black,fill=black] (5,3.0) circle (0.05);
    
    % axes
    \draw[thick,color=black,->] (0,0) -- (5,0); % x axis
    \node at (0,0) {\phantom{$x$}};
%    \node[rotate=0] at (2.5,-0.4){$1 - \mbox{specificity}$};
    \draw[thick,color=black,->] (0,0) -- (0,4); % y axis
%    \node[rotate=90] at (-0.4,2.5){$\mbox{sensitivity}$};

    \draw[ultra thick,color=red,opacity=0.75,->] (3.0,2.125) -- (4.0,2.9375);
    \draw[ultra thick,color=red,opacity=0.75,->] (3.0,2.125) -- (2.75,2.4325);
%    \draw[ultra thick,color=red,opacity=0.75,->] (3.0,2.15) -- (4.15,3.0);
%    \draw[ultra thick,color=red,opacity=0.75,->] (3.0,2.15) -- (2.75,2.48);

    % grid (aid to plotting)
%    \draw[step=0.5,gray,very thin] (0,0) grid (5,5);

    \end{tikzpicture}
  \end{center}
  \vglue-0.125in
  \caption{A better basis}\label{fig:PCA_basis}
\end{figure}

Geometrically, PCA aligns a basis with the (orthogonal) directions 
having the largest variances.
These directions are defined to be the principal components.
A simple illustration of such a change of basis 
appears in Figure~\ref{fig:PCA_basis}.

Intuitively, larger variances correspond to more informative data---if the variance is small,
the training data is clumped tightly around the mean and we have
limited ability to distinguish between samples. In contrast, if the 
variance is large, there is a much better chance of separating the samples based
on the characteristic (or characteristics) under consideration. 
Consequently, once we have aligned the
basis with the variances, we can ignore those directions that correspond
to small variances without losing significant information. In fact, small variances
often contribute only noise, in which cases we can actually improve our results 
by neglecting those directions that correspond to small variances.

The linear algebra behind PCA training (i.e., deriving a new-and-improved basis)
is fairly deep, involving eigenvalue analysis. Yet, the scoring phase is simplicity itself, 
requiring little more than the computation of a few dot products, 
which makes scoring extremely efficient and practical.

Note that we treat singular value decomposition (SVD) 
as a special case of PCA, in the sense that SVD provides a method for
determining the principal components.
It is possible to take the opposite perspective, where PCA is viewed as a special case
of the general change of basis technique provided by SVD. In any case, for our purposes, 
PCA and SVD can be considered as essentially synonymous.

\subsection{Classifiers}\label{sect:classifiers}

In the research presented in this paper, we consider four different classifiers,
namely, $k$-nearest neighbors ($\kNN{k}$),
multilayer perceptron (MLP), random forest (RF), and support vector machine (SVM).
We have already discussed MLPs above, so in this section, we give 
a brief overview of $\kNN{k}$, RF, and SVM.

\subsubsection{$k$-Nearest Neighbors}

Perhaps the simplest possible machine learning algorithm is $k$-nearest neighbors ($\kNN{k}$). 
In the scoring phase, $\kNN{k}$ consists of classifying based on the $k$ nearest samples
in the training set, typically using a simple majority vote. 
Since all computation is deferred to the scoring phase, $\kNN{k}$ 
is considered to be a ``lazy learner.''

\begin{figure}[!htb]
\centering
  \begin{tabular}{ccc}
	\begin{tikzpicture}[scale=0.7, every node/.style={scale=0.9}]
        
    % circles
    \draw[thick,color=red,fill=red] (0.5,3) circle (0.08);
    \draw[thick,color=red,fill=red] (1.0,4.25) circle (0.08);
    \draw[thick,color=red,fill=red] (1.5,2.0) circle (0.08);
    \draw[thick,color=red,fill=red] (2.0,2.75) circle (0.08);%
%    \node[color=black] at (1.75,3.0){$r_1$};
    \draw[thick,color=red,fill=red] (2.5,1.65) circle (0.08);
    \draw[thick,color=red,fill=red] (3.0,2.7) circle (0.08);%
%    \node[color=black] at (3.25,3.0){$r_2$};
    \draw[thick,color=red,fill=red] (3.5,1.0) circle (0.08);
    \draw[thick,color=red,fill=red] (4.0,2.5) circle (0.08);
    \draw[thick,color=red,fill=red] (4.5,2.1) circle (0.08);
    \draw[thick,color=red,fill=red] (5.0,2.75) circle (0.08);

    % squares
%    \draw[thick,color=blue] (0.5,1.75) rectangle (0.65,1.9);%%%
%    \draw[thick,color=blue] (1.0,1.25) rectangle (1.15,1.4);
%    \draw[thick,color=blue] (1.5,1.5) rectangle (1.65,1.65);%%%
    \draw[thick,color=blue] (2.5,4.0) rectangle (2.65,4.15);%
    \node[color=black] at (2.575,4.45){$b$};
    \draw[thick,color=blue] (2.5,2.1) rectangle (2.65,2.25);
%    \draw[thick,color=blue] (3.0,1.5) rectangle (3.15,1.65);%%%
    \draw[thick,color=blue] (3.5,1.85) rectangle (3.65,2.0);
    \draw[thick,color=blue] (4.0,3.5) rectangle (4.15,3.65);
%    \draw[thick,color=blue] (4.5,0.95) rectangle (4.65,1.1);
%    \draw[thick,color=blue] (5.0,1.45) rectangle (5.15,1.6);%%%

    % diamond
    \draw[thick,color=green,fill=green,rotate around={45:(2.5,3.5)}] (2.5,3.5) rectangle (2.7,3.7);
    \node[color=black] at (2.125,3.65){$X$};

    \draw[thick,color=green,dashed] (2.5,3.6) circle (0.675);

    % axes
    \draw[thick,color=black,->] (0,0) -- (6,0); % x axis
    \draw[thick,color=black,->] (0,0) -- (0,5); % y axis
    \node at (0,0) {\phantom{$x$}};

\end{tikzpicture}
	& \ \ \ \ \ &
	\begin{tikzpicture}[scale=0.7, every node/.style={scale=0.9}]
        
    % circles
    \draw[thick,color=red,fill=red] (0.5,3) circle (0.08);
    \draw[thick,color=red,fill=red] (1.0,4.25) circle (0.08);
    \draw[thick,color=red,fill=red] (1.5,2.0) circle (0.08);
    \draw[thick,color=red,fill=red] (2.0,2.75) circle (0.08);%
    \node[color=black] at (1.75,3.0){$r_1$};
    \draw[thick,color=red,fill=red] (2.5,1.65) circle (0.08);
    \draw[thick,color=red,fill=red] (3.0,2.7) circle (0.08);%
    \node[color=black] at (3.25,3.0){$r_2$};
    \draw[thick,color=red,fill=red] (3.5,1.0) circle (0.08);
    \draw[thick,color=red,fill=red] (4.0,2.5) circle (0.08);
    \draw[thick,color=red,fill=red] (4.5,2.1) circle (0.08);
    \draw[thick,color=red,fill=red] (5.0,2.75) circle (0.08);

    % squares
%    \draw[thick,color=blue] (0.5,1.75) rectangle (0.65,1.9);%%%
%    \draw[thick,color=blue] (1.0,1.25) rectangle (1.15,1.4);
%    \draw[thick,color=blue] (1.5,1.5) rectangle (1.65,1.65);%%%
    \draw[thick,color=blue] (2.5,4.0) rectangle (2.65,4.15);%
    \node[color=black] at (2.575,4.45){$b$};
    \draw[thick,color=blue] (2.5,2.1) rectangle (2.65,2.25);
%    \draw[thick,color=blue] (3.0,1.5) rectangle (3.15,1.65);%%%
    \draw[thick,color=blue] (3.5,1.85) rectangle (3.65,2.0);
    \draw[thick,color=blue] (4.0,3.5) rectangle (4.15,3.65);
%    \draw[thick,color=blue] (4.5,0.95) rectangle (4.65,1.1);
%    \draw[thick,color=blue] (5.0,1.45) rectangle (5.15,1.6);%%%

    % diamond
    \draw[thick,color=green,fill=green,rotate around={45:(2.5,3.5)}] (2.5,3.5) rectangle (2.7,3.7);
    \node[color=black] at (2.125,3.65){$X$};

    \draw[thick,color=green,dashed] (2.5,3.6) circle (1.25);

    % axes
    \draw[thick,color=black,->] (0,0) -- (6,0); % x axis
    \draw[thick,color=black,->] (0,0) -- (0,5); % y axis
    \node at (0,0) {\phantom{$x$}};

\end{tikzpicture}
	\\
	(a) 1-nearest neighbor & & (b) 3-nearest neighbors
  \end{tabular}	
\caption{Examples of $\kNN{k}$ classification~\cite{StampML2017}}\label{fig:knn}
\end{figure}
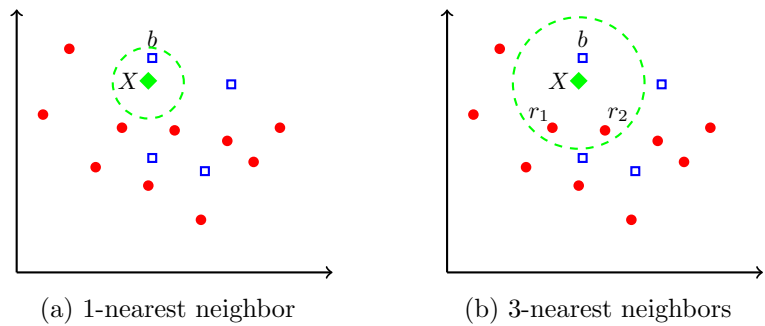

Figure~\ref{fig:knn} shows examples of $\kNN{k}$, where the training 
data consists of two classes, represented by the open blue squares and the
solid red circles, with the green diamond (the point labeled~$X$) being
a point that we want to classify. 
Figure~\ref{fig:knn}~(a), shows that if we use the 1-nearest
neighbor, we would classify the green diamond 
as being of same type as the open blue squares, whereas Figure~\ref{fig:knn}~(b)
shows that~$X$ would be classified as the solid red circle type if using the
3-nearest neighbors. 
%The \kNN\ technique is said to be a ``lazy'' learner, since 
%there is no training phase---all work is deferred to the scoring phase.

\subsubsection{Random Forest}

A random forest (RF) generalizes a simple decision tree algorithm. 
A decision tree is constructed by building a tree, based on 
features from the training data. It is easy to construct such trees,
and trivial to classify samples once a tree has been constructed. However, 
decision trees tend to overfit the input data. 

An RF combines multiple decision trees to generalize the training data. 
To do so, RFs use different subsets of the training data as well as 
different subsets of features, a process known as bagging~\cite{StampML2017}.
A simple majority vote of the decision trees comprising the RF is
typically used for classification~\cite{liaw2002classification}.

\subsubsection{Support Vector Machine}

Support vector machines (SVM) are a class of 
supervised learning methods that are based on four major ideas,
namely, a separating hyperplane, maximizing the ``margin'' (i.e., separation between classes), 
working in a higher-dimensional space, and the so-called kernel trick. The goal in SVM is to
use a hyperplane to separate labeled data into two classes. If it exists, such a hyperplane 
is chosen to maximize the margin~\cite{StampML2017}.
  
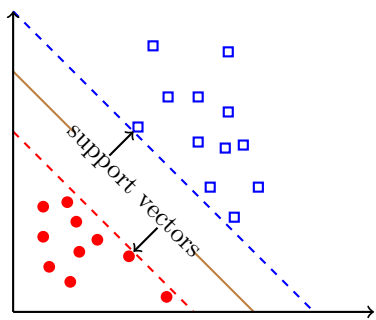
\begin{figure}[!htb]
\centering
\begin{tikzpicture}[scale=0.8]
    
    % squares
    \draw[thick,color=blue] (4,2) rectangle (4.15,2.15);
    \draw[thick,color=blue] (3.5,4.25) rectangle (3.65,4.4);
    \draw[thick,color=blue] (3.2,2.0) rectangle (3.35,2.15);
    \draw[thick,color=blue] (3.0,2.75) rectangle (3.15,2.9);
    \draw[thick,color=blue] (3.45,2.65) rectangle (3.6,2.8);
    \draw[thick,color=blue] (3.75,2.7) rectangle (3.9,2.85);
    \draw[thick,color=blue] (3.5,3.25) rectangle (3.65,3.4);
    \draw[thick,color=blue] (3.0,3.5) rectangle (3.15,3.65);
    \draw[thick,color=blue] (2,3) rectangle (2.15,3.15);
    \draw[thick,color=blue] (2.5,3.5) rectangle (2.65,3.65);
    \draw[thick,color=blue] (2.25,4.35) rectangle (2.4,4.5);
    \draw[thick,color=blue] (3.6,1.5) rectangle (3.75,1.65);
    
    % circles
    \draw[thick,color=red,fill=red] (1.4,1.2) circle (0.08);
    \draw[thick,color=red,fill=red] (0.6,0.75) circle (0.08);
    \draw[thick,color=red,fill=red] (0.95,0.5) circle (0.08);
    \draw[thick,color=red,fill=red] (1.925,0.925) circle (0.08);
    \draw[thick,color=red,fill=red] (0.5,1.25) circle (0.08);
    \draw[thick,color=red,fill=red] (1.1,1.0) circle (0.08);
    \draw[thick,color=red,fill=red] (2.55,0.25) circle (0.08);
    \draw[thick,color=red,fill=red] (1.05,1.5) circle (0.08);
    \draw[thick,color=red,fill=red] (0.9,1.825) circle (0.08);
    \draw[thick,color=red,fill=red] (0.5,1.75) circle (0.08);
    
    % observations
%    \node at (2.5,2.5){$\O_0$};

    % grid (aid to plotting)
%    \draw[step=0.5,gray,very thin] (0,0) grid (5,5);
    
    % diagonal lines
%    \draw[thick,color=brown] (0,4) -- (4,0); % good
    \draw[thick,color=brown] (0,4) -- (1,3); % good
    \draw[thick,color=brown] (3,1) -- (4,0); % good
%    \draw[thick,dashed,color=green] (0,2.2) -- (5.5,1.0); % bad
    \draw[thick,dashed,color=blue] (0,5) -- (5,0); % upper
    \draw[thick,dashed,color=red] (0,3) -- (3,0); % lower

    % margin
    \draw[thick,color=black,->] (2.4,1.4) -- (2,1); % lower
    \draw[thick,color=black,->] (1.6,2.6) -- (2,3); % upper
    
%    \draw[very thick,color=black,<->] (0.25,2.75) -- (1.25,3.75); % m line segment
%    \node at (0.8,3.65){$m$};

    \node[rotate=-45] at (2,2){support vectors};
%    \node at (-2.25,0.0){support vectors};
%    \draw[thick,color=gray,smooth,->] plot coordinates {(-0.9,0.125) (0,2.5) (1.99,3.075)}; 
%    \draw[thick,color=gray,smooth,->] plot coordinates { (-0.825,-0.025) (0.85,-0.2) (1.885,0.885)};    
    
    % axes
     \draw[thick,color=black,->] (0,0) -- (6,0); % x axis
     \draw[thick,color=black,->] (0,0) -- (0,5); % y axis
     \node at (0,0) {\phantom{$x$}};

\end{tikzpicture}
\caption{Support vectors in SVM~\cite{StampML2017}}\label{fig:svm}
\end{figure}

An example of a trained SVM is illustrated in Figure~\ref{fig:svm}. Note that the
points that actually minimize the distance to the separating hyperplane
correspond to support vectors. In general, the number of support vectors
will be small relative to the number of training data points,
and this is the key to the efficiency of SVM in the classification phase.

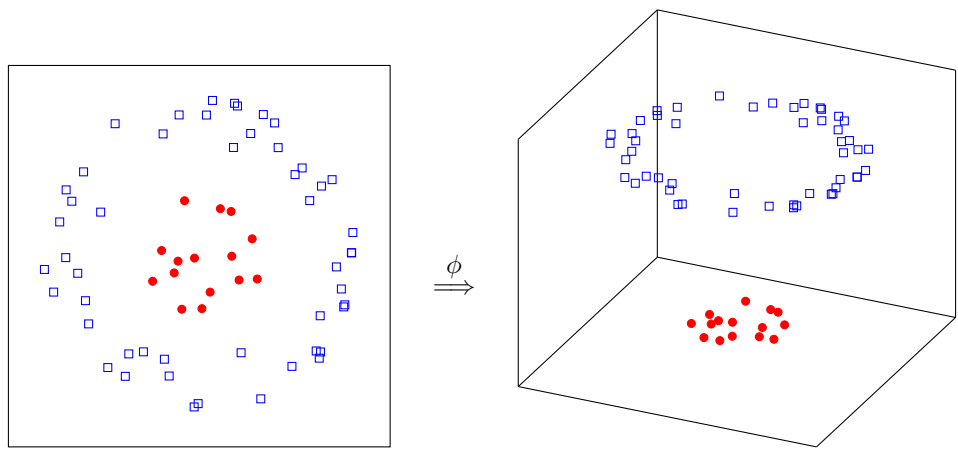
\begin{figure}[!htb]
\centering
%\begin{figure}[!htb]
%  \begin{center}
  \begin{tabular}{ccc}
  \begin{tikzpicture} %[declare function={a(\x)=0.5*\x-2;},declare function={b(\x)=0.5*\x+2;}]
    \begin{axis}[width=0.45\textwidth,height=0.45\textwidth,xmin=-3,xmax=3,ymin=-3,ymax=3] 
      \pgfplotsset{ticks=none}
      
%      \addplot [color=red, only marks, mark=*, samples=10]
%          {0.5*(a(x)+b(x)) + 0.5*rand*(a(x)-b(x))};
       \addplot[color=red, mark=*,only marks,mark size=1.5] coordinates { 
(0.331578, 0.746603)
(-0.394443, -0.263352)
(0.625403, -0.375284)
(0.911628, -0.359273)
(0.500405, 0.703271)
(-0.231736, 0.872177)
(0.170166, -0.565556)
(-0.274232, -0.834498)
(-0.334136, -0.078180)
(-0.592072, 0.089388)
(-0.075179, -0.030616)
(0.041852, -0.825840)
(-0.732711, -0.394369)
(0.511456, -0.000657)
(0.832060, 0.272873)
      };
      \addplot[color=blue, mark=square,fill=blue,only marks,mark size=1.5] coordinates { 
(2.390971, 0.052211)
(-1.437509, -1.752351)
(1.886234, -1.604676)
(2.395163, 0.054876)
(2.288330, -0.765636)
(0.807904, 1.929355)
(0.207518, 2.449499)
(1.900863, -0.937399)
(-0.548196, -1.621026)
(-1.322488, 2.083358)
(1.006897, 2.228109)
(1.735974, 0.874161)
(1.838050, -1.491165)
(0.600720, 2.362412)
(1.456580, -1.733536)
(-0.316657, 2.221768)
(2.087640, 1.203409)
(1.616520, 1.387743)
(-2.433935, -0.209267)
(-1.788833, -0.700340)
(-0.877470, -1.504627)
(-1.107672, -1.537246)
(1.184006, 2.095965)
(-0.080668, -2.371692)
(2.414325, 0.371814)
(0.554228, 2.404566)
(1.505664, 1.283217)
(-1.162894, -1.890363)
(-2.098915, -0.022764)
(1.920753, 1.102422)
(-2.288927, -0.565147)
(1.905982, -1.506967)
(2.233564, -0.516066)
(-2.003422, 0.863837)
(-1.910277, -0.269332)
(0.966902, -2.244612)
(-0.568292, 1.923142)
(0.112558, 2.219818)
(-1.549785, 0.692908)
(-2.193879, 0.538788)
(-1.818708, 1.326454)
(2.270850, -0.798137)
(-2.092595, 1.043138)
(2.155481, -0.168913)
(-0.017986, -2.319378)
(1.239888, 1.707677)
(-0.473226, -1.885332)
(0.537416, 1.709329)
(-1.739303, -1.065284)
(0.657574, -1.519209)
      };
    \end{axis}
  \end{tikzpicture}  
%  & $\displaystyle\genfrac{}{}{0pt}{}{\phi}{\implies}$ &
%  & $\displaystyle\overset{\displaystyle\phi}{\displaystyle\implies}$ &
%  & ${{\displaystyle\phi} \above 0pt {\displaystyle\implies}}$ &
  & \raisebox{0.85in}{${{\displaystyle\phi} \atop {\displaystyle\implies}}$} &
    \begin{tikzpicture}
    \begin{axis}[width=0.5\textwidth,height=0.5\textwidth,xmin=-3,xmax=3,ymin=-3,ymax=3,zmin=0,zmax=5] 
      \pgfplotsset{ticks=none}
%      \addplot3[mesh,color=blue,opacity=0.2]  table {Part_1/Chapter_5/out5.dat};
%      \addplot3[black,very thick]  table {Part_1/Chapter_5/out.dat};
      \addplot3[color=red, mark=*,only marks,mark size=1.5] coordinates { 
(0.331578, 0.746603,0.6)
(-0.394443, -0.263352,0.6)
(0.625403, -0.375284,0.6)
(0.911628, -0.359273,0.6)
(0.500405, 0.703271,0.6)
(-0.231736, 0.872177,0.6)
(0.170166, -0.565556,0.6)
(-0.274232, -0.834498,0.6)
(-0.334136, -0.078180,0.6)
(-0.592072, 0.089388,0.6)
(-0.075179, -0.030616,0.6)
(0.041852, -0.825840,0.6)
(-0.732711, -0.394369,0.6)
(0.511456, -0.000657,0.6)
(0.832060, 0.272873,0.6)
      };
      \addplot3[color=blue, mark=square,fill=blue,only marks,mark size=1.5] coordinates { 
(2.390971, 0.052211,4.0)
(-1.437509, -1.752351,4.0)
(1.886234, -1.604676,4.0)
(2.395163, 0.054876,4.0)
(2.288330, -0.765636,4.0)
(0.807904, 1.929355,4.0)
(0.207518, 2.449499,4.0)
(1.900863, -0.937399,4.0)
(-0.548196, -1.621026,4.0)
(-1.322488, 2.083358,4.0)
(1.006897, 2.228109,4.0)
(1.735974, 0.874161,4.0)
(1.838050, -1.491165,4.0)
(0.600720, 2.362412,4.0)
(1.456580, -1.733536,4.0)
(-0.316657, 2.221768,4.0)
(2.087640, 1.203409,4.0)
(1.616520, 1.387743,4.0)
(-2.433935, -0.209267,4.0)
(-1.788833, -0.700340,4.0)
(-0.877470, -1.504627,4.0)
(-1.107672, -1.537246,4.0)
(1.184006, 2.095965,4.0)
(-0.080668, -2.371692,4.0)
(2.414325, 0.371814,4.0)
(0.554228, 2.404566,4.0)
(1.505664, 1.283217,4.0)
(-1.162894, -1.890363,4.0)
(-2.098915, -0.022764,4.0)
(1.920753, 1.102422,4.0)
(-2.288927, -0.565147,4.0)
(1.905982, -1.506967,4.0)
(2.233564, -0.516066,4.0)
(-2.003422, 0.863837,4.0)
(-1.910277, -0.269332,4.0)
(0.966902, -2.244612,4.0)
(-0.568292, 1.923142,4.0)
(0.112558, 2.219818,4.0)
(-1.549785, 0.692908,4.0)
(-2.193879, 0.538788,4.0)
(-1.818708, 1.326454,4.0)
(2.270850, -0.798137,4.0)
(-2.092595, 1.043138,4.0)
(2.155481, -0.168913,4.0)
(-0.017986, -2.319378,4.0)
(1.239888, 1.707677,4.0)
(-0.473226, -1.885332,4.0)
(0.537416, 1.709329,4.0)
(-1.739303, -1.065284,4.0)
(0.657574, -1.519209,4.0)
      };
    \end{axis}
  \end{tikzpicture}  
  \end{tabular}
%      \includegraphics[width=0.9\textwidth]{Part_1/Chapter_5/x_data_2d_to_3d.jpg}
%  \end{center}
%  \caption{Transformation from~2-d to~3-d\label{fig:SVM3d}}
%\end{figure}
\caption{A function~$\phi$ illustrating the kernel trick~\cite{StampML2017}}\label{fig:svm2}
\end{figure}

Of course, there is no assurance that the training data will be linearly separable. 
In such cases,
a nonlinear kernel function can be embedded into the SVM process
in such a way that the input data is, in effect, 
transformed to a higher dimensional ``feature space.'' 
In this higher dimensional space, it is far more likely that the 
transformed data will be linearly separable. This is the essence of the 
kernel trick---an example of which
is illustrated in Figure~\ref{fig:svm2}. That we can transform our training data 
in such a manner is not surprising, 
but the fact that we can do so 
without paying any significant penalty in terms of computational efficiency 
makes the kernel trick a very powerful ``trick'' indeed. However,
the kernel function must be specified by the user, and selecting a (near) optimal
kernel can be challenging.

\subsubsection{Last Word on Classification Techniques}\label{sect:lastWord}

We note in passing that MLP and SVM are related techniques,
as both of these approaches generate nonlinear decision 
boundaries (assuming a nonlinear kernel). 
For SVM, the nonlinear boundary is based 
on a user-specified kernel function,
whereas the equivalent aspect of an MLP 
is learned as part of the training 
process---in effect, the ``kernel'' is learned when training
an MLP. This suggests that MLPs have an advantage, since
there are limitations on SVM kernels, and selecting
an optimal kernel is more art than science.
However, the tradeoff is that more data and more computation will 
generally be required to train a comparable MLP, since the MLP has
more to learn, in comparison to an SVM.

It is also the case that~$\kNN{k}$ and RF are closely related. In fact, both are 
neighborhood-based algorithms, but with neighborhood 
structures that are somewhat different~\cite{LinJeon}.

Thus, we generally expect that the results obtain using SVM and MLP 
will be qualitatively similar, and the same is true when comparing results obtained
using~$\kNN{k}$ and RF. 
%Consequently, the results obtained with
%SVM and MLP would be expected to, in general, differ more
%with the results obtained with~$\kNN{k}$ and RF than they would
%differ with each other, and conversely.
By using these four classifiers, we obtain
a ``sanity check'' on the results. If, for example,
our SVM and MLP results differ dramatically, this
would indicate that we should investigate further.
On the other hand, if, say, our MLP and~RF results differ significantly,
this would not raise the same level of concern.

\section{Word Embedding Techniques}\label{sect:wordEmbed}

Word embeddings are often used in natural language processing
as they provide a way to quantify relationships between words. Here, we use word
embeddings to generate higher-level features for malware classification.

In this section, we discuss three distinct word embedding techniques.
First, we consider word embeddings derived from trained HMMs, which we
refer to as HMM2Vec. Then we consider a word embedding technique
based on PCA, which we refer to as PCA2Vec. Finally, we discuss the 
popular neural network based technique known as Word2Vec.

\subsection{HMM2Vec}\label{sect:HMM2Vec}

Before discussing the basic ideas behind Word2Vec, we consider a somewhat
analogous approach to generating vector representations
based on hidden Markov models. To begin with
we consider individual letters, as opposed to words---we call
this simpler version Letter2Vec.

Recall that an HMM is defined by the three matrices~$A$, $B$, and~$\pi$, 
and is denoted as~$\lambda=(A,B,\pi)$. The~$\pi$ matrix contains
the initial state probabilities, $A$
contains the hidden state transition probabilities, 
and~$B$ consists of the observation 
probability distributions corresponding to the hidden states.
Each of these matrices is row stochastic, that is, each row satisfies the
requirements of a discrete probability distribution.
Notation-wise, $N$ is the number of hidden states,
$M$ is the number of distinct observation symbols, and~$T$ is the length
of the observation (i.e., training) sequence. 
Note that~$M$ and~$T$ are determined by the training data,
while~$N$ is a user-defined parameter. 

Suppose that we train an HMM on a sequence of letters extracted from English text,
where we convert all upper-case letters to lower-case and discard any character that is
not an alphabetic letter or word-space. Then~$M=27$, and we select~$N=2$ hidden states,
and we use~$T=50{,}000$ observations for training. 
Note that each observation is one of the~$M=27$
symbols (letters plus word-space). For the example discussed below,
the sequence of~$T=50{,}000$ observations was obtained from 
the Brown corpus of English~\cite{BrownCorpus}. Of course,
any source of English text could be used.

In one specific case, an HMM trained with the parameters listed in the previous
paragraph yields the~$B$ matrix in Table~\ref{tab:bt}.
Observe that this~$B$ matrix gives us two probability distributions over the observation
symbols---one for each of the hidden states. We observe that one hidden state
essentially corresponds to vowels, while the other corresponds to consonants.
This simple example nicely illustrates the concept of machine learning,
as no assumption was made a priori concerning consonants and vowels,
and the only parameter we selected was the number of hidden states~$N$.
Thanks to this training process, the model has learned a crucial aspect of English 
directly from the data.
%This illustrative example is discussed in more detail in~\cite{Stamp04arevealing}
%and originally appeared in Cave and Neuwirth's classic paper~\cite{CaveNeuwirth}.

%\begin{table}[!htb]
%  \caption{Final $B^T$ for HMM}\label{tab:initFinal_B}
%  \centering
%%  {\small
%  { %\footnotesize
%  \begin{tabular}{c|cc} \midrule\midrule
%Letter & State~0 & State~1\\ \midrule
%a  &   0.13537  &  0.00364\\
%b  &  0.00023  &  0.02307\\
%c  &  0.00039  &  0.05605\\
%d  &  0.00025  &  0.06873\\
%e  &  0.21176  &  0.00223\\
%f   &  0.00018  &  0.03556\\
%g  &  0.00041  &  0.02751\\
%h  &  0.00526  &  0.06808\\
%i  &  0.12193  &  0.00077\\
%j  &  0.00014  &  0.00326\\
%k  &   0.00112  &  0.00759\\
%l  &  0.00143  &  0.07227\\
%m  &  0.00027  &  0.03897\\
%n   &  0.00035  &  0.11429\\
%o  &  0.13081  &  0.00143\\
%p  &  0.00073  &  0.03637\\
%q  &  0.00019  &  0.00134\\
%r  &  0.00041  &  0.10128\\
%s  &  0.00032  &  0.11069\\
%t  &  0.00158  &  0.15238\\
%u  &   0.04352  &  0.00098\\
%v  &  0.00019  &  0.01608\\
%w  &  0.00017  &  0.02301\\
%x  &  0.00030  &  0.00426\\
%y  &  0.00028  &  0.02542\\
%z  &  0.00017  &  0.00100\\
%space & 0.34226 & 0.00375\\ \midrule\midrule 
%  \end{tabular}
%  }
%\end{table} 

Suppose that for a given letter~$\ell$, 
we define its Letter2Vec representation~$V(\ell)$ to be
the corresponding row of the converged matrix~$B^T$ in 
the last two columns of Table~\ref{tab:bt}.
Then, for example,
\begin{align}\label{eq:V_aest}
\begin{split}
  V(\ma) &= \rowvecc{0.13845}{0.00075}\ \ \ \ 
  V(\me) = \rowvecc{0.21404}{0.00000}\\
  V(\ms) &= \rowvecc{0.00000}{0.11042}\ \ \ \ 
  V(\mt) = \rowvecc{0.01102}{0.14392}
%  V_{\ma} &= \rowvecc{0.13537}{0.00364}\\
%  V_{\me} &= \rowvecc{0.21176}{0.00223}\\
%  V_{\ms} &= \rowvecc{0.00032}{0.11069}\\
%  V_{\mt} &= \rowvecc{0.00158}{0.15238}
\end{split}
\end{align}
Next, we consider the distance between these Letter2Vec
embeddings. However, instead of using Euclidean distance, we
measure distance based on cosine similarity.

The cosine similarity of vectors~$X$ and~$Y$ is the cosine of the angle between
the two vectors. %This is easily computed as
Let~$X=(X_0,X_1,\ldots,X_{n-1})$ and~$Y=(Y_0,Y_1,\ldots,Y_{n-1})$.
Then the cosine similarity is given by
$$
  \Csim(X,Y) = \frac{\displaystyle\sum_{i=0}^{n-1} X_i Y_i}{
  	\sqrt{\displaystyle\sum_{i=0}^{n-1} X_i^2}\sqrt{\displaystyle\sum_{i=0}^{n-1} Y_i^2}}
$$
In general, $-1\leq \Csim(X,Y)\leq 1$, but since our Letter2Vec 
encoding vectors consist of probabilities---and hence are non-negative---we
have~$0\leq \Csim(X,Y)\leq 1$ for the~$X$ and~$Y$ under consideration.

For the vector encodings in~\eref{eq:V_aest},
we find that for the vowels ``a'' and ``e'', the cosine simarity
is~$\Csim(V(\ma),V(\me))=0.9999$. In contrast, the cosine similarity
between the vowel ``a'' and the consonant ``t'' is~$\Csim(V(\ma),V(\mt))=0.0817$. 
These results indicate that these Letter2Vec embeddings---which are derived from a trained 
HMM---provide useful information on the similarity (or not) of pairs of letters.

%Our HMM-based approach to generate Letter2Vec embeddings could be 
%extended to~$N>2$
%hidden states. This is explored in the suggested exercises at the end of this chapter.

Analogous to our Letter2Vec embeddings,
we could train an HMM on words (or other features) and then use 
the columns of the resulting~$B$ matrix 
(equivalently, the rows of~$B^T$) to define word (feature) embeddings. 

The state of the art for Word2Vec based on words from English text
is trained on a dataset corresponding to~$M=10{,}000$, $N=300$
and~$T=10^9$. Training an HMM with such parameters would be
decidedly non-trivial, as the work factor for Baum-Welch re-estimation
is on the order of~$N^2 T$.

While the word embedding technique discussed in the previous
paragraph---we call it HMM2Vec---is plausible, it has some potential limitations. 
Perhaps the biggest issue with HMM2Vec
is that we typically train an HMM based on a Markov model of order one.
That is, the current state only depends on the immediately-preceding state. 
By basing our word embeddings on such a model, the resulting vectors 
would likely provide only a very limited sense of context. While we can 
train HMMs using models of higher order, the work factor would be prohibitive.

\subsection{PCA2Vec}\label{sect:p2v}

Another option for generating embedding vectors is to apply PCA to
a matrix of pointwise mutual information (PMI). To construct a PMI matrix,
based on a specified window size~$W$, 
we compute~$P(w_i,w_j)$ for all pairs of words~$(w_i,w_j)$ 
that occur within a window~$W$ of each other within out dataset,
and we also compute~$P(w_i)$ for each individual word~$w_i$.
Then we define the PMI matrix as
$$
  X = \{x_{ij}\} = \log\frac{P(w_j,w_i)}{P(w_i)P(w_j)}
$$
We treat column~$i$ of~$X$, denoted~$X_i$, as the feature vector for word~$w_i$
Next, we perform PCA (using a singular value decomposition) based on these~$X_i$ 
feature vectors, and we project the feature vectors~$X_i$ onto the resulting eigenspace. 
Finally, by choosing the~$N$ dominant eigenvalues for this projection, 
we obtain embedding vectors of length~$N$.

It is shown in~\cite{Moody} that these embedding vectors have many
similar properties as Word2Vec embeddings, with the author 
providing examples analogous to those we give in the next section.
Interestingly, it may be beneficial in certain applications to omit some
of the dominant eigenvectors when determining the PCA2Vec embedding
vectors~\cite{LGD}.

For more details on using PCA to generate word embeddings,
see~\cite{LGD}. The aforecited blog~\cite{Moody}
gives an intuitive introduction to the topic.

\subsection{Word2Vec}\label{sect:w2v}

Word2Vec is a technique for embedding ``words''---or more 
generally, any features---into a high-dimensional space.
In Word2Vec, the embeddings
are obtained by training a shallow neural network.
After the training process, 
words that are more similar in context will tend to be 
closer together in the Word2Vec space.

Perhaps surprisingly, certain algebraic properties also hold for Word2Vec
embeddings. For example, according to~\cite{w2v}, if we let
$$
  w_0=\mbox{``king''}, w_1=\mbox{``man''}, w_2=\mbox{``woman''}, w_3=\mbox{``queen''}
$$
and we define~$V(w_i)$ to be the Word2Vec embedding of~$w_i$, then~$V(w_3)$
is the vector that is closest to
$$
  V(w_0) - V(w_1) + V(w_2)
$$
where ``closest'' is in terms of cosine similarity.
Results such as this indicate that Word2Vec embeddings capture meaningful
aspects of the semantics of the language.

Word2Vec uses a similar approach as the HMM2Vec concept outlined above.
But, instead of using an HMM, Word2Vec embeddings are obtained from 
shallow (one hidden layer) neural network. 
Analogous to HMM2Vec, in Word2Vec, we are not interested in
the resulting model itself, but instead we make use the learning that is represented by
the trained model to define word embeddings. Next, we consider the basic ideas
behind Word2Vec. Our approach is similar to that found
in the excellent tutorial~\cite{McCormick}.
Here, we describe the process in terms of words, but these
``words'' can be general features.

Suppose that we have a vocabulary of size~$M$.
We encode each word as a ``one-hot'' vector
of length~$M$. For example, suppose that our vocabulary consists of
the set of~$M=8$ words
\begin{align*}
   W &= (w_0,w_1,w_2,w_3,w_4,w_5,w_6,w_7) \\
        &= (\mbox{``for''}, \mbox{``giant''}, \mbox{``leap''}, \mbox{``man''}, \mbox{``mankind''},
   		\mbox{``one''}, \mbox{``small''}, \mbox{``step''})   	   
\end{align*}
Then we encode ``for'' and ``man'' as
$$
  E(w_0)=E(\mbox{``for''}) = 10000000 
  	\mbox{\ \ and\ \ } E(w_6)=E(\mbox{``man''}) = 00010000
$$
respectively.

Now, suppose that our training data consists of the phrase
\begin{equation}\label{eq:oneSmall}
  \mbox{``one small step for man one giant leap for mankind''}
\end{equation}
To obtain our training samples, we specify a window size~$W$, and for each offset 
we consider pairs of words within the specified window. For this example, 
we select~$W=2$, so that we consider words at a distance
of one or two, in either direction. For the sentence in~\eref{eq:oneSmall},
a window size of two gives us the training
pairs in Table~\ref{tab:train_Word2Vec}.

\begin{table}[!htb]
  \caption{Training data}\label{tab:train_Word2Vec}
  \vglue 0.125in
  \centering\def\vp{\vphantom{step}}
  {\footnotesize
  \begin{tabular}{l|l} \midrule\midrule
    \multicolumn{1}{c|}{\textbf{Offset}\hspace*{0.5in}} 
    	& \multicolumn{1}{c}{\textbf{Training pairs}\hspace*{0.8in}} \\ \midrule
``\fbox{one\vp} small step $\ldots$'' &   (one,small), (one,step) \\
``one \fbox{small\vp} step for $\ldots$'' &  (small,one), (small,step), (small,for) \\
``one small \fbox{step\vp} for man $\ldots$'' &   (step,one), (step,small), (step,for), (step,man) \\
``$\ldots$ small step \fbox{for\vp} man one $\ldots$'' & (for,small), (for,step), (for,man), (for,one) \\
``$\ldots$ step for \fbox{man\vp} one giant $\ldots$'' & (man,step), (man,for), (man,one), (man,giant) \\
``$\ldots$ for man \fbox{one\vp} giant leap $\ldots$'' &  (one,for), (one,man), (one,giant), (one,leap) \\
``$\ldots$ man one \fbox{giant\vp} leap for $\ldots$'' & (giant,man), (giant,one), (giant,leap), (giant,for) \\
``$\ldots$ one giant \fbox{leap\vp} for mankind'' & (leap,one), (leap,giant), (leap,for), (leap,mankind) \\
``$\ldots$ giant leap \fbox{for\vp} mankind'' & (for,giant), (for,leap), (for,mankind) \\
``$\ldots$ leap for \fbox{mankind\vp}'' & (mankind,leap), (mankind,for) \\
  \midrule\midrule 
  \end{tabular}
  }
\end{table}

Consider the pair ``(for,man)'' from the fourth row in 
Table~\ref{tab:train_Word2Vec}. As one-hot vectors, 
this training pair corresponds to the input vector~10000000 
and output vector~00010000.

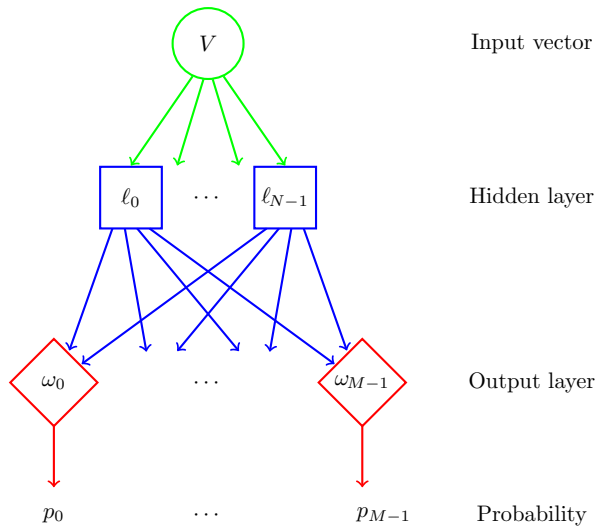
\begin{figure}[!htb]
  \centering
    %    \begin{tikzpicture}[thick,scale=0.9]
\begin{tikzpicture}[scale=0.82,every node/.style={scale=0.85}]
    
    % circles
    \draw[thick,color=green] (4.25,8.5) circle (0.575);

    % squares (top)
    \draw[thick,color=blue] (2.5,5.5) rectangle (3.5,6.5);
    \node at (4.25,6.0) {$\cdots$};
    \draw[thick,color=blue] (5.0,5.5) rectangle (6.0,6.5);
        
    % diamonds
    \draw[thick,color=red,rotate around={45:(1.75,3.0)}] (1.25,2.5) rectangle (2.25,3.5);
%    \draw[thick,color=red,rotate around={45:(4.25,3.5)}] (3.75,3.0) rectangle (4.75,4.0);
    \draw[thick,color=red,rotate around={45:(6.75,3.0)}] (6.25,2.5) rectangle (7.25,3.5);

    % circle to square
    \draw[thick,color=green,->] (4.0,7.97) -- (3.0,6.52);
    \draw[thick,color=green,->] (4.175,7.92) -- (3.75,6.52);
    \draw[thick,color=green,->] (4.325,7.92) -- (4.75,6.52);
    \draw[thick,color=green,->] (4.5,7.97) -- (5.5,6.52);
    
    % square to diamond
    \draw[thick,color=blue,->] (2.7,5.5) -- (2.0,3.52);
    \draw[thick,color=blue,->] (2.9,5.5) -- (3.25,3.5);
    \draw[thick,color=blue,->] (3.1,5.5) -- (4.75,3.51);
    \draw[thick,color=blue,->] (3.3,5.5) -- (6.3,3.3);

    \draw[thick,color=blue,->] (5.2,5.5) -- (2.2,3.3);
    \draw[thick,color=blue,->] (5.4,5.5) -- (3.75,3.51);
    \draw[thick,color=blue,->] (5.6,5.5) -- (5.25,3.5);
    \draw[thick,color=blue,->] (5.8,5.5) -- (6.5,3.52);

    % output
    \draw[thick,color=red,->] (1.75,2.3) -- (1.75,1.3);
    \draw[thick,color=red,->] (6.75,2.3) -- (6.75,1.3);

    % labels for circles
    \node at (4.25,8.5) {$V$};

    % labels for squares (top)
    \node at (3.0,6.0) {$\ell_0$};
    \node at (5.5,6.0) {$\ell_{\kern-1pt N-1}$};

    % labels for diamonds
    \node at (1.75,3.0) {$\omega_0$};
    \node at (4.25,3.0) {$\cdots$};
    \node at (6.75,3.0) {$\omega_{\kern-1pt M-1}$};

    % labels for output
    \node at (1.75,0.85) {$p_0$};
    \node at (4.25,0.85) {$\cdots$};
%    \node at (6.75,0.85) {$p_{M-1}$};
    \node at (7.1,0.85) {$p_{M-1}$};
    
    % labels
    \node at (9.5,8.5) {Input vector};
    \node at (9.5,6.0) {Hidden layer};
    \node at (9.5,3.0) {Output layer};
    \node at (9.5,0.85) {Probability};

\end{tikzpicture}
%  \vglue-0.1in
  \caption{Neural network for Word2Vec embeddings}\label{fig:w2v}
\end{figure}

A neural network similar to that illustrated in Figure~\ref{fig:w2v} is used to generate
Word2Vec embeddings. The input is a one-hot vector of length~$M$
representing the first element of a training pair, such as those in 
Table~\ref{tab:train_Word2Vec}. The network is trained to output
the second element of each ordered pair which, again, is represented
as a one-hot vector. The hidden layer consists of~$N$
linear neurons and the output layer uses a softmax function to generate~$M$
probabilities, where~$p_i$ is the probability of the output vector corresponding to~$w_i$
for the given input. 

Observe that the Word2Vec network in Figure~\ref{fig:w2v}
has~$NM$ weights that are to be determined via training,
and these weights are represented by the blue lines from the hidden layer to the 
output layer. For each output
node~$\omega_i$, there are~$N$ edges (i.e., weights) from the hidden layer.
The~$N$ weights that connect to output node~$\omega_i$ form 
the Word2Vec embedding~$V(w_i)$ of the word~$w_i$. 

The state of the art in Word2Vec
for English text are trained on
a vocabulary of some~$M=10{,}000$ words,
and embedding vectors of length~$N=300$,
training on about~$10^9$ samples. Clearly, training a 
model of this magnitude is an extremely challenging
computational task, as there are~$3\times 10^6$ weights
to be determined, not to mention a huge number of training samples to deal with. 
Most of the complexity of Word2Vec comes from tricks that
are used to make it feasible to train such
a large network with such a massive amount of data.

One trick that is used to speed training in Word2Vec is ``subsampling'' of frequent words.
Common words such as ``a'' and ``the'' contribute little to the model,
so these words can appear in training pairs at a much lower rate
than they are present in the training text.

Another key trick that is used in Word2Vec is ``negative sampling.''
When training a neural network, each training sample potentially affects
all of the weights of the model. Instead of adjusting all of the weights,
in Word2Vec, only a small number of ``negative'' 
samples have their weights modified per training sample. 
For example, suppose that the
output vector of a training pair corresponds to word~$w_0$. 
Then the ``positive'' weights are those connected to the output node~$\omega_0$, 
and these weights are modified. 
In addition, a small subset of the~$M-1$
``negative'' words (i.e., every word in the dataset except~$w_0$) are selected and 
their corresponding weights are adjusted. 
The distribution used to select negative
cases is biased towards more frequent words.

A general discussion of Word2Vec can be found in~\cite{w2v_baby}, while
an intuitive---yet reasonably detailed---introduction
is given in~\cite{McCormick}. The original paper describing
Word2Vec is~\cite{w2v} and an immediate followup paper
discusses a variety of improvements that mostly serve to make
training practical for large datasets~\cite{w2v2}.

\section{Experiments and Results}\label{sect:exp}

In this section, we summarize our experimental results. These results are based on
HMM2Vec, PCA2Vec, and Word2Vec experiments. But, first we discuss the
dataset that we have used for all of the experiments reported in this section.

\subsection{Dataset}

The experimental results discussed in this section are based on the families
in Table~\ref{tab:fams}, with the number of available samples listed. 
From to keep the test set balanced, 
from each of these families, we randomly selected~1000 samples,
for a total of~7000 samples in our classification experiments.
These families have been used in many recent studies, including~\cite{samanvitha} 
and~\cite{wadkar}, for example.

\begin{table}[!htb]
  \caption{Malware families and the number of samples}\label{tab:fams}
  \vglue 0.125in
  \centering
  \begin{tabular}{lcc} \midrule\midrule
    \textbf{Family} & \textbf{Type} & \textbf{Samples}\\ \midrule
%    Adload & Trojan downloader & \n1219\\ %%%%%
    BHO & Trojan & \n1396\\
    CeeInject & VirTool & \n1077\\
    FakeRean & Rogue & \n1017\\
    OnLineGames & Password stealer & \n1508\\
    Renos & Trojan downloader & \n1567 \\
%    VBinject & VirTool & \n2431\\ %%%%%
    Vobfus & Worm & \n1107\\
    Winwebsec & Rogue & \n2302\\
    \midrule
%    Total & --- & 13624\\    
    Total & --- & \n9974\\    
    \midrule\midrule
  \end{tabular}
\end{table}

The malware families in Table~\ref{tab:fams} are of a wide variety of different
types. Next, we briefly discuss each of these families.

\begin{description}
%\item[Adload] downloads an executable, remotely stores it, then executes it 
%and disables proxy settings~\cite{adload}. 
\item[\bf BHO]\hspace*{-12pt} can perform a wide variety of malicious actions, 
as specified by an attacker~\cite{bho}. 
\item[\bf CeeInject]\hspace*{-12pt} is designed to conceal itself from detection, 
and hence various families use it as a shield to prevent detection. 
For example, CeeInject can obfuscate a bitcoin mining client,
which might have been installed on a system without the user's 
knowledge or consent~\cite{ceeinjectmicrosoft}.	
\item[\bf FakeRean]\hspace*{-12pt} pretends to scan the system, notifies the user of nonexistent issues, 
and asks the user to pay to clean the system~\cite{fakerean}. 
\item[\bf OnLineGames]\hspace*{-12pt} steals login information of online games and tracks user 
keystroke activity~\cite{onlinegames}. 
\item[\bf Renos]\hspace*{-12pt} will claim that the system has spyware and ask for a payment to 
remove the supposed spyware~\cite{renos}. 
%\item[VBinject] is used to make other malware difficult to detect. 
%VBinject utilizes encryption and compression to obscure its contents, 
%and hence it is difficult to detect malware that it is 
%concealing~\cite{vbinjectmicrosoft}.
\item[\bf Vobfus]\hspace*{-12pt} is a family that downloads other malware onto a user's computer
and makes changes to the device configuration that cannot be restored 
by simply removing the downloaded malware~\cite{vobfusmicrosoft}.
\item[\bf Winwebsec]\hspace*{-12pt} is a trojan that presents itself as antivirus software---it displays
misleading messages stating that the device has been 
infected and attempts to persuade the user to pay a fee to free the 
system of malware~\cite{winwebsecmicrosoft}.
\end{description}

In the remainder of this section, we present our experimental results. First,
we discuss the selection of parameters for the various classifiers. Then
we give results from a series of experiments for malware classification, 
based on each of the three word embedding techniques discussed 
in Section~\ref{sect:wordEmbed}, namely, HMM2Vec, PCA2Vec, and Word2Vec.
Note that all of our experiments were performed using
\texttt{scikit-learn}~\cite{scikit}.

\begin{table}[!htb]%%%%% What ML package is being used?????
\caption{Classifier hyperparameters tested}\label{tab:classTest}
\vglue 0.125in
\centering
\resizebox{0.75\textwidth}{!}{%
\begin{tabular}{c|l|c}\midrule\midrule
\textbf{Classifier} & \textbf{\ \ \ \ Hyperparameter} & \textbf{Tested values} \\ \midrule
\multirow{5}{*}{MLP}
& \texttt{learning\un rate} & \texttt{constant}, \texttt{invscaling}, \texttt{adaptive} \\
& \texttt{hidden\un layer\un size} & $[(30,30,30), (10,10,10)]$ \\
& \texttt{solver} & \texttt{sgd}, \texttt{adam} \\
& \texttt{activation} & \texttt{relu}, \texttt{logistic}, \texttt{tanh} \\
& \texttt{max\un iter} & $[10000]$ \\ 
\midrule
\multirow{3}{*}{SVM}
& \texttt{kernel} & \texttt{rbf}, \texttt{linear} \\
& \texttt{C} & $[1, 10, 100, 1000]$ \\
& \texttt{gamma} (\texttt{rbf} only) & $[0.001, 0.0001]$ \\
\midrule
\multirow{3}{*}{$\kNN{k}$}
& \texttt{n\un neighbors} & $[3, 5, 11, 19]$ \\
& \texttt{weights} & \texttt{uniform}, \texttt{distance} \\
%& \texttt{p} &  \texttt{\tq manhatten distance\tq}, \texttt{\tq euclidean distance\tq} \\
%& \texttt{p} &  \texttt{\tq manhatten\tq}, \texttt{\tq euclidean\tq} \\
& \texttt{p} &  $[1, 2, 3]$ \\
\midrule
\multirow{4}{*}{RF}
& \texttt{n\un estimators} & $[30, 100, 500, 1000]$ \\
& \texttt{max\un depth} & $[5, 8, 15, 25, 30]$ \\
& \texttt{min\un samples\un split} & $[2, 5, 10, 15, 100]$ \\
& \texttt{min\un samples\un leaf} & $[1, 2, 5, 10]$ \\
\midrule\midrule
\end{tabular}
}
\end{table}

\begin{table}[!htb]
\caption{Classifier hyperparameters selected}\label{tab:classSelect}
\vglue 0.125in
\centering
\resizebox{0.85\textwidth}{!}{%
\begin{tabular}{c|l|cccc}\midrule\midrule
\multirow{2}{*}{\textbf{Classifier}} & \multirow{2}{*}{\textbf{\ \ \ \ Hyperparameter}} & 
\multirow{2}{*}{\textbf{HMM2Vec}} & \multirow{2}{*}{\textbf{Word2Vec}} & \multirow{2}{*}{\textbf{PCA2Vec}}
	& \textbf{Baseline} \\ 
	&  &  &  & 
	& \textbf{HMM} \\ 
	\midrule
\multirow{5}{*}{MLP}
& \texttt{learning\un rate} & \texttt{invscaling} & \texttt{constant} & \texttt{adaptive} %\\
	& \texttt{constant} \\
& \texttt{hidden\un layer\un size} & $(30,30,30)$ & $(30,30,30)$ & $(30,30,30)$ %\\
	& $(30,30,30)$ \\
& \texttt{solver} & \texttt{adam} & \texttt{adam} & \texttt{sgd} %\\
	& \texttt{adam} \\
& \texttt{activation} & \texttt{relu} & \texttt{relu} & \texttt{relu} %\\
	& \texttt{relu} \\
& \texttt{max\un iter} & $10000$ & $10000$ & $10000$ %\\
	& $10000$ \\
\midrule
\multirow{3}{*}{SVM}
& \texttt{kernel} & \texttt{linear} & \texttt{rbf} & \texttt{rbf} %\\
	& \texttt{rbf} \\
& \texttt{C} & $1000$ & $1000$ & $1000$ %\\
	& $10$ \\
& \texttt{gamma} & NA & $0.001$ & $0.001$ %\\
	& $0.0001$ \\
\midrule
\multirow{3}{*}{$\kNN{k}$}
& \texttt{n\un neighbors} & $3$ & $3$ & $3$ %\\
	& $3$ \\
& \texttt{weights} & \texttt{distance} & \texttt{distance} & \texttt{distance} %\\
	& \texttt{distance} \\
%& \texttt{p} & \texttt{\tq manhatten\tq} & \texttt{\tq euclidean\tq} & \texttt{\tq manhatten\tq} %\\
%	& \texttt{\tq euclidean\tq} \\
& \texttt{p} & $1$ & $2$ & $1$ %\\
	& $3$ \\
\midrule
\multirow{4}{*}{RF}
& \texttt{n\un estimators} & $100$ & $500$ & $1000$ %\\
	& $1000$ \\
& \texttt{max\un depth} & $\n25$ & $\n30$ & $\n\n30$ %\\
	& $\n\n30$ \\
& \texttt{min\un samples\un split} & $\n\n2$ & $\n\n2$ & $\n\n\n2$ %\\
	& $\n\n\n2$ \\
& \texttt{min\un samples\un leaf} & $\n\n1$ & $\n\n1$ & $\n\n\n1$ %\\
	& $\n\n\n1$ \\
\midrule\midrule
\end{tabular}
}
\end{table}

\subsection{Classifier Parameters}

For each of our word embedding classification experiments, 
we test the three classifiers discussed in Section~\ref{sect:classifiers},
namely, $k$-nearest neighbors ($\kNN{k}$),
random forest (RF), and support vector machine (SVM),
along with the multilayer perceptron (MLP), which is discussed
in Section~\ref{sect:percept}. The features considered are the 
word embeddings from HMM2Vec, PCA2Vec, and Word2Vec.
Note that this gives us a total of~12 distinct experiments.
 
For each case, we performed a grid search over a set of hyperparameters
using \texttt{GridSearchCV}~\cite{grid} in \texttt{scikit-learn}. 
\texttt{GridSearchCV} performs five-fold cross validation to determine 
the best parameters for each embedding technique.
The parameters tested are listed in Table~\ref{tab:classTest}.
Observe that for each of the three different word embedding techniques,
we tested~$36$ combinations of parameters for~MLP, 
we tested~$12$ combinations for~SVM,
we tested~$16$ combinations for~$\kNN{k}$,
and we tested~$400$ RF combinations.
Overall, we conducted
$$
  3\cdot (36+12+16+400) = 1392
$$
experiments to determine the parameters for the remaining experiments.

The optimal parameters selected for each classifier and for each
embedding technique are listed in Table~\ref{tab:classSelect}.
We note that overall there is considerable agreement between the parameters for
the different word embedding techniques, but in two cases 
(\texttt{learning\un rate} and \texttt{n\un estimators}), a different
parameter is selected for each of the three embedding techniques.

\subsection{Baseline Results}

First, we consider experiments based on opcode sequences and HMMs.
These results serve as a baseline for comparison with the 
vector embedding techniques that are the primary focus of this research.
We choose these HMM-based experiments for the baseline, as HMM
trained on opcode features have proven popular and highly successful 
in the field of malware 
analysis~\cite{AnnachhatreAS15,AustinFJS13,KalbhorAFJS15,RaghavanTS19,WongS06}.

Specifically, we train an HMM for each of the seven families in our dataset,
using~$N=2$ hidden states in each case. For classification,
we score a sample against all seven of these HMMs, and
the resulting score vector (of length seven) serves 
as our feature vector. We use the same classification
algorithms as in our word embedding experiments, namely,
$\kNN{k}$, MLP, RF, and SVM.

Note that ee use the same opcode sequences here as in our vector embedding
experiments. Specifically, the top~20 most frequent opcodes are used, 
with all remaining opcodes deleted. 

\begin{figure}[!htb]
\centering
\begin{tabular}{cc}
%\begin{tikzpicture}[scale=0.8,every node/.style={scale=0.8}]
\begin{tikzpicture}[scale=0.4]
    \begin{axis}[%colorbar/width=2.5mm,
        width=10cm,
        height=10cm,
%        colormap={blackwhite}{gray(0cm)=(1); gray(1cm)=(0.5)},
%	colormap={bluewhite}{color=(white) color=(blue)},
%	colormap={bluewhite}{color=(white) rgb255=(0,191,255)},
	colormap={bluewhite}{color=(white) rgb255=(100,149,237)},
        xticklabels={BHO,OnLineGames,Renos,Winwebsec,CeeInject,FakeRean,Vobfus},
        xtick={0,...,6},
        xtick style={draw=none},
	xticklabel style={anchor=east,rotate=45,yshift=-5pt},
        yticklabels={BHO,OnLineGames,Renos,Winwebsec,CeeInject,FakeRean,Vobfus},
        ytick={0,...,6},
        ytick style={draw=none},
        enlargelimits=false,
        colorbar,
        colorbar style={
%     	  	width=0.05*\pgfkeysvalueof{/pgfplots/parent axis width},%%% added this
%     	  	height=0.5*\pgfkeysvalueof{/pgfplots/parent axis height},
%		plot graphics/node/.style={scale=1.33,anchor=south west,inner sep=0pt,}, %%% scale colorbar fill %%%
            ytick={0.0,0.2,0.4,0.6,0.8,1.0},
            yticklabels={0.0,0.2,0.4,0.6,0.8,1.0},
            yticklabel={\pgfmathprintnumber\tick},
            yticklabel style={%font=\footnotesize,
            		/pgf/number format/fixed,
			/pgf/number format/precision=1}
        },
%        point meta min=0,
%        point meta max=100,
        point meta min=0.0,
        point meta max=1.0,
        nodes near coords={\pgfmathprintnumber\pgfplotspointmeta},
        % ---------------------------------------------------------------------
        % show `nodes near coords' but adapt the style so that values
        % above a threshold get another style
        % (adapted from <http://tex.stackexchange.com/a/141006/95441>)
        % #1: the THRESHOLD after which we switch to a special display.
        nodes near coords black white/.style={
            % define the style of the nodes with "small" values
            small value/.style={
                yshift=-7pt,
%                text=white,
                text=black,
                /pgf/number format/fixed,
                /pgf/number format/precision=2
%                /pgf/number format/precision=0
            },
            % define the style of the nodes with "large" values
            large value/.style={
                yshift=-7pt,
%                text=black,
                text=white,
                /pgf/number format/fixed,
                /pgf/number format/precision=2
%                /pgf/number format/precision=0
            },
            every node near coord/.style={
                check for zero/.code={
                    \pgfmathfloatifflags{\pgfplotspointmeta}{0}{
                        % If meta=0, make the node a coordinate
                        % (which doesn't have text)
                        \pgfkeys{/tikz/coordinate}
                    }{
                        \begingroup
                        % this group is merely to switch to FPU locally.
                        % Might be unnecessary, but who knows.
                        \pgfkeys{/pgf/fpu}
                        \pgfmathparse{\pgfplotspointmeta<#1}
                        \global\let\result=\pgfmathresult
                        \endgroup
                        %
                        % simplifies debugging:
                        %\show\result
                        %
                        \pgfmathfloatcreate{1}{1.0}{0}
                        \let\ONE=\pgfmathresult
                        \ifx\result\ONE
                            % AH: our condition 'y < #1' is met.
                            \pgfkeysalso{/pgfplots/small value}
                        \else
                            % ok, proceed as usual.
                            \pgfkeysalso{/pgfplots/large value}
                        \fi
                    }
                },
                check for zero,
            },
        },
        % asign a value to the new style which is the threshold at which
        % the two style `small value' or `large value' are used
%        nodes near coords black white=50,
        nodes near coords black white=0.5,
        % -----------------------------------------------------------------
    ]
        \addplot[
            matrix plot,
            mesh/cols=7,
            point meta=explicit,draw=gray
        ] table [meta=C] {
            x y C
 0  0 0.99
 1  0 0.00
 2  0 0.00
 3  0 0.00
 4  0 0.01
 5  0 0.00
 6  0 0.00
 0  1 0.00
 1  1 0.87
 2  1 0.02
 3  1 0.03
 4  1 0.06
 5  1 0.02
 6  1 0.00
 0  2 0.00
 1  2 0.00
 2  2 0.93
 3  2 0.02
 4  2 0.04
 5  2 0.00
 6  2 0.01
 0  3 0.00
 1  3 0.00
 2  3 0.02
 3  3 0.93
 4  3 0.03
 5  3 0.02
 6  3 0.00
 0  4 0.00
 1  4 0.05
 2  4 0.03
 3  4 0.02
 4  4 0.89
 5  4 0.00
 6  4 0.01
 0  5 0.01
 1  5 0.01
 2  5 0.01
 3  5 0.00
 4  5 0.05
 5  5 0.90
 6  5 0.02
 0  6 0.01
 1  6 0.01
 2  6 0.00
 3  6 0.04
 4  6 0.00
 5  6 0.01
 6  6 0.93
        };
    \end{axis}
\end{tikzpicture}
%
%\caption{I'm confused~5!}\label{tab:CM5}
%\end{figure*}
& 
%\begin{tikzpicture}[scale=0.8,every node/.style={scale=0.8}]
\begin{tikzpicture}[scale=0.4]
    \begin{axis}[%colorbar/width=2.5mm,
        width=10cm,
        height=10cm,
%        colormap={blackwhite}{gray(0cm)=(1); gray(1cm)=(0.5)},
%	colormap={bluewhite}{color=(white) color=(blue)},
%	colormap={bluewhite}{color=(white) rgb255=(0,191,255)},
	colormap={bluewhite}{color=(white) rgb255=(100,149,237)},
        xticklabels={BHO,OnLineGames,Renos,Winwebsec,CeeInject,FakeRean,Vobfus},
        xtick={0,...,6},
        xtick style={draw=none},
	xticklabel style={anchor=east,rotate=45,yshift=-5pt},
        yticklabels={BHO,OnLineGames,Renos,Winwebsec,CeeInject,FakeRean,Vobfus},
        ytick={0,...,6},
        ytick style={draw=none},
        enlargelimits=false,
        colorbar,
        colorbar style={
%     	  	width=0.05*\pgfkeysvalueof{/pgfplots/parent axis width},%%% added this
%     	  	height=0.5*\pgfkeysvalueof{/pgfplots/parent axis height},
%		plot graphics/node/.style={scale=1.33,anchor=south west,inner sep=0pt,}, %%% scale colorbar fill %%%
            ytick={0.0,0.2,0.4,0.6,0.8,1.0},
            yticklabels={0.0,0.2,0.4,0.6,0.8,1.0},
            yticklabel={\pgfmathprintnumber\tick},
            yticklabel style={%font=\footnotesize,
            		/pgf/number format/fixed,
			/pgf/number format/precision=1}
        },
%        point meta min=0,
%        point meta max=100,
        point meta min=0.0,
        point meta max=1.0,
        nodes near coords={\pgfmathprintnumber\pgfplotspointmeta},
        % ---------------------------------------------------------------------
        % show `nodes near coords' but adapt the style so that values
        % above a threshold get another style
        % (adapted from <http://tex.stackexchange.com/a/141006/95441>)
        % #1: the THRESHOLD after which we switch to a special display.
        nodes near coords black white/.style={
            % define the style of the nodes with "small" values
            small value/.style={
                yshift=-7pt,
%                text=white,
                text=black,
                /pgf/number format/fixed,
                /pgf/number format/precision=2
%                /pgf/number format/precision=0
            },
            % define the style of the nodes with "large" values
            large value/.style={
                yshift=-7pt,
%                text=black,
                text=white,
                /pgf/number format/fixed,
                /pgf/number format/precision=2
%                /pgf/number format/precision=0
            },
            every node near coord/.style={
                check for zero/.code={
                    \pgfmathfloatifflags{\pgfplotspointmeta}{0}{
                        % If meta=0, make the node a coordinate
                        % (which doesn't have text)
                        \pgfkeys{/tikz/coordinate}
                    }{
                        \begingroup
                        % this group is merely to switch to FPU locally.
                        % Might be unnecessary, but who knows.
                        \pgfkeys{/pgf/fpu}
                        \pgfmathparse{\pgfplotspointmeta<#1}
                        \global\let\result=\pgfmathresult
                        \endgroup
                        %
                        % simplifies debugging:
                        %\show\result
                        %
                        \pgfmathfloatcreate{1}{1.0}{0}
                        \let\ONE=\pgfmathresult
                        \ifx\result\ONE
                            % AH: our condition 'y < #1' is met.
                            \pgfkeysalso{/pgfplots/small value}
                        \else
                            % ok, proceed as usual.
                            \pgfkeysalso{/pgfplots/large value}
                        \fi
                    }
                },
                check for zero,
            },
        },
        % asign a value to the new style which is the threshold at which
        % the two style `small value' or `large value' are used
%        nodes near coords black white=50,
        nodes near coords black white=0.5,
        % -----------------------------------------------------------------
    ]
        \addplot[
            matrix plot,
            mesh/cols=7,
            point meta=explicit,draw=gray
        ] table [meta=C] {
            x y C
 0  0 0.97
 1  0 0.00
 2  0 0.00
 3  0 0.00
 4  0 0.00
 5  0 0.00
 6  0 0.03
 0  1 0.34
 1  1 0.00
 2  1 0.19
 3  1 0.05
 4  1 0.25
 5  1 0.01
 6  1 0.16
 0  2 0.07
 1  2 0.00
 2  2 0.23
 3  2 0.07
 4  2 0.27
 5  2 0.12
 6  2 0.24
 0  3 0.20
 1  3 0.00
 2  3 0.01
 3  3 0.07
 4  3 0.40
 5  3 0.19
 6  3 0.13
 0  4 0.19
 1  4 0.00
 2  4 0.01
 3  4 0.00
 4  4 0.54
 5  4 0.03
 6  4 0.23
 0  5 0.04
 1  5 0.00
 2  5 0.26
 3  5 0.01
 4  5 0.06
 5  5 0.32
 6  5 0.31
 0  6 0.02
 1  6 0.00
 2  6 0.00
 3  6 0.01
 4  6 0.01
 5  6 0.04
 6  6 0.92
        };
    \end{axis}
\end{tikzpicture}
%
%\caption{I'm confused~5!}\label{tab:CM5}
%\end{figure*}
\\[-1ex]
(a) $\kNN{k}$ & (b) MLP \\
\\[-1ex]
%\begin{tikzpicture}[scale=0.8,every node/.style={scale=0.8}]
\begin{tikzpicture}[scale=0.4]
    \begin{axis}[%colorbar/width=2.5mm,
        width=10cm,
        height=10cm,
%        colormap={blackwhite}{gray(0cm)=(1); gray(1cm)=(0.5)},
%	colormap={bluewhite}{color=(white) color=(blue)},
%	colormap={bluewhite}{color=(white) rgb255=(0,191,255)},
	colormap={bluewhite}{color=(white) rgb255=(100,149,237)},
        xticklabels={BHO,OnLineGames,Renos,Winwebsec,CeeInject,FakeRean,Vobfus},
        xtick={0,...,6},
        xtick style={draw=none},
	xticklabel style={anchor=east,rotate=45,yshift=-5pt},
        yticklabels={BHO,OnLineGames,Renos,Winwebsec,CeeInject,FakeRean,Vobfus},
        ytick={0,...,6},
        ytick style={draw=none},
        enlargelimits=false,
        colorbar,
        colorbar style={
%     	  	width=0.05*\pgfkeysvalueof{/pgfplots/parent axis width},%%% added this
%     	  	height=0.5*\pgfkeysvalueof{/pgfplots/parent axis height},
%		plot graphics/node/.style={scale=1.33,anchor=south west,inner sep=0pt,}, %%% scale colorbar fill %%%
            ytick={0.0,0.2,0.4,0.6,0.8,1.0},
            yticklabels={0.0,0.2,0.4,0.6,0.8,1.0},
            yticklabel={\pgfmathprintnumber\tick},
            yticklabel style={%font=\footnotesize,
            		/pgf/number format/fixed,
			/pgf/number format/precision=1}
        },
%        point meta min=0,
%        point meta max=100,
        point meta min=0.0,
        point meta max=1.0,
        nodes near coords={\pgfmathprintnumber\pgfplotspointmeta},
        % ---------------------------------------------------------------------
        % show `nodes near coords' but adapt the style so that values
        % above a threshold get another style
        % (adapted from <http://tex.stackexchange.com/a/141006/95441>)
        % #1: the THRESHOLD after which we switch to a special display.
        nodes near coords black white/.style={
            % define the style of the nodes with "small" values
            small value/.style={
                yshift=-7pt,
%                text=white,
                text=black,
                /pgf/number format/fixed,
                /pgf/number format/precision=2
%                /pgf/number format/precision=0
            },
            % define the style of the nodes with "large" values
            large value/.style={
                yshift=-7pt,
%                text=black,
                text=white,
                /pgf/number format/fixed,
                /pgf/number format/precision=2
%                /pgf/number format/precision=0
            },
            every node near coord/.style={
                check for zero/.code={
                    \pgfmathfloatifflags{\pgfplotspointmeta}{0}{
                        % If meta=0, make the node a coordinate
                        % (which doesn't have text)
                        \pgfkeys{/tikz/coordinate}
                    }{
                        \begingroup
                        % this group is merely to switch to FPU locally.
                        % Might be unnecessary, but who knows.
                        \pgfkeys{/pgf/fpu}
                        \pgfmathparse{\pgfplotspointmeta<#1}
                        \global\let\result=\pgfmathresult
                        \endgroup
                        %
                        % simplifies debugging:
                        %\show\result
                        %
                        \pgfmathfloatcreate{1}{1.0}{0}
                        \let\ONE=\pgfmathresult
                        \ifx\result\ONE
                            % AH: our condition 'y < #1' is met.
                            \pgfkeysalso{/pgfplots/small value}
                        \else
                            % ok, proceed as usual.
                            \pgfkeysalso{/pgfplots/large value}
                        \fi
                    }
                },
                check for zero,
            },
        },
        % asign a value to the new style which is the threshold at which
        % the two style `small value' or `large value' are used
%        nodes near coords black white=50,
        nodes near coords black white=0.5,
        % -----------------------------------------------------------------
    ]
        \addplot[
            matrix plot,
            mesh/cols=7,
            point meta=explicit,draw=gray
        ] table [meta=C] {
            x y C
 0  0 0.99
 1  0 0.00
 2  0 0.00
 3  0 0.00
 4  0 0.01
 5  0 0.00
 6  0 0.00
 0  1 0.00
 1  1 0.83
 2  1 0.03
 3  1 0.03
 4  1 0.08
 5  1 0.03
 6  1 0.00
 0  2 0.00
 1  2 0.00
 2  2 0.93
 3  2 0.02
 4  2 0.03
 5  2 0.00
 6  2 0.02
 0  3 0.00
 1  3 0.01
 2  3 0.02
 3  3 0.92
 4  3 0.02
 5  3 0.01
 6  3 0.02
 0  4 0.00
 1  4 0.04
 2  4 0.04
 3  4 0.03
 4  4 0.87
 5  4 0.01
 6  4 0.01
 0  5 0.02
 1  5 0.01
 2  5 0.01
 3  5 0.00
 4  5 0.05
 5  5 0.89
 6  5 0.02
 0  6 0.01
 1  6 0.01
 2  6 0.01
 3  6 0.05
 4  6 0.00
 5  6 0.00
 6  6 0.92
        };
    \end{axis}
\end{tikzpicture}
%
%\caption{I'm confused~5!}\label{tab:CM5}
%\end{figure*}
& 
%\begin{tikzpicture}[scale=0.8,every node/.style={scale=0.8}]
\begin{tikzpicture}[scale=0.4]
    \begin{axis}[%colorbar/width=2.5mm,
        width=10cm,
        height=10cm,
%        colormap={blackwhite}{gray(0cm)=(1); gray(1cm)=(0.5)},
%	colormap={bluewhite}{color=(white) color=(blue)},
%	colormap={bluewhite}{color=(white) rgb255=(0,191,255)},
	colormap={bluewhite}{color=(white) rgb255=(100,149,237)},
        xticklabels={BHO,OnLineGames,Renos,Winwebsec,CeeInject,FakeRean,Vobfus},
        xtick={0,...,6},
        xtick style={draw=none},
	xticklabel style={anchor=east,rotate=45,yshift=-5pt},
        yticklabels={BHO,OnLineGames,Renos,Winwebsec,CeeInject,FakeRean,Vobfus},
        ytick={0,...,6},
        ytick style={draw=none},
        enlargelimits=false,
        colorbar,
        colorbar style={
%     	  	width=0.05*\pgfkeysvalueof{/pgfplots/parent axis width},%%% added this
%     	  	height=0.5*\pgfkeysvalueof{/pgfplots/parent axis height},
%		plot graphics/node/.style={scale=1.33,anchor=south west,inner sep=0pt,}, %%% scale colorbar fill %%%
            ytick={0.0,0.2,0.4,0.6,0.8,1.0},
            yticklabels={0.0,0.2,0.4,0.6,0.8,1.0},
            yticklabel={\pgfmathprintnumber\tick},
            yticklabel style={%font=\footnotesize,
            		/pgf/number format/fixed,
			/pgf/number format/precision=1}
        },
%        point meta min=0,
%        point meta max=100,
        point meta min=0.0,
        point meta max=1.0,
        nodes near coords={\pgfmathprintnumber\pgfplotspointmeta},
        % ---------------------------------------------------------------------
        % show `nodes near coords' but adapt the style so that values
        % above a threshold get another style
        % (adapted from <http://tex.stackexchange.com/a/141006/95441>)
        % #1: the THRESHOLD after which we switch to a special display.
        nodes near coords black white/.style={
            % define the style of the nodes with "small" values
            small value/.style={
                yshift=-7pt,
%                text=white,
                text=black,
                /pgf/number format/fixed,
                /pgf/number format/precision=2
%                /pgf/number format/precision=0
            },
            % define the style of the nodes with "large" values
            large value/.style={
                yshift=-7pt,
%                text=black,
                text=white,
                /pgf/number format/fixed,
                /pgf/number format/precision=2
%                /pgf/number format/precision=0
            },
            every node near coord/.style={
                check for zero/.code={
                    \pgfmathfloatifflags{\pgfplotspointmeta}{0}{
                        % If meta=0, make the node a coordinate
                        % (which doesn't have text)
                        \pgfkeys{/tikz/coordinate}
                    }{
                        \begingroup
                        % this group is merely to switch to FPU locally.
                        % Might be unnecessary, but who knows.
                        \pgfkeys{/pgf/fpu}
                        \pgfmathparse{\pgfplotspointmeta<#1}
                        \global\let\result=\pgfmathresult
                        \endgroup
                        %
                        % simplifies debugging:
                        %\show\result
                        %
                        \pgfmathfloatcreate{1}{1.0}{0}
                        \let\ONE=\pgfmathresult
                        \ifx\result\ONE
                            % AH: our condition 'y < #1' is met.
                            \pgfkeysalso{/pgfplots/small value}
                        \else
                            % ok, proceed as usual.
                            \pgfkeysalso{/pgfplots/large value}
                        \fi
                    }
                },
                check for zero,
            },
        },
        % asign a value to the new style which is the threshold at which
        % the two style `small value' or `large value' are used
%        nodes near coords black white=50,
        nodes near coords black white=0.5,
        % -----------------------------------------------------------------
    ]
        \addplot[
            matrix plot,
            mesh/cols=7,
            point meta=explicit,draw=gray
        ] table [meta=C] {
            x y C
 0  0 0.93
 1  0 0.00
 2  0 0.00
 3  0 0.00
 4  0 0.00
 5  0 0.07
 6  0 0.00
 0  1 0.00
 1  1 0.72
 2  1 0.00
 3  1 0.00
 4  1 0.00
 5  1 0.28
 6  1 0.00
 0  2 0.00
 1  2 0.00
 2  2 0.69
 3  2 0.01
 4  2 0.00
 5  2 0.29
 6  2 0.01
 0  3 0.00
 1  3 0.00
 2  3 0.00
 3  3 0.61
 4  3 0.01
 5  3 0.38
 6  3 0.00
 0  4 0.00
 1  4 0.00
 2  4 0.00
 3  4 0.01
 4  4 0.79
 5  4 0.20
 6  4 0.00
 0  5 0.00
 1  5 0.00
 2  5 0.00
 3  5 0.00
 4  5 0.05
 5  5 0.95
 6  5 0.00
 0  6 0.00
 1  6 0.00
 2  6 0.00
 3  6 0.00
 4  6 0.00
 5  6 0.21
 6  6 0.79
        };
    \end{axis}
\end{tikzpicture}
%
%\caption{I'm confused~5!}\label{tab:CM5}
%\end{figure*}
\\[-1ex]
(c) RF & (d) SVM 
\\[-0.5ex]
\end{tabular}
\caption{Confusion matrices for baseline HMM experiments}\label{fig:raw}
\end{figure}
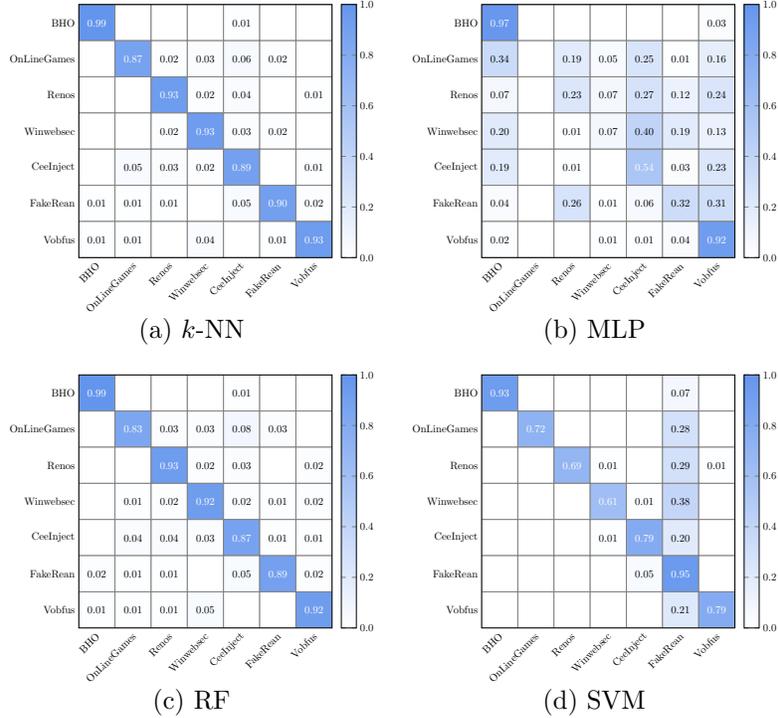

The confusion matrices for these baseline HMM experiments are given in
Figure~\ref{fig:raw}. The accuracies obtained
for~$\kNN{k}$, MLP, RF, and SVM
are~0.92, 0.44, 0.91, and~0.78, respectively. We see that 
MLP and SVM both perform poorly, whereas the neighborhood
based techniques, namely, $\kNN{k}$ and RF, 
are both strong, considering that we have
seven classes. In addition, $\kNN{k}$ and RF 
give very similar results.

\subsection{HMM2Vec Results}

For these experiments, we train an HMM on each sample in our dataset. Recall that
our dataset consists of~1000 samples from each of the seven
families listed in Table~\ref{tab:fams}. We train each of these~7000 models
with~$N=2$ hidden states, using the~$M=20$ most
frequent opcodes over all malware samples. Opcodes outside the
top~20 are ignored. 

As mentioned in Section~\ref{sect:HMM3probs},
we often train multiple HMMs with different initial conditions,
and select the best scoring model.
This becomes more important as the length of the
observation sequence decreases. Hence, when training
our HMMs, we perform multiple random restarts---the 
number of restarts is determined by the length of the 
training sequence, as indicated in Table~\ref{tab:random}.
%We select the highest scoring model from those trained
%with the specified number of random restarts.
%The top~20 opcodes represent well 
%in excess of~98\%\ of the opcodes.

\begin{table}[!htb]
  \caption{Number of random restarts}\label{tab:random}
  \vglue 0.125in
  \centering
  \begin{tabular}{lcc} \midrule\midrule
    \textbf{Observations} & \textbf{Restarts}\\ \midrule
    Greater than~30,000 & \n10\\
    10,000-30,000 & \n30\\
    5000-10,000 & 100\\
    Less than~500 & 500\\
    \midrule\midrule
  \end{tabular}
\end{table}

Each~$B$ matrix is~$2\times 20$, where each row corresponds to
one of the hidden states of the model. From each of these matrices, we construct
a vector of length~40 by appending the two rows. Since the order of the
hidden states can vary between models, we select the order of the 
rows so as to obtain a consistency with respect to the most common 
opcode. That is, the row corresponding to the state that accumulates the
highest probability for \texttt{MOV} is the first half of the 
feature vector, with the other row of the~$B$ matrix becoming the last~20
elements of the feature vector. This accounts for any cases
where the hidden states differ.

\begin{figure}[!htb]
\centering
\begin{tabular}{cc}
%\begin{tikzpicture}[scale=0.8,every node/.style={scale=0.8}]
%\begin{tikzpicture}[scale=0.375,every node/.style={scale=1.25}]
\begin{tikzpicture}[scale=0.4]
    \begin{axis}[%colorbar/width=2.5mm,
        width=10cm,
        height=10cm,
%        colormap={blackwhite}{gray(0cm)=(1); gray(1cm)=(0.5)},
%	colormap={bluewhite}{color=(white) color=(blue)},
%	colormap={bluewhite}{color=(white) rgb255=(0,191,255)},
	colormap={bluewhite}{color=(white) rgb255=(100,149,237)},
        xticklabels={BHO,OnLineGames,Renos,Winwebsec,CeeInject,FakeRean,Vobfus},
        xtick={0,...,6},
        xtick style={draw=none},
	xticklabel style={anchor=east,rotate=45,yshift=-5pt},
        yticklabels={BHO,OnLineGames,Renos,Winwebsec,CeeInject,FakeRean,Vobfus},
        ytick={0,...,6},
        ytick style={draw=none},
        enlargelimits=false,
        colorbar,
        colorbar style={
%     	  	width=0.05*\pgfkeysvalueof{/pgfplots/parent axis width},%%% added this
%     	  	height=0.5*\pgfkeysvalueof{/pgfplots/parent axis height},
%		plot graphics/node/.style={scale=1.33,anchor=south west,inner sep=0pt,}, %%% scale colorbar fill %%%
            ytick={0,0.20,0.40,0.60,0.80,1.00},
            yticklabels={0,0.20,0.40,0.60,0.80,1.00},
            yticklabel={\pgfmathprintnumber\tick},
            yticklabel style={%font=\footnotesize,
            		/pgf/number format/fixed,
			/pgf/number format/precision=2}
        },
%        point meta min=0,
%        point meta max=100,
        point meta min=0.0,
        point meta max=1.0,
        nodes near coords={\pgfmathprintnumber\pgfplotspointmeta},
        % ---------------------------------------------------------------------
        % show `nodes near coords' but adapt the style so that values
        % above a threshold get another style
        % (adapted from <http://tex.stackexchange.com/a/141006/95441>)
        % #1: the THRESHOLD after which we switch to a special display.
        nodes near coords black white/.style={
            % define the style of the nodes with "small" values
            small value/.style={
                yshift=-7pt,
%                text=white,
                text=black,
                /pgf/number format/fixed,
                /pgf/number format/precision=2
%                /pgf/number format/precision=0
            },
            % define the style of the nodes with "large" values
            large value/.style={
                yshift=-7pt,
%                text=black,
                text=white,
                /pgf/number format/fixed,
                /pgf/number format/precision=2
%                /pgf/number format/precision=0
            },
            every node near coord/.style={
                check for zero/.code={
                    \pgfmathfloatifflags{\pgfplotspointmeta}{0}{
                        % If meta=0, make the node a coordinate
                        % (which doesn't have text)
                        \pgfkeys{/tikz/coordinate}
                    }{
                        \begingroup
                        % this group is merely to switch to FPU locally.
                        % Might be unnecessary, but who knows.
                        \pgfkeys{/pgf/fpu}
                        \pgfmathparse{\pgfplotspointmeta<#1}
                        \global\let\result=\pgfmathresult
                        \endgroup
                        %
                        % simplifies debugging:
                        %\show\result
                        %
                        \pgfmathfloatcreate{1}{1.0}{0}
                        \let\ONE=\pgfmathresult
                        \ifx\result\ONE
                            % AH: our condition 'y < #1' is met.
                            \pgfkeysalso{/pgfplots/small value}
                        \else
                            % ok, proceed as usual.
                            \pgfkeysalso{/pgfplots/large value}
                        \fi
                    }
                },
                check for zero,
            },
        },
        % asign a value to the new style which is the threshold at which
        % the two style `small value' or `large value' are used
%        nodes near coords black white=50,
        nodes near coords black white=0.5,
        % -----------------------------------------------------------------
    ]
        \addplot[
            matrix plot,
            mesh/cols=7,
            point meta=explicit,draw=gray
        ] table [meta=C] {
            x y C
 0  0 0.98
 1  0 0.00
 2  0 0.00
 3  0 0.00
 4  0 0.01
 5  0 0.01
 6  0 0.00
 0  1 0.00
 1  1 0.94
 2  1 0.01
 3  1 0.00
 4  1 0.03
 5  1 0.02
 6  1 0.00
 0  2 0.00
 1  2 0.01
 2  2 0.95
 3  2 0.01
 4  2 0.00
 5  2 0.03
 6  2 0.00
 0  3 0.00
 1  3 0.01
 2  3 0.01
 3  3 0.86
 4  3 0.02
 5  3 0.10
 6  3 0.00
 0  4 0.00
 1  4 0.01
 2  4 0.01
 3  4 0.03
 4  4 0.90
 5  4 0.04
 6  4 0.01
 0  5 0.01
 1  5 0.02
 2  5 0.02
 3  5 0.05
 4  5 0.00
 5  5 0.90
 6  5 0.00
 0  6 0.00
 1  6 0.00
 2  6 0.01
 3  6 0.01
 4  6 0.01
 5  6 0.00
 6  6 0.97
        };
    \end{axis}
\end{tikzpicture}
%
%\caption{I'm confused~5!}\label{tab:CM5}
%\end{figure*}
& 
%\begin{tikzpicture}[scale=0.8,every node/.style={scale=0.8}]
%\begin{tikzpicture}[scale=0.375,every node/.style={scale=1.25}]
\begin{tikzpicture}[scale=0.4]
    \begin{axis}[%colorbar/width=2.5mm,
        width=10cm,
        height=10cm,
%        colormap={blackwhite}{gray(0cm)=(1); gray(1cm)=(0.5)},
%	colormap={bluewhite}{color=(white) color=(blue)},
%	colormap={bluewhite}{color=(white) rgb255=(0,191,255)},
	colormap={bluewhite}{color=(white) rgb255=(100,149,237)},
        xticklabels={BHO,OnLineGames,Renos,Winwebsec,CeeInject,FakeRean,Vobfus},
        xtick={0,...,6},
        xtick style={draw=none},
	xticklabel style={anchor=east,rotate=45,yshift=-5pt},
        yticklabels={BHO,OnLineGames,Renos,Winwebsec,CeeInject,FakeRean,Vobfus},
        ytick={0,...,6},
        ytick style={draw=none},
        enlargelimits=false,
        colorbar,
        colorbar style={
%     	  	width=0.05*\pgfkeysvalueof{/pgfplots/parent axis width},%%% added this
%     	  	height=0.5*\pgfkeysvalueof{/pgfplots/parent axis height},
%		plot graphics/node/.style={scale=1.33,anchor=south west,inner sep=0pt,}, %%% scale colorbar fill %%%
            ytick={0,0.20,0.40,0.60,0.80,1.00},
            yticklabels={0,0.20,0.40,0.60,0.80,1.00},
            yticklabel={\pgfmathprintnumber\tick},
            yticklabel style={%font=\footnotesize,
            		/pgf/number format/fixed,
			/pgf/number format/precision=2}
        },
%        point meta min=0,
%        point meta max=100,
        point meta min=0.0,
        point meta max=1.0,
        nodes near coords={\pgfmathprintnumber\pgfplotspointmeta},
        % ---------------------------------------------------------------------
        % show `nodes near coords' but adapt the style so that values
        % above a threshold get another style
        % (adapted from <http://tex.stackexchange.com/a/141006/95441>)
        % #1: the THRESHOLD after which we switch to a special display.
        nodes near coords black white/.style={
            % define the style of the nodes with "small" values
            small value/.style={
                yshift=-7pt,
%                text=white,
                text=black,
                /pgf/number format/fixed,
                /pgf/number format/precision=2
%                /pgf/number format/precision=0
            },
            % define the style of the nodes with "large" values
            large value/.style={
                yshift=-7pt,
%                text=black,
                text=white,
                /pgf/number format/fixed,
                /pgf/number format/precision=2
%                /pgf/number format/precision=0
            },
            every node near coord/.style={
                check for zero/.code={
                    \pgfmathfloatifflags{\pgfplotspointmeta}{0}{
                        % If meta=0, make the node a coordinate
                        % (which doesn't have text)
                        \pgfkeys{/tikz/coordinate}
                    }{
                        \begingroup
                        % this group is merely to switch to FPU locally.
                        % Might be unnecessary, but who knows.
                        \pgfkeys{/pgf/fpu}
                        \pgfmathparse{\pgfplotspointmeta<#1}
                        \global\let\result=\pgfmathresult
                        \endgroup
                        %
                        % simplifies debugging:
                        %\show\result
                        %
                        \pgfmathfloatcreate{1}{1.0}{0}
                        \let\ONE=\pgfmathresult
                        \ifx\result\ONE
                            % AH: our condition 'y < #1' is met.
                            \pgfkeysalso{/pgfplots/small value}
                        \else
                            % ok, proceed as usual.
                            \pgfkeysalso{/pgfplots/large value}
                        \fi
                    }
                },
                check for zero,
            },
        },
        % asign a value to the new style which is the threshold at which
        % the two style `small value' or `large value' are used
%        nodes near coords black white=50,
        nodes near coords black white=0.5,
        % -----------------------------------------------------------------
    ]
        \addplot[
            matrix plot,
            mesh/cols=7,
            point meta=explicit,draw=gray
        ] table [meta=C] {
            x y C
 0  0 0.98
 1  0 0.01
 2  0 0.00
 3  0 0.00
 4  0 0.01
 5  0 0.00
 6  0 0.00
 0  1 0.00
 1  1 0.89
 2  1 0.01
 3  1 0.03
 4  1 0.05
 5  1 0.01
 6  1 0.01
 0  2 0.00
 1  2 0.00
 2  2 0.95
 3  2 0.02
 4  2 0.02
 5  2 0.01
 6  2 0.00
 0  3 0.00
 1  3 0.00
 2  3 0.01
 3  3 0.87
 4  3 0.02
 5  3 0.08
 6  3 0.02
 0  4 0.01
 1  4 0.00
 2  4 0.02
 3  4 0.04
 4  4 0.91
 5  4 0.02
 6  4 0.00
 0  5 0.00
 1  5 0.01
 2  5 0.03
 3  5 0.08
 4  5 0.04
 5  5 0.84
 6  5 0.00
 0  6 0.01
 1  6 0.01
 2  6 0.00
 3  6 0.01
 4  6 0.00
 5  6 0.00
 6  6 0.97
        };
    \end{axis}
\end{tikzpicture}
%
%\caption{I'm confused~5!}\label{tab:CM5}
%\end{figure*}
\\[-0.5ex]
(a) $\kNN{k}$ & (b) MLP \\
\\[-0.5ex]
%\begin{tikzpicture}[scale=0.8,every node/.style={scale=0.8}]
%\begin{tikzpicture}[scale=0.375,every node/.style={scale=1.25}]
\begin{tikzpicture}[scale=0.4]
    \begin{axis}[%colorbar/width=2.5mm,
        width=10cm,
        height=10cm,
%        colormap={blackwhite}{gray(0cm)=(1); gray(1cm)=(0.5)},
%	colormap={bluewhite}{color=(white) color=(blue)},
%	colormap={bluewhite}{color=(white) rgb255=(0,191,255)},
	colormap={bluewhite}{color=(white) rgb255=(100,149,237)},
        xticklabels={BHO,OnLineGames,Renos,Winwebsec,CeeInject,FakeRean,Vobfus},
        xtick={0,...,6},
        xtick style={draw=none},
	xticklabel style={anchor=east,rotate=45,yshift=-5pt},
        yticklabels={BHO,OnLineGames,Renos,Winwebsec,CeeInject,FakeRean,Vobfus},
        ytick={0,...,6},
        ytick style={draw=none},
        enlargelimits=false,
        colorbar,
        colorbar style={
%     	  	width=0.05*\pgfkeysvalueof{/pgfplots/parent axis width},%%% added this
%     	  	height=0.5*\pgfkeysvalueof{/pgfplots/parent axis height},
%		plot graphics/node/.style={scale=1.33,anchor=south west,inner sep=0pt,}, %%% scale colorbar fill %%%
            ytick={0,0.20,0.40,0.60,0.80,1.00},
            yticklabels={0,0.20,0.40,0.60,0.80,1.00},
            yticklabel={\pgfmathprintnumber\tick},
            yticklabel style={%font=\footnotesize,
            		/pgf/number format/fixed,
			/pgf/number format/precision=2}
        },
%        point meta min=0,
%        point meta max=100,
        point meta min=0.0,
        point meta max=1.0,
        nodes near coords={\pgfmathprintnumber\pgfplotspointmeta},
        % ---------------------------------------------------------------------
        % show `nodes near coords' but adapt the style so that values
        % above a threshold get another style
        % (adapted from <http://tex.stackexchange.com/a/141006/95441>)
        % #1: the THRESHOLD after which we switch to a special display.
        nodes near coords black white/.style={
            % define the style of the nodes with "small" values
            small value/.style={
                yshift=-7pt,
%                text=white,
                text=black,
                /pgf/number format/fixed,
                /pgf/number format/precision=2
%                /pgf/number format/precision=0
            },
            % define the style of the nodes with "large" values
            large value/.style={
                yshift=-7pt,
%                text=black,
                text=white,
                /pgf/number format/fixed,
                /pgf/number format/precision=2
%                /pgf/number format/precision=0
            },
            every node near coord/.style={
                check for zero/.code={
                    \pgfmathfloatifflags{\pgfplotspointmeta}{0}{
                        % If meta=0, make the node a coordinate
                        % (which doesn't have text)
                        \pgfkeys{/tikz/coordinate}
                    }{
                        \begingroup
                        % this group is merely to switch to FPU locally.
                        % Might be unnecessary, but who knows.
                        \pgfkeys{/pgf/fpu}
                        \pgfmathparse{\pgfplotspointmeta<#1}
                        \global\let\result=\pgfmathresult
                        \endgroup
                        %
                        % simplifies debugging:
                        %\show\result
                        %
                        \pgfmathfloatcreate{1}{1.0}{0}
                        \let\ONE=\pgfmathresult
                        \ifx\result\ONE
                            % AH: our condition 'y < #1' is met.
                            \pgfkeysalso{/pgfplots/small value}
                        \else
                            % ok, proceed as usual.
                            \pgfkeysalso{/pgfplots/large value}
                        \fi
                    }
                },
                check for zero,
            },
        },
        % asign a value to the new style which is the threshold at which
        % the two style `small value' or `large value' are used
%        nodes near coords black white=50,
        nodes near coords black white=0.5,
        % -----------------------------------------------------------------
    ]
        \addplot[
            matrix plot,
            mesh/cols=7,
            point meta=explicit,draw=gray
        ] table [meta=C] {
            x y C
 0  0 0.98
 1  0 0.00
 2  0 0.01
 3  0 0.00
 4  0 0.00
 5  0 0.01
 6  0 0.00
 0  1 0.00
 1  1 0.92
 2  1 0.01
 3  1 0.01
 4  1 0.03
 5  1 0.03
 6  1 0.00
 0  2 0.00
 1  2 0.00
 2  2 0.95
 3  2 0.01
 4  2 0.01
 5  2 0.03
 6  2 0.00
 0  3 0.00
 1  3 0.00
 2  3 0.01
 3  3 0.86
 4  3 0.01
 5  3 0.12
 6  3 0.00
 0  4 0.01
 1  4 0.00
 2  4 0.01
 3  4 0.04
 4  4 0.91
 5  4 0.03
 6  4 0.00
 0  5 0.00
 1  5 0.01
 2  5 0.01
 3  5 0.03
 4  5 0.02
 5  5 0.93
 6  5 0.00
 0  6 0.00
 1  6 0.00
 2  6 0.00
 3  6 0.01
 4  6 0.00
 5  6 0.02
 6  6 0.97
        };
    \end{axis}
\end{tikzpicture}
%
%\caption{I'm confused~5!}\label{tab:CM5}
%\end{figure*}
& 
%\begin{tikzpicture}[scale=0.8,every node/.style={scale=0.8}]
%\begin{tikzpicture}[scale=0.375,every node/.style={scale=1.25}]
\begin{tikzpicture}[scale=0.4]
    \begin{axis}[%colorbar/width=2.5mm,
        width=10cm,
        height=10cm,
%        colormap={blackwhite}{gray(0cm)=(1); gray(1cm)=(0.5)},
%	colormap={bluewhite}{color=(white) color=(blue)},
%	colormap={bluewhite}{color=(white) rgb255=(0,191,255)},
	colormap={bluewhite}{color=(white) rgb255=(100,149,237)},
        xticklabels={BHO,OnLineGames,Renos,Winwebsec,CeeInject,FakeRean,Vobfus},
        xtick={0,...,6},
        xtick style={draw=none},
	xticklabel style={anchor=east,rotate=45,yshift=-5pt},
        yticklabels={BHO,OnLineGames,Renos,Winwebsec,CeeInject,FakeRean,Vobfus},
        ytick={0,...,6},
        ytick style={draw=none},
        enlargelimits=false,
        colorbar,
        colorbar style={
%     	  	width=0.05*\pgfkeysvalueof{/pgfplots/parent axis width},%%% added this
%     	  	height=0.5*\pgfkeysvalueof{/pgfplots/parent axis height},
%		plot graphics/node/.style={scale=1.33,anchor=south west,inner sep=0pt,}, %%% scale colorbar fill %%%
            ytick={0,0.20,0.40,0.60,0.80,1.00},
            yticklabels={0,0.20,0.40,0.60,0.80,1.00},
            yticklabel={\pgfmathprintnumber\tick},
            yticklabel style={%font=\footnotesize,
            		/pgf/number format/fixed,
			/pgf/number format/precision=2}
        },
%        point meta min=0,
%        point meta max=100,
        point meta min=0.0,
        point meta max=1.0,
        nodes near coords={\pgfmathprintnumber\pgfplotspointmeta},
        % ---------------------------------------------------------------------
        % show `nodes near coords' but adapt the style so that values
        % above a threshold get another style
        % (adapted from <http://tex.stackexchange.com/a/141006/95441>)
        % #1: the THRESHOLD after which we switch to a special display.
        nodes near coords black white/.style={
            % define the style of the nodes with "small" values
            small value/.style={
                yshift=-7pt,
%                text=white,
                text=black,
                /pgf/number format/fixed,
                /pgf/number format/precision=2
%                /pgf/number format/precision=0
            },
            % define the style of the nodes with "large" values
            large value/.style={
                yshift=-7pt,
%                text=black,
                text=white,
                /pgf/number format/fixed,
                /pgf/number format/precision=2
%                /pgf/number format/precision=0
            },
            every node near coord/.style={
                check for zero/.code={
                    \pgfmathfloatifflags{\pgfplotspointmeta}{0}{
                        % If meta=0, make the node a coordinate
                        % (which doesn't have text)
                        \pgfkeys{/tikz/coordinate}
                    }{
                        \begingroup
                        % this group is merely to switch to FPU locally.
                        % Might be unnecessary, but who knows.
                        \pgfkeys{/pgf/fpu}
                        \pgfmathparse{\pgfplotspointmeta<#1}
                        \global\let\result=\pgfmathresult
                        \endgroup
                        %
                        % simplifies debugging:
                        %\show\result
                        %
                        \pgfmathfloatcreate{1}{1.0}{0}
                        \let\ONE=\pgfmathresult
                        \ifx\result\ONE
                            % AH: our condition 'y < #1' is met.
                            \pgfkeysalso{/pgfplots/small value}
                        \else
                            % ok, proceed as usual.
                            \pgfkeysalso{/pgfplots/large value}
                        \fi
                    }
                },
                check for zero,
            },
        },
        % asign a value to the new style which is the threshold at which
        % the two style `small value' or `large value' are used
%        nodes near coords black white=50,
        nodes near coords black white=0.5,
        % -----------------------------------------------------------------
    ]
        \addplot[
            matrix plot,
            mesh/cols=7,
            point meta=explicit,draw=gray
        ] table [meta=C] {
            x y C
 0  0 0.97
 1  0 0.00
 2  0 0.01
 3  0 0.01
 4  0 0.01
 5  0 0.00
 6  0 0.00
 0  1 0.00
 1  1 0.87
 2  1 0.01
 3  1 0.03
 4  1 0.04
 5  1 0.04
 6  1 0.01
 0  2 0.00
 1  2 0.01
 2  2 0.93
 3  2 0.01
 4  2 0.02
 5  2 0.03
 6  2 0.00
 0  3 0.00
 1  3 0.01
 2  3 0.01
 3  3 0.82
 4  3 0.04
 5  3 0.12
 6  3 0.00
 0  4 0.01
 1  4 0.02
 2  4 0.02
 3  4 0.05
 4  4 0.86
 5  4 0.03
 6  4 0.01
 0  5 0.01
 1  5 0.01
 2  5 0.02
 3  5 0.07
 4  5 0.03
 5  5 0.86
 6  5 0.00
 0  6 0.00
 1  6 0.01
 2  6 0.01
 3  6 0.02
 4  6 0.01
 5  6 0.02
 6  6 0.93
        };
    \end{axis}
\end{tikzpicture}
%
%\caption{I'm confused~5!}\label{tab:CM5}
%\end{figure*}
\\[-0.5ex]
(c) RF & (d) SVM 
\\[-0.5ex]
\end{tabular}
\caption{Confusion matrices for HMM2Vec experiments}\label{fig:H2V}
\end{figure}
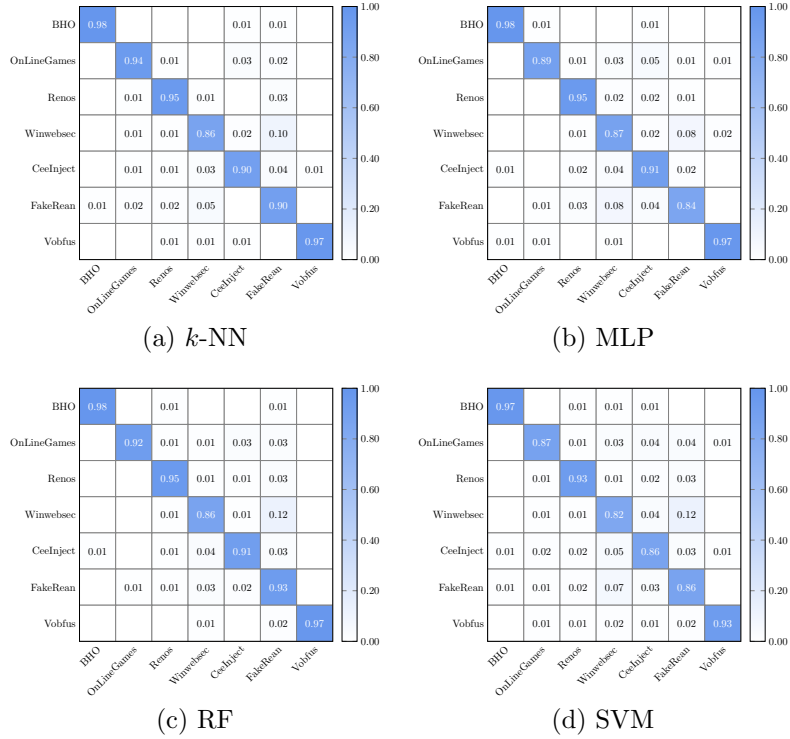

Based on the resulting feature vectors, we use the parameters in 
the HMM2Vec column of Table~\ref{tab:classSelect} to classify
the samples using~$\kNN{k}$, MLP, RF, and~SVM. The confusion
matrices for each of these cases is give in Figure~\ref{fig:H2V}.

The accuracies obtained
for~$\kNN{k}$, MLP, RF, and SVM based on HMM2Vec features
are~0.93, 0.91, 0.93, and~0.89, respectively. 
From the confusion matrices in Figure~\ref{fig:H2V}, we see that the greatest
source of misclassifications is between FakeRean and Winwebsec families.
In most---but not all---of our subsequent experiments, these two families
will prove to be the most challenging to distinguish.

%\begin{figure}[!htb]
%\centering
%\begin{tabular}{cc}
%\input figures/HMM-PCA/KNN.tex
%& 
%\input figures/HMM-PCA/MLP.tex
%\\[-1ex]
%(a) $\kNN{k}$ & (b) MLP \\
%\\[-1ex]
%\input figures/HMM-PCA/RF.tex
%& 
%\input figures/HMM-PCA/SVM.tex
%\\[-1ex]
%(c) RF & (d) SVM 
%\\[-0.5ex]
%\end{tabular}
%\caption{Confusion matrices for HMM2Vec-PCA experiments}\label{fig:H2V-PCA}
%\end{figure}

\subsection{PCA2Vec Results}

\begin{figure}[!htb]
\centering
\begin{tabular}{cc}
%\begin{tikzpicture}[scale=0.8,every node/.style={scale=0.8}]
\begin{tikzpicture}[scale=0.4]
    \begin{axis}[%colorbar/width=2.5mm,
        width=10cm,
        height=10cm,
%        colormap={blackwhite}{gray(0cm)=(1); gray(1cm)=(0.5)},
%	colormap={bluewhite}{color=(white) color=(blue)},
%	colormap={bluewhite}{color=(white) rgb255=(0,191,255)},
	colormap={bluewhite}{color=(white) rgb255=(100,149,237)},
        xticklabels={BHO,OnLineGames,Renos,Winwebsec,CeeInject,FakeRean,Vobfus},
        xtick={0,...,6},
        xtick style={draw=none},
	xticklabel style={anchor=east,rotate=45,yshift=-5pt},
        yticklabels={BHO,OnLineGames,Renos,Winwebsec,CeeInject,FakeRean,Vobfus},
        ytick={0,...,6},
        ytick style={draw=none},
        enlargelimits=false,
        colorbar,
        colorbar style={
%     	  	width=0.05*\pgfkeysvalueof{/pgfplots/parent axis width},%%% added this
%     	  	height=0.5*\pgfkeysvalueof{/pgfplots/parent axis height},
%		plot graphics/node/.style={scale=1.33,anchor=south west,inner sep=0pt,}, %%% scale colorbar fill %%%
            ytick={0,0.20,0.40,0.60,0.80,1.00},
            yticklabels={0,0.20,0.40,0.60,0.80,1.00},
            yticklabel={\pgfmathprintnumber\tick},
            yticklabel style={%font=\footnotesize,
            		/pgf/number format/fixed,
			/pgf/number format/precision=2}
        },
%        point meta min=0,
%        point meta max=100,
        point meta min=0.0,
        point meta max=1.0,
        nodes near coords={\pgfmathprintnumber\pgfplotspointmeta},
        % ---------------------------------------------------------------------
        % show `nodes near coords' but adapt the style so that values
        % above a threshold get another style
        % (adapted from <http://tex.stackexchange.com/a/141006/95441>)
        % #1: the THRESHOLD after which we switch to a special display.
        nodes near coords black white/.style={
            % define the style of the nodes with "small" values
            small value/.style={
                yshift=-7pt,
%                text=white,
                text=black,
                /pgf/number format/fixed,
                /pgf/number format/precision=2
%                /pgf/number format/precision=0
            },
            % define the style of the nodes with "large" values
            large value/.style={
                yshift=-7pt,
%                text=black,
                text=white,
                /pgf/number format/fixed,
                /pgf/number format/precision=2
%                /pgf/number format/precision=0
            },
            every node near coord/.style={
                check for zero/.code={
                    \pgfmathfloatifflags{\pgfplotspointmeta}{0}{
                        % If meta=0, make the node a coordinate
                        % (which doesn't have text)
                        \pgfkeys{/tikz/coordinate}
                    }{
                        \begingroup
                        % this group is merely to switch to FPU locally.
                        % Might be unnecessary, but who knows.
                        \pgfkeys{/pgf/fpu}
                        \pgfmathparse{\pgfplotspointmeta<#1}
                        \global\let\result=\pgfmathresult
                        \endgroup
                        %
                        % simplifies debugging:
                        %\show\result
                        %
                        \pgfmathfloatcreate{1}{1.0}{0}
                        \let\ONE=\pgfmathresult
                        \ifx\result\ONE
                            % AH: our condition 'y < #1' is met.
                            \pgfkeysalso{/pgfplots/small value}
                        \else
                            % ok, proceed as usual.
                            \pgfkeysalso{/pgfplots/large value}
                        \fi
                    }
                },
                check for zero,
            },
        },
        % asign a value to the new style which is the threshold at which
        % the two style `small value' or `large value' are used
%        nodes near coords black white=50,
        nodes near coords black white=0.5,
        % -----------------------------------------------------------------
    ]
        \addplot[
            matrix plot,
            mesh/cols=7,
            point meta=explicit,draw=gray
        ] table [meta=C] {
            x y C
 0  0 0.98
 1  0 0.00
 2  0 0.01
 3  0 0.00
 4  0 0.00
 5  0 0.01
 6  0 0.00
 0  1 0.00
 1  1 0.81
 2  1 0.04
 3  1 0.07
 4  1 0.03
 5  1 0.05
 6  1 0.00
 0  2 0.00
 1  2 0.01
 2  2 0.81
 3  2 0.05
 4  2 0.06
 5  2 0.07
 6  2 0.00
 0  3 0.00
 1  3 0.02
 2  3 0.05
 3  3 0.68
 4  3 0.04
 5  3 0.20
 6  3 0.01
 0  4 0.01
 1  4 0.00
 2  4 0.05
 3  4 0.05
 4  4 0.84
 5  4 0.05
 6  4 0.00
 0  5 0.00
 1  5 0.03
 2  5 0.02
 3  5 0.05
 4  5 0.01
 5  5 0.89
 6  5 0.00
 0  6 0.00
 1  6 0.02
 2  6 0.02
 3  6 0.00
 4  6 0.04
 5  6 0.04
 6  6 0.88
        };
    \end{axis}
\end{tikzpicture}
%
%\caption{I'm confused~5!}\label{tab:CM5}
%\end{figure*}
& 
%\begin{tikzpicture}[scale=0.8,every node/.style={scale=0.8}]
\begin{tikzpicture}[scale=0.4]
    \begin{axis}[%colorbar/width=2.5mm,
        width=10cm,
        height=10cm,
%        colormap={blackwhite}{gray(0cm)=(1); gray(1cm)=(0.5)},
%	colormap={bluewhite}{color=(white) color=(blue)},
%	colormap={bluewhite}{color=(white) rgb255=(0,191,255)},
	colormap={bluewhite}{color=(white) rgb255=(100,149,237)},
        xticklabels={BHO,OnLineGames,Renos,Winwebsec,CeeInject,FakeRean,Vobfus},
        xtick={0,...,6},
        xtick style={draw=none},
	xticklabel style={anchor=east,rotate=45,yshift=-5pt},
        yticklabels={BHO,OnLineGames,Renos,Winwebsec,CeeInject,FakeRean,Vobfus},
        ytick={0,...,6},
        ytick style={draw=none},
        enlargelimits=false,
        colorbar,
        colorbar style={
%     	  	width=0.05*\pgfkeysvalueof{/pgfplots/parent axis width},%%% added this
%     	  	height=0.5*\pgfkeysvalueof{/pgfplots/parent axis height},
%		plot graphics/node/.style={scale=1.33,anchor=south west,inner sep=0pt,}, %%% scale colorbar fill %%%
            ytick={0,0.20,0.40,0.60,0.80,1.00},
            yticklabels={0,0.20,0.40,0.60,0.80,1.00},
            yticklabel={\pgfmathprintnumber\tick},
            yticklabel style={%font=\footnotesize,
            		/pgf/number format/fixed,
			/pgf/number format/precision=2}
        },
%        point meta min=0,
%        point meta max=100,
        point meta min=0.0,
        point meta max=1.0,
        nodes near coords={\pgfmathprintnumber\pgfplotspointmeta},
        % ---------------------------------------------------------------------
        % show `nodes near coords' but adapt the style so that values
        % above a threshold get another style
        % (adapted from <http://tex.stackexchange.com/a/141006/95441>)
        % #1: the THRESHOLD after which we switch to a special display.
        nodes near coords black white/.style={
            % define the style of the nodes with "small" values
            small value/.style={
                yshift=-7pt,
%                text=white,
                text=black,
                /pgf/number format/fixed,
                /pgf/number format/precision=2
%                /pgf/number format/precision=0
            },
            % define the style of the nodes with "large" values
            large value/.style={
                yshift=-7pt,
%                text=black,
                text=white,
                /pgf/number format/fixed,
                /pgf/number format/precision=2
%                /pgf/number format/precision=0
            },
            every node near coord/.style={
                check for zero/.code={
                    \pgfmathfloatifflags{\pgfplotspointmeta}{0}{
                        % If meta=0, make the node a coordinate
                        % (which doesn't have text)
                        \pgfkeys{/tikz/coordinate}
                    }{
                        \begingroup
                        % this group is merely to switch to FPU locally.
                        % Might be unnecessary, but who knows.
                        \pgfkeys{/pgf/fpu}
                        \pgfmathparse{\pgfplotspointmeta<#1}
                        \global\let\result=\pgfmathresult
                        \endgroup
                        %
                        % simplifies debugging:
                        %\show\result
                        %
                        \pgfmathfloatcreate{1}{1.0}{0}
                        \let\ONE=\pgfmathresult
                        \ifx\result\ONE
                            % AH: our condition 'y < #1' is met.
                            \pgfkeysalso{/pgfplots/small value}
                        \else
                            % ok, proceed as usual.
                            \pgfkeysalso{/pgfplots/large value}
                        \fi
                    }
                },
                check for zero,
            },
        },
        % asign a value to the new style which is the threshold at which
        % the two style `small value' or `large value' are used
%        nodes near coords black white=50,
        nodes near coords black white=0.5,
        % -----------------------------------------------------------------
    ]
        \addplot[
            matrix plot,
            mesh/cols=7,
            point meta=explicit,draw=gray
        ] table [meta=C] {
            x y C
 0  0 0.95
 1  0 0.01
 2  0 0.02
 3  0 0.02
 4  0 0.00
 5  0 0.00
 6  0 0.00
 0  1 0.01
 1  1 0.80
 2  1 0.01
 3  1 0.08
 4  1 0.04
 5  1 0.05
 6  1 0.01
 0  2 0.01
 1  2 0.08
 2  2 0.75
 3  2 0.07
 4  2 0.04
 5  2 0.04
 6  2 0.01
 0  3 0.01
 1  3 0.08
 2  3 0.11
 3  3 0.68
 4  3 0.02
 5  3 0.07
 6  3 0.03
 0  4 0.01
 1  4 0.02
 2  4 0.07
 3  4 0.06
 4  4 0.74
 5  4 0.08
 6  4 0.02
 0  5 0.01
 1  5 0.05
 2  5 0.05
 3  5 0.07
 4  5 0.04
 5  5 0.73
 6  5 0.05
 0  6 0.01
 1  6 0.01
 2  6 0.04
 3  6 0.05
 4  6 0.04
 5  6 0.03
 6  6 0.82
        };
    \end{axis}
\end{tikzpicture}
%
%\caption{I'm confused~5!}\label{tab:CM5}
%\end{figure*}
\\[-0.5ex]
(a) $\kNN{k}$ & (b) MLP \\
\\[-0.5ex]
%\begin{tikzpicture}[scale=0.8,every node/.style={scale=0.8}]
\begin{tikzpicture}[scale=0.4]
    \begin{axis}[%colorbar/width=2.5mm,
        width=10cm,
        height=10cm,
%        colormap={blackwhite}{gray(0cm)=(1); gray(1cm)=(0.5)},
%	colormap={bluewhite}{color=(white) color=(blue)},
%	colormap={bluewhite}{color=(white) rgb255=(0,191,255)},
	colormap={bluewhite}{color=(white) rgb255=(100,149,237)},
        xticklabels={BHO,OnLineGames,Renos,Winwebsec,CeeInject,FakeRean,Vobfus},
        xtick={0,...,6},
        xtick style={draw=none},
	xticklabel style={anchor=east,rotate=45,yshift=-5pt},
        yticklabels={BHO,OnLineGames,Renos,Winwebsec,CeeInject,FakeRean,Vobfus},
        ytick={0,...,6},
        ytick style={draw=none},
        enlargelimits=false,
        colorbar,
        colorbar style={
%     	  	width=0.05*\pgfkeysvalueof{/pgfplots/parent axis width},%%% added this
%     	  	height=0.5*\pgfkeysvalueof{/pgfplots/parent axis height},
%		plot graphics/node/.style={scale=1.33,anchor=south west,inner sep=0pt,}, %%% scale colorbar fill %%%
            ytick={0,0.20,0.40,0.60,0.80,1.00},
            yticklabels={0,0.20,0.40,0.60,0.80,1.00},
            yticklabel={\pgfmathprintnumber\tick},
            yticklabel style={%font=\footnotesize,
            		/pgf/number format/fixed,
			/pgf/number format/precision=2}
        },
%        point meta min=0,
%        point meta max=100,
        point meta min=0.0,
        point meta max=1.0,
        nodes near coords={\pgfmathprintnumber\pgfplotspointmeta},
        % ---------------------------------------------------------------------
        % show `nodes near coords' but adapt the style so that values
        % above a threshold get another style
        % (adapted from <http://tex.stackexchange.com/a/141006/95441>)
        % #1: the THRESHOLD after which we switch to a special display.
        nodes near coords black white/.style={
            % define the style of the nodes with "small" values
            small value/.style={
                yshift=-7pt,
%                text=white,
                text=black,
                /pgf/number format/fixed,
                /pgf/number format/precision=2
%                /pgf/number format/precision=0
            },
            % define the style of the nodes with "large" values
            large value/.style={
                yshift=-7pt,
%                text=black,
                text=white,
                /pgf/number format/fixed,
                /pgf/number format/precision=2
%                /pgf/number format/precision=0
            },
            every node near coord/.style={
                check for zero/.code={
                    \pgfmathfloatifflags{\pgfplotspointmeta}{0}{
                        % If meta=0, make the node a coordinate
                        % (which doesn't have text)
                        \pgfkeys{/tikz/coordinate}
                    }{
                        \begingroup
                        % this group is merely to switch to FPU locally.
                        % Might be unnecessary, but who knows.
                        \pgfkeys{/pgf/fpu}
                        \pgfmathparse{\pgfplotspointmeta<#1}
                        \global\let\result=\pgfmathresult
                        \endgroup
                        %
                        % simplifies debugging:
                        %\show\result
                        %
                        \pgfmathfloatcreate{1}{1.0}{0}
                        \let\ONE=\pgfmathresult
                        \ifx\result\ONE
                            % AH: our condition 'y < #1' is met.
                            \pgfkeysalso{/pgfplots/small value}
                        \else
                            % ok, proceed as usual.
                            \pgfkeysalso{/pgfplots/large value}
                        \fi
                    }
                },
                check for zero,
            },
        },
        % asign a value to the new style which is the threshold at which
        % the two style `small value' or `large value' are used
%        nodes near coords black white=50,
        nodes near coords black white=0.5,
        % -----------------------------------------------------------------
    ]
        \addplot[
            matrix plot,
            mesh/cols=7,
            point meta=explicit,draw=gray
        ] table [meta=C] {
            x y C
 0  0 0.98
 1  0 0.00
 2  0 0.00
 3  0 0.02
 4  0 0.00
 5  0 0.00
 6  0 0.00
 0  1 0.00
 1  1 0.86
 2  1 0.06
 3  1 0.04
 4  1 0.01
 5  1 0.01
 6  1 0.02
 0  2 0.00
 1  2 0.01
 2  2 0.90
 3  2 0.04
 4  2 0.04
 5  2 0.01
 6  2 0.00
 0  3 0.00
 1  3 0.01
 2  3 0.10
 3  3 0.81
 4  3 0.03
 5  3 0.03
 6  3 0.02
 0  4 0.00
 1  4 0.01
 2  4 0.06
 3  4 0.06
 4  4 0.85
 5  4 0.02
 6  4 0.00
 0  5 0.00
 1  5 0.00
 2  5 0.06
 3  5 0.02
 4  5 0.02
 5  5 0.88
 6  5 0.02
 0  6 0.00
 1  6 0.02
 2  6 0.03
 3  6 0.01
 4  6 0.02
 5  6 0.02
 6  6 0.90
        };
    \end{axis}
\end{tikzpicture}
%
%\caption{I'm confused~5!}\label{tab:CM5}
%\end{figure*}
& 
%\begin{tikzpicture}[scale=0.8,every node/.style={scale=0.8}]
\begin{tikzpicture}[scale=0.4]
    \begin{axis}[%colorbar/width=2.5mm,
        width=10cm,
        height=10cm,
%        colormap={blackwhite}{gray(0cm)=(1); gray(1cm)=(0.5)},
%	colormap={bluewhite}{color=(white) color=(blue)},
%	colormap={bluewhite}{color=(white) rgb255=(0,191,255)},
	colormap={bluewhite}{color=(white) rgb255=(100,149,237)},
        xticklabels={BHO,OnLineGames,Renos,Winwebsec,CeeInject,FakeRean,Vobfus},
        xtick={0,...,6},
        xtick style={draw=none},
	xticklabel style={anchor=east,rotate=45,yshift=-5pt},
        yticklabels={BHO,OnLineGames,Renos,Winwebsec,CeeInject,FakeRean,Vobfus},
        ytick={0,...,6},
        ytick style={draw=none},
        enlargelimits=false,
        colorbar,
        colorbar style={
%     	  	width=0.05*\pgfkeysvalueof{/pgfplots/parent axis width},%%% added this
%     	  	height=0.5*\pgfkeysvalueof{/pgfplots/parent axis height},
%		plot graphics/node/.style={scale=1.33,anchor=south west,inner sep=0pt,}, %%% scale colorbar fill %%%
            ytick={0,0.20,0.40,0.60,0.80,1.00},
            yticklabels={0,0.20,0.40,0.60,0.80,1.00},
            yticklabel={\pgfmathprintnumber\tick},
            yticklabel style={%font=\footnotesize,
            		/pgf/number format/fixed,
			/pgf/number format/precision=2}
        },
%        point meta min=0,
%        point meta max=100,
        point meta min=0.0,
        point meta max=1.0,
        nodes near coords={\pgfmathprintnumber\pgfplotspointmeta},
        % ---------------------------------------------------------------------
        % show `nodes near coords' but adapt the style so that values
        % above a threshold get another style
        % (adapted from <http://tex.stackexchange.com/a/141006/95441>)
        % #1: the THRESHOLD after which we switch to a special display.
        nodes near coords black white/.style={
            % define the style of the nodes with "small" values
            small value/.style={
                yshift=-7pt,
%                text=white,
                text=black,
                /pgf/number format/fixed,
                /pgf/number format/precision=2
%                /pgf/number format/precision=0
            },
            % define the style of the nodes with "large" values
            large value/.style={
                yshift=-7pt,
%                text=black,
                text=white,
                /pgf/number format/fixed,
                /pgf/number format/precision=2
%                /pgf/number format/precision=0
            },
            every node near coord/.style={
                check for zero/.code={
                    \pgfmathfloatifflags{\pgfplotspointmeta}{0}{
                        % If meta=0, make the node a coordinate
                        % (which doesn't have text)
                        \pgfkeys{/tikz/coordinate}
                    }{
                        \begingroup
                        % this group is merely to switch to FPU locally.
                        % Might be unnecessary, but who knows.
                        \pgfkeys{/pgf/fpu}
                        \pgfmathparse{\pgfplotspointmeta<#1}
                        \global\let\result=\pgfmathresult
                        \endgroup
                        %
                        % simplifies debugging:
                        %\show\result
                        %
                        \pgfmathfloatcreate{1}{1.0}{0}
                        \let\ONE=\pgfmathresult
                        \ifx\result\ONE
                            % AH: our condition 'y < #1' is met.
                            \pgfkeysalso{/pgfplots/small value}
                        \else
                            % ok, proceed as usual.
                            \pgfkeysalso{/pgfplots/large value}
                        \fi
                    }
                },
                check for zero,
            },
        },
        % asign a value to the new style which is the threshold at which
        % the two style `small value' or `large value' are used
%        nodes near coords black white=50,
        nodes near coords black white=0.5,
        % -----------------------------------------------------------------
    ]
        \addplot[
            matrix plot,
            mesh/cols=7,
            point meta=explicit,draw=gray
        ] table [meta=C] {
            x y C
 0  0 0.97
 1  0 0.00
 2  0 0.01
 3  0 0.02
 4  0 0.00
 5  0 0.00
 6  0 0.00
 0  1 0.00
 1  1 0.71
 2  1 0.03
 3  1 0.14
 4  1 0.03
 5  1 0.08
 6  1 0.01
 0  2 0.02
 1  2 0.04
 2  2 0.64
 3  2 0.15
 4  2 0.05
 5  2 0.06
 6  2 0.04
 0  3 0.00
 1  3 0.06
 2  3 0.05
 3  3 0.82
 4  3 0.03
 5  3 0.04
 6  3 0.00
 0  4 0.00
 1  4 0.00
 2  4 0.04
 3  4 0.12
 4  4 0.78
 5  4 0.03
 6  4 0.03
 0  5 0.01
 1  5 0.05
 2  5 0.04
 3  5 0.37
 4  5 0.03
 5  5 0.48
 6  5 0.02
 0  6 0.01
 1  6 0.02
 2  6 0.01
 3  6 0.02
 4  6 0.02
 5  6 0.01
 6  6 0.91
        };
    \end{axis}
\end{tikzpicture}
%
%\caption{I'm confused~5!}\label{tab:CM5}
%\end{figure*}
\\[-0.5ex]
(c) RF & (d) SVM 
\\[-0.5ex]
\end{tabular}
\caption{Confusion matrices for PCA2Vec experiments}\label{fig:P2V}
\end{figure}
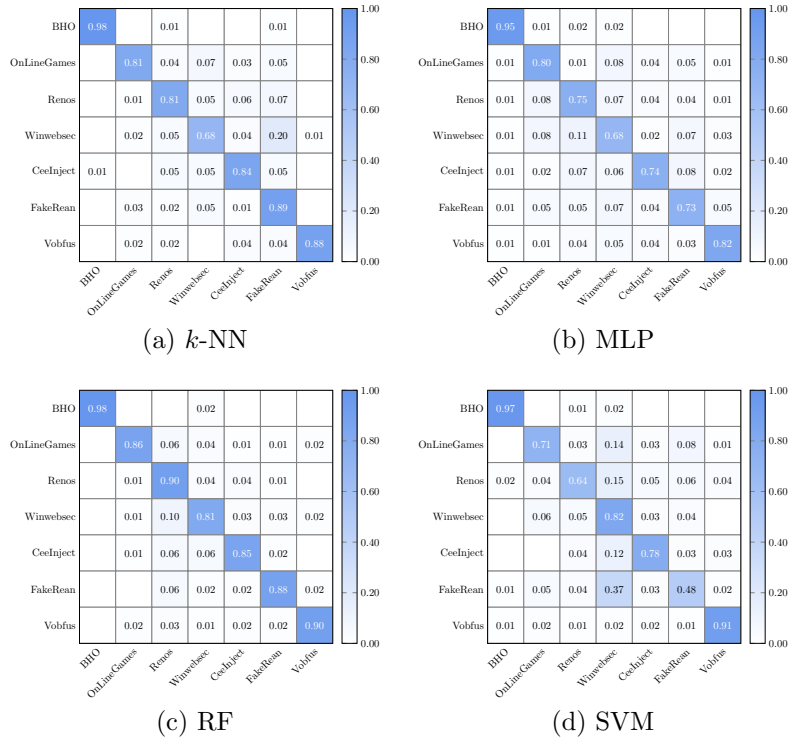

For our PCA2Vec experiments, we generate embedding vectors for each of the~7000
samples in our training set, as discussed in Section~\ref{sect:p2v}. 
%%%%% ????? Training/testing split ?????
We then train and classify the~7000 malware samples using these 
PCA2Vec feature vectors. %, and using the same four classifiers as above. 
The confusion matrices for these experiments are summarized in
Figure~\ref{fig:P2V}.

As above, each model is based on the~20 most frequent opcodes, which
gives us a~$20\times 20$ PMI matrix. For consistency
with the HMM2Vec experiments above, 
we consider the two most dominant eigenvectors, and
for consistency with the Word2Vec models below,
we use a window size of~$W=10$ when constructing the PMI matrix.
The resulting projection into the eigenspace is~$2\times 20$, which we
vectorize to obtain a feature vector of length~40.

The accuracies obtained
for~$\kNN{k}$, MLP, RF, and SVM based on PCA2Vec features
are~0.84, 0.78, 0.88, and~0.76, respectively. 
From these numbers, we see that PCA2Vec performed
poorly for each of the classifiers considered, as compared to HMM2Vec.

%\begin{figure}[!htb]
%\centering
%\begin{tabular}{cc}
%\input figures/P2V-PCA/KNN.tex
%& 
%\input figures/P2V-PCA/MLP.tex
%\\[-1ex]
%(a) $\kNN{k}$ & (b) MLP \\
%\\[-1ex]
%\input figures/P2V-PCA/RF.tex
%& 
%\input figures/P2V-PCA/SVM.tex
%\\[-1ex]
%(c) RF & (d) SVM 
%\\[-0.5ex]
%\end{tabular}
%\caption{Confusion matrices for PCA2Vec-PCA experiments}\label{fig:P2V-PCA}
%\end{figure}

\subsection{Word2Vec Results}

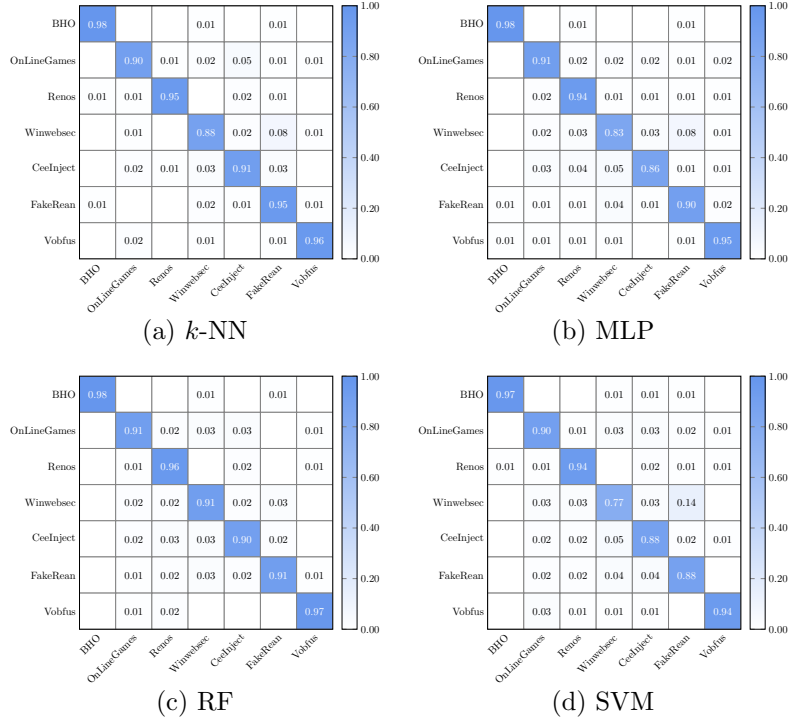
\begin{figure}[!htb]
\centering
\begin{tabular}{cc}
%\begin{tikzpicture}[scale=0.8,every node/.style={scale=0.8}]
\begin{tikzpicture}[scale=0.4]
    \begin{axis}[%colorbar/width=2.5mm,
        width=10cm,
        height=10cm,
%        colormap={blackwhite}{gray(0cm)=(1); gray(1cm)=(0.5)},
%	colormap={bluewhite}{color=(white) color=(blue)},
%	colormap={bluewhite}{color=(white) rgb255=(0,191,255)},
	colormap={bluewhite}{color=(white) rgb255=(100,149,237)},
        xticklabels={BHO,OnLineGames,Renos,Winwebsec,CeeInject,FakeRean,Vobfus},
        xtick={0,...,6},
        xtick style={draw=none},
	xticklabel style={anchor=east,rotate=45,yshift=-5pt},
        yticklabels={BHO,OnLineGames,Renos,Winwebsec,CeeInject,FakeRean,Vobfus},
        ytick={0,...,6},
        ytick style={draw=none},
        enlargelimits=false,
        colorbar,
        colorbar style={
%     	  	width=0.05*\pgfkeysvalueof{/pgfplots/parent axis width},%%% added this
%     	  	height=0.5*\pgfkeysvalueof{/pgfplots/parent axis height},
%		plot graphics/node/.style={scale=1.33,anchor=south west,inner sep=0pt,}, %%% scale colorbar fill %%%
            ytick={0,0.20,0.40,0.60,0.80,1.00},
            yticklabels={0,0.20,0.40,0.60,0.80,1.00},
            yticklabel={\pgfmathprintnumber\tick},
            yticklabel style={%font=\footnotesize,
            		/pgf/number format/fixed,
			/pgf/number format/precision=2}
        },
%        point meta min=0,
%        point meta max=100,
        point meta min=0.0,
        point meta max=1.0,
        nodes near coords={\pgfmathprintnumber\pgfplotspointmeta},
        % ---------------------------------------------------------------------
        % show `nodes near coords' but adapt the style so that values
        % above a threshold get another style
        % (adapted from <http://tex.stackexchange.com/a/141006/95441>)
        % #1: the THRESHOLD after which we switch to a special display.
        nodes near coords black white/.style={
            % define the style of the nodes with "small" values
            small value/.style={
                yshift=-7pt,
%                text=white,
                text=black,
                /pgf/number format/fixed,
                /pgf/number format/precision=2
%                /pgf/number format/precision=0
            },
            % define the style of the nodes with "large" values
            large value/.style={
                yshift=-7pt,
%                text=black,
                text=white,
                /pgf/number format/fixed,
                /pgf/number format/precision=2
%                /pgf/number format/precision=0
            },
            every node near coord/.style={
                check for zero/.code={
                    \pgfmathfloatifflags{\pgfplotspointmeta}{0}{
                        % If meta=0, make the node a coordinate
                        % (which doesn't have text)
                        \pgfkeys{/tikz/coordinate}
                    }{
                        \begingroup
                        % this group is merely to switch to FPU locally.
                        % Might be unnecessary, but who knows.
                        \pgfkeys{/pgf/fpu}
                        \pgfmathparse{\pgfplotspointmeta<#1}
                        \global\let\result=\pgfmathresult
                        \endgroup
                        %
                        % simplifies debugging:
                        %\show\result
                        %
                        \pgfmathfloatcreate{1}{1.0}{0}
                        \let\ONE=\pgfmathresult
                        \ifx\result\ONE
                            % AH: our condition 'y < #1' is met.
                            \pgfkeysalso{/pgfplots/small value}
                        \else
                            % ok, proceed as usual.
                            \pgfkeysalso{/pgfplots/large value}
                        \fi
                    }
                },
                check for zero,
            },
        },
        % asign a value to the new style which is the threshold at which
        % the two style `small value' or `large value' are used
%        nodes near coords black white=50,
        nodes near coords black white=0.5,
        % -----------------------------------------------------------------
    ]
        \addplot[
            matrix plot,
            mesh/cols=7,
            point meta=explicit,draw=gray
        ] table [meta=C] {
            x y C
 0  0 0.98
 1  0 0.00
 2  0 0.00
 3  0 0.01
 4  0 0.00
 5  0 0.01
 6  0 0.00
 0  1 0.00
 1  1 0.90
 2  1 0.01
 3  1 0.02
 4  1 0.05
 5  1 0.01
 6  1 0.01
 0  2 0.01
 1  2 0.01
 2  2 0.95
 3  2 0.00
 4  2 0.02
 5  2 0.01
 6  2 0.00
 0  3 0.00
 1  3 0.01
 2  3 0.00
 3  3 0.88
 4  3 0.02
 5  3 0.08
 6  3 0.01
 0  4 0.00
 1  4 0.02
 2  4 0.01
 3  4 0.03
 4  4 0.91
 5  4 0.03
 6  4 0.00
 0  5 0.01
 1  5 0.00
 2  5 0.00
 3  5 0.02
 4  5 0.01
 5  5 0.95
 6  5 0.01
 0  6 0.00
 1  6 0.02
 2  6 0.00
 3  6 0.01
 4  6 0.00
 5  6 0.01
 6  6 0.96
        };
    \end{axis}
\end{tikzpicture}
%
%\caption{I'm confused~5!}\label{tab:CM5}
%\end{figure*}
& 
%\begin{tikzpicture}[scale=0.8,every node/.style={scale=0.8}]
\begin{tikzpicture}[scale=0.4]
    \begin{axis}[%colorbar/width=2.5mm,
        width=10cm,
        height=10cm,
%        colormap={blackwhite}{gray(0cm)=(1); gray(1cm)=(0.5)},
%	colormap={bluewhite}{color=(white) color=(blue)},
%	colormap={bluewhite}{color=(white) rgb255=(0,191,255)},
	colormap={bluewhite}{color=(white) rgb255=(100,149,237)},
        xticklabels={BHO,OnLineGames,Renos,Winwebsec,CeeInject,FakeRean,Vobfus},
        xtick={0,...,6},
        xtick style={draw=none},
	xticklabel style={anchor=east,rotate=45,yshift=-5pt},
        yticklabels={BHO,OnLineGames,Renos,Winwebsec,CeeInject,FakeRean,Vobfus},
        ytick={0,...,6},
        ytick style={draw=none},
        enlargelimits=false,
        colorbar,
        colorbar style={
%     	  	width=0.05*\pgfkeysvalueof{/pgfplots/parent axis width},%%% added this
%     	  	height=0.5*\pgfkeysvalueof{/pgfplots/parent axis height},
%		plot graphics/node/.style={scale=1.33,anchor=south west,inner sep=0pt,}, %%% scale colorbar fill %%%
            ytick={0,0.20,0.40,0.60,0.80,1.00},
            yticklabels={0,0.20,0.40,0.60,0.80,1.00},
            yticklabel={\pgfmathprintnumber\tick},
            yticklabel style={%font=\footnotesize,
            		/pgf/number format/fixed,
			/pgf/number format/precision=2}
        },
%        point meta min=0,
%        point meta max=100,
        point meta min=0.0,
        point meta max=1.0,
        nodes near coords={\pgfmathprintnumber\pgfplotspointmeta},
        % ---------------------------------------------------------------------
        % show `nodes near coords' but adapt the style so that values
        % above a threshold get another style
        % (adapted from <http://tex.stackexchange.com/a/141006/95441>)
        % #1: the THRESHOLD after which we switch to a special display.
        nodes near coords black white/.style={
            % define the style of the nodes with "small" values
            small value/.style={
                yshift=-7pt,
%                text=white,
                text=black,
                /pgf/number format/fixed,
                /pgf/number format/precision=2
%                /pgf/number format/precision=0
            },
            % define the style of the nodes with "large" values
            large value/.style={
                yshift=-7pt,
%                text=black,
                text=white,
                /pgf/number format/fixed,
                /pgf/number format/precision=2
%                /pgf/number format/precision=0
            },
            every node near coord/.style={
                check for zero/.code={
                    \pgfmathfloatifflags{\pgfplotspointmeta}{0}{
                        % If meta=0, make the node a coordinate
                        % (which doesn't have text)
                        \pgfkeys{/tikz/coordinate}
                    }{
                        \begingroup
                        % this group is merely to switch to FPU locally.
                        % Might be unnecessary, but who knows.
                        \pgfkeys{/pgf/fpu}
                        \pgfmathparse{\pgfplotspointmeta<#1}
                        \global\let\result=\pgfmathresult
                        \endgroup
                        %
                        % simplifies debugging:
                        %\show\result
                        %
                        \pgfmathfloatcreate{1}{1.0}{0}
                        \let\ONE=\pgfmathresult
                        \ifx\result\ONE
                            % AH: our condition 'y < #1' is met.
                            \pgfkeysalso{/pgfplots/small value}
                        \else
                            % ok, proceed as usual.
                            \pgfkeysalso{/pgfplots/large value}
                        \fi
                    }
                },
                check for zero,
            },
        },
        % asign a value to the new style which is the threshold at which
        % the two style `small value' or `large value' are used
%        nodes near coords black white=50,
        nodes near coords black white=0.5,
        % -----------------------------------------------------------------
    ]
        \addplot[
            matrix plot,
            mesh/cols=7,
            point meta=explicit,draw=gray
        ] table [meta=C] {
            x y C
 0  0 0.98
 1  0 0.00
 2  0 0.01
 3  0 0.00
 4  0 0.00
 5  0 0.01
 6  0 0.00
 0  1 0.00
 1  1 0.91
 2  1 0.02
 3  1 0.02
 4  1 0.02
 5  1 0.01
 6  1 0.02
 0  2 0.00
 1  2 0.02
 2  2 0.94
 3  2 0.01
 4  2 0.01
 5  2 0.01
 6  2 0.01
 0  3 0.00
 1  3 0.02
 2  3 0.03
 3  3 0.83
 4  3 0.03
 5  3 0.08
 6  3 0.01
 0  4 0.00
 1  4 0.03
 2  4 0.04
 3  4 0.05
 4  4 0.86
 5  4 0.01
 6  4 0.01
 0  5 0.01
 1  5 0.01
 2  5 0.01
 3  5 0.04
 4  5 0.01
 5  5 0.90
 6  5 0.02
 0  6 0.01
 1  6 0.01
 2  6 0.01
 3  6 0.01
 4  6 0.00
 5  6 0.01
 6  6 0.95
        };
    \end{axis}
\end{tikzpicture}
%
%\caption{I'm confused~5!}\label{tab:CM5}
%\end{figure*}
\\[-1ex]
(a) $\kNN{k}$ & (b) MLP \\
\\[-1ex]
%\begin{tikzpicture}[scale=0.8,every node/.style={scale=0.8}]
\begin{tikzpicture}[scale=0.4]
    \begin{axis}[%colorbar/width=2.5mm,
        width=10cm,
        height=10cm,
%        colormap={blackwhite}{gray(0cm)=(1); gray(1cm)=(0.5)},
%	colormap={bluewhite}{color=(white) color=(blue)},
%	colormap={bluewhite}{color=(white) rgb255=(0,191,255)},
	colormap={bluewhite}{color=(white) rgb255=(100,149,237)},
        xticklabels={BHO,OnLineGames,Renos,Winwebsec,CeeInject,FakeRean,Vobfus},
        xtick={0,...,6},
        xtick style={draw=none},
	xticklabel style={anchor=east,rotate=45,yshift=-5pt},
        yticklabels={BHO,OnLineGames,Renos,Winwebsec,CeeInject,FakeRean,Vobfus},
        ytick={0,...,6},
        ytick style={draw=none},
        enlargelimits=false,
        colorbar,
        colorbar style={
%     	  	width=0.05*\pgfkeysvalueof{/pgfplots/parent axis width},%%% added this
%     	  	height=0.5*\pgfkeysvalueof{/pgfplots/parent axis height},
%		plot graphics/node/.style={scale=1.33,anchor=south west,inner sep=0pt,}, %%% scale colorbar fill %%%
            ytick={0,0.20,0.40,0.60,0.80,1.00},
            yticklabels={0,0.20,0.40,0.60,0.80,1.00},
            yticklabel={\pgfmathprintnumber\tick},
            yticklabel style={%font=\footnotesize,
            		/pgf/number format/fixed,
			/pgf/number format/precision=2}
        },
%        point meta min=0,
%        point meta max=100,
        point meta min=0.0,
        point meta max=1.0,
        nodes near coords={\pgfmathprintnumber\pgfplotspointmeta},
        % ---------------------------------------------------------------------
        % show `nodes near coords' but adapt the style so that values
        % above a threshold get another style
        % (adapted from <http://tex.stackexchange.com/a/141006/95441>)
        % #1: the THRESHOLD after which we switch to a special display.
        nodes near coords black white/.style={
            % define the style of the nodes with "small" values
            small value/.style={
                yshift=-7pt,
%                text=white,
                text=black,
                /pgf/number format/fixed,
                /pgf/number format/precision=2
%                /pgf/number format/precision=0
            },
            % define the style of the nodes with "large" values
            large value/.style={
                yshift=-7pt,
%                text=black,
                text=white,
                /pgf/number format/fixed,
                /pgf/number format/precision=2
%                /pgf/number format/precision=0
            },
            every node near coord/.style={
                check for zero/.code={
                    \pgfmathfloatifflags{\pgfplotspointmeta}{0}{
                        % If meta=0, make the node a coordinate
                        % (which doesn't have text)
                        \pgfkeys{/tikz/coordinate}
                    }{
                        \begingroup
                        % this group is merely to switch to FPU locally.
                        % Might be unnecessary, but who knows.
                        \pgfkeys{/pgf/fpu}
                        \pgfmathparse{\pgfplotspointmeta<#1}
                        \global\let\result=\pgfmathresult
                        \endgroup
                        %
                        % simplifies debugging:
                        %\show\result
                        %
                        \pgfmathfloatcreate{1}{1.0}{0}
                        \let\ONE=\pgfmathresult
                        \ifx\result\ONE
                            % AH: our condition 'y < #1' is met.
                            \pgfkeysalso{/pgfplots/small value}
                        \else
                            % ok, proceed as usual.
                            \pgfkeysalso{/pgfplots/large value}
                        \fi
                    }
                },
                check for zero,
            },
        },
        % asign a value to the new style which is the threshold at which
        % the two style `small value' or `large value' are used
%        nodes near coords black white=50,
        nodes near coords black white=0.5,
        % -----------------------------------------------------------------
    ]
        \addplot[
            matrix plot,
            mesh/cols=7,
            point meta=explicit,draw=gray
        ] table [meta=C] {
            x y C
 0  0 0.98
 1  0 0.00
 2  0 0.00
 3  0 0.01
 4  0 0.00
 5  0 0.01
 6  0 0.00
 0  1 0.00
 1  1 0.91
 2  1 0.02
 3  1 0.03
 4  1 0.03
 5  1 0.00
 6  1 0.01
 0  2 0.00
 1  2 0.01
 2  2 0.96
 3  2 0.00
 4  2 0.02
 5  2 0.00
 6  2 0.01
 0  3 0.00
 1  3 0.02
 2  3 0.02
 3  3 0.91
 4  3 0.02
 5  3 0.03
 6  3 0.00
 0  4 0.00
 1  4 0.02
 2  4 0.03
 3  4 0.03
 4  4 0.90
 5  4 0.02
 6  4 0.00
 0  5 0.00
 1  5 0.01
 2  5 0.02
 3  5 0.03
 4  5 0.02
 5  5 0.91
 6  5 0.01
 0  6 0.00
 1  6 0.01
 2  6 0.02
 3  6 0.00
 4  6 0.00
 5  6 0.00
 6  6 0.97
        };
    \end{axis}
\end{tikzpicture}
%
%\caption{I'm confused~5!}\label{tab:CM5}
%\end{figure*}
& 
%\begin{tikzpicture}[scale=0.8,every node/.style={scale=0.8}]
\begin{tikzpicture}[scale=0.4]
    \begin{axis}[%colorbar/width=2.5mm,
        width=10cm,
        height=10cm,
%        colormap={blackwhite}{gray(0cm)=(1); gray(1cm)=(0.5)},
%	colormap={bluewhite}{color=(white) color=(blue)},
%	colormap={bluewhite}{color=(white) rgb255=(0,191,255)},
	colormap={bluewhite}{color=(white) rgb255=(100,149,237)},
        xticklabels={BHO,OnLineGames,Renos,Winwebsec,CeeInject,FakeRean,Vobfus},
        xtick={0,...,6},
        xtick style={draw=none},
	xticklabel style={anchor=east,rotate=45,yshift=-5pt},
        yticklabels={BHO,OnLineGames,Renos,Winwebsec,CeeInject,FakeRean,Vobfus},
        ytick={0,...,6},
        ytick style={draw=none},
        enlargelimits=false,
        colorbar,
        colorbar style={
%     	  	width=0.05*\pgfkeysvalueof{/pgfplots/parent axis width},%%% added this
%     	  	height=0.5*\pgfkeysvalueof{/pgfplots/parent axis height},
%		plot graphics/node/.style={scale=1.33,anchor=south west,inner sep=0pt,}, %%% scale colorbar fill %%%
            ytick={0,0.20,0.40,0.60,0.80,1.00},
            yticklabels={0,0.20,0.40,0.60,0.80,1.00},
            yticklabel={\pgfmathprintnumber\tick},
            yticklabel style={%font=\footnotesize,
            		/pgf/number format/fixed,
			/pgf/number format/precision=2}
        },
%        point meta min=0,
%        point meta max=100,
        point meta min=0.0,
        point meta max=1.0,
        nodes near coords={\pgfmathprintnumber\pgfplotspointmeta},
        % ---------------------------------------------------------------------
        % show `nodes near coords' but adapt the style so that values
        % above a threshold get another style
        % (adapted from <http://tex.stackexchange.com/a/141006/95441>)
        % #1: the THRESHOLD after which we switch to a special display.
        nodes near coords black white/.style={
            % define the style of the nodes with "small" values
            small value/.style={
                yshift=-7pt,
%                text=white,
                text=black,
                /pgf/number format/fixed,
                /pgf/number format/precision=2
%                /pgf/number format/precision=0
            },
            % define the style of the nodes with "large" values
            large value/.style={
                yshift=-7pt,
%                text=black,
                text=white,
                /pgf/number format/fixed,
                /pgf/number format/precision=2
%                /pgf/number format/precision=0
            },
            every node near coord/.style={
                check for zero/.code={
                    \pgfmathfloatifflags{\pgfplotspointmeta}{0}{
                        % If meta=0, make the node a coordinate
                        % (which doesn't have text)
                        \pgfkeys{/tikz/coordinate}
                    }{
                        \begingroup
                        % this group is merely to switch to FPU locally.
                        % Might be unnecessary, but who knows.
                        \pgfkeys{/pgf/fpu}
                        \pgfmathparse{\pgfplotspointmeta<#1}
                        \global\let\result=\pgfmathresult
                        \endgroup
                        %
                        % simplifies debugging:
                        %\show\result
                        %
                        \pgfmathfloatcreate{1}{1.0}{0}
                        \let\ONE=\pgfmathresult
                        \ifx\result\ONE
                            % AH: our condition 'y < #1' is met.
                            \pgfkeysalso{/pgfplots/small value}
                        \else
                            % ok, proceed as usual.
                            \pgfkeysalso{/pgfplots/large value}
                        \fi
                    }
                },
                check for zero,
            },
        },
        % asign a value to the new style which is the threshold at which
        % the two style `small value' or `large value' are used
%        nodes near coords black white=50,
        nodes near coords black white=0.5,
        % -----------------------------------------------------------------
    ]
        \addplot[
            matrix plot,
            mesh/cols=7,
            point meta=explicit,draw=gray
        ] table [meta=C] {
            x y C
 0  0 0.97
 1  0 0.00
 2  0 0.00
 3  0 0.01
 4  0 0.01
 5  0 0.01
 6  0 0.00
 0  1 0.00
 1  1 0.90
 2  1 0.01
 3  1 0.03
 4  1 0.03
 5  1 0.02
 6  1 0.01
 0  2 0.01
 1  2 0.01
 2  2 0.94
 3  2 0.00
 4  2 0.02
 5  2 0.01
 6  2 0.01
 0  3 0.00
 1  3 0.03
 2  3 0.03
 3  3 0.77
 4  3 0.03
 5  3 0.14
 6  3 0.00
 0  4 0.00
 1  4 0.02
 2  4 0.02
 3  4 0.05
 4  4 0.88
 5  4 0.02
 6  4 0.01
 0  5 0.00
 1  5 0.02
 2  5 0.02
 3  5 0.04
 4  5 0.04
 5  5 0.88
 6  5 0.00
 0  6 0.00
 1  6 0.03
 2  6 0.01
 3  6 0.01
 4  6 0.01
 5  6 0.00
 6  6 0.94
        };
    \end{axis}
\end{tikzpicture}
%
%\caption{I'm confused~5!}\label{tab:CM5}
%\end{figure*}
\\[-1ex]
(c) RF & (d) SVM
\\[-0.5ex]
\end{tabular}
\caption{Confusion matrices for Word2Vec experiments}\label{fig:W2V}
\end{figure}

Analogous to the HMM2Vec and PCA2Vec experiments above,
we classify the samples using the same four classifiers % ($\kNN{k}$, MLP, RF, and SVM),
but with Word2Vec embeddings as features.
The confusion matrices for these experiments are given in Figure~\ref{fig:W2V}.

As with the PCA2Vec experiments above,
to generate our Word2Vec models, we use a window size of~$W=10$. And, to
be consistent with both the HMM2Vec and PCA2Vec models discussed above,
we use a vector length of two, giving us feature vectors of length~40.
We use the so-called continuous bag of words (CBOW) model, which is
the model that we described in Section~\ref{sect:w2v}.

The accuracies obtained
for~$\kNN{k}$, MLP, RF, and SVM based on Word2Vec features
are~0.93, 0.91, 0.93, and~0.89, respectively. 
These results match those obtained using HMM2Vec.

In Section~\ref{sect:discuss}, 
we compare the accuracies obtained 
in our baseline HMM, HMM2Vec, PCA2Vec, and Word2Vec experiments.
But first we discuss possible overfitting issues with respect to 
the~$\kNN{k}$ and~RF classifiers discussed above.

\subsection{Overfitting}

As discussed above in Section~\ref{sect:lastWord},
both~$\kNN{k}$ and random forest are neighborhood-based 
classification algorithms, but with different neighborhood 
structure. Thus, we expect that
these two classification algorithms will generally perform
in a somewhat similar manner, at least in a qualitative sense. 

For~$\kNN{k}$, small values of~$k$ tend to result
in overfitting. To avoid overfitting,
the rule of thumb is that we should 
choose~$k\approx\sqrt{N}$, where~$N$ is the number of samples
in the training set~\cite{JirinaJ10}. 
Since we use an~80-20 split for training-testing
and we have~7000 samples, for our~$\kNN{k}$ experiments,
this rule of thumb gives us~$k=\sqrt{5600}\approx 75$.
However, for each feature set considered, our grid search yielded
an optimal value of~$k\leq 3$.

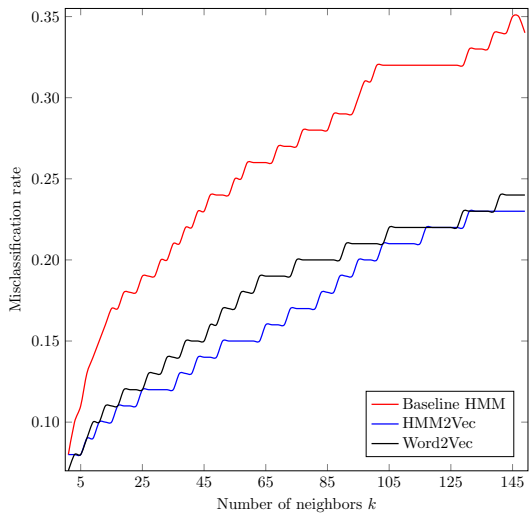
\begin{figure}[!htb]
\centering
    %\begin{tikzpicture}[scale=0.75]
\begin{tikzpicture}[scale=0.6]
\begin{axis}[no markers,smooth,
		   width=0.8\textwidth,
		   height=0.8\textwidth,
%		   /pgf/number format/1000 sep={},
%   	           symbolic x coords={1, 3, 5, 7, 9, 11, 13, 15, 17, 19, 21, 23, 25, 27, 29, 31, 33, 35, 37, 39, 41, 
%	           43, 45, 47, 49, 51, 53, 55, 57, 59, 61, 63, 65, 67, 69, 71, 73, 75, 77, 79, 81, 83, 85, 87, 89, 
%	           91, 93, 95, 97, 99, 101, 103, 105, 107, 109, 111, 113, 115, 117, 119, 121, 123, 125, 127, 
%	           129, 131, 133, 135, 137, 139, 141, 143, 145, 147, 149},
	 	   x tick label style={
   		 	/pgf/number format/.cd,
			/pgf/number format/1000 sep={},
   			fixed,
   			fixed zerofill,
    			precision=0
		   },
	 	   y tick label style={
    		 	/pgf/number format/.cd,
   			fixed,
   			fixed zerofill,
    			precision=2
		    },
                    xmin=0,xmax=150,
                    ymin=0.07,ymax=0.355,
                    legend pos=south east,
                    legend cell align={left},
                    xtick={5,25,45,65,85,105,125,145},
                    ytick={0.1,0.15,0.2,0.25,0.3,0.35},
                    xlabel={Number of neighbors $k$},
                    ylabel={Misclassification rate}] 
\addplot[color=red,thick] coordinates { % baseline HMM
(1, 0.08)
(3, 0.1)
(5, 0.11)
(7, 0.13)
(9, 0.14)
(11, 0.15)
(13, 0.16)
(15, 0.17)
(17, 0.17)
(19, 0.18)
(21, 0.18)
(23, 0.18)
(25, 0.19)
(27, 0.19)
(29, 0.19)
(31, 0.2)
(33, 0.2)
(35, 0.21)
(37, 0.21)
(39, 0.22)
(41, 0.22)
(43, 0.23)
(45, 0.23)
(47, 0.24)
(49, 0.24)
(51, 0.24)
(53, 0.24)
(55, 0.25)
(57, 0.25)
(59, 0.26)
(61, 0.26)
(63, 0.26)
(65, 0.26)
(67, 0.26)
(69, 0.27)
(71, 0.27)
(73, 0.27)
(75, 0.27)
(77, 0.28)
(79, 0.28)
(81, 0.28)
(83, 0.28)
(85, 0.28)
(87, 0.29)
(89, 0.29)
(91, 0.29)
(93, 0.29)
(95, 0.3)
(97, 0.31)
(99, 0.31)
(101, 0.32)
(103, 0.32)
(105, 0.32)
(107, 0.32)
(109, 0.32)
(111, 0.32)
(113, 0.32)
(115, 0.32)
(117, 0.32)
(119, 0.32)
(121, 0.32)
(123, 0.32)
(125, 0.32)
(127, 0.32)
(129, 0.32)
(131, 0.33)
(133, 0.33)
(135, 0.33)
(137, 0.33)
(139, 0.34)
(141, 0.34)
(143, 0.34)
(145, 0.35)
(147, 0.35)
(149, 0.34)
};
\addplot[color=blue,thick] coordinates { % HMM2Vec
(1, 0.08)
(3, 0.08)
(5, 0.08)
(7, 0.09)
(9, 0.09)
(11, 0.1)
(13, 0.1)
(15, 0.1)
(17, 0.11)
(19, 0.11)
(21, 0.11)
(23, 0.11)
(25, 0.12)
(27, 0.12)
(29, 0.12)
(31, 0.12)
(33, 0.12)
(35, 0.12)
(37, 0.13)
(39, 0.13)
(41, 0.13)
(43, 0.14)
(45, 0.14)
(47, 0.14)
(49, 0.14)
(51, 0.15)
(53, 0.15)
(55, 0.15)
(57, 0.15)
(59, 0.15)
(61, 0.15)
(63, 0.15)
(65, 0.16)
(67, 0.16)
(69, 0.16)
(71, 0.16)
(73, 0.17)
(75, 0.17)
(77, 0.17)
(79, 0.17)
(81, 0.17)
(83, 0.18)
(85, 0.18)
(87, 0.18)
(89, 0.19)
(91, 0.19)
(93, 0.19)
(95, 0.2)
(97, 0.2)
(99, 0.2)
(101, 0.2)
(103, 0.21)
(105, 0.21)
(107, 0.21)
(109, 0.21)
(111, 0.21)
(113, 0.21)
(115, 0.21)
(117, 0.22)
(119, 0.22)
(121, 0.22)
(123, 0.22)
(125, 0.22)
(127, 0.22)
(129, 0.22)
(131, 0.23)
(133, 0.23)
(135, 0.23)
(137, 0.23)
(139, 0.23)
(141, 0.23)
(143, 0.23)
(145, 0.23)
(147, 0.23)
(149, 0.23)
};
\addplot[color=black,thick] coordinates { % Word2Vec
(1, 0.07)
(3, 0.08)
(5, 0.08)
(7, 0.09)
(9, 0.1)
(11, 0.1)
(13, 0.11)
(15, 0.11)
(17, 0.11)
(19, 0.12)
(21, 0.12)
(23, 0.12)
(25, 0.12)
(27, 0.13)
(29, 0.13)
(31, 0.13)
(33, 0.14)
(35, 0.14)
(37, 0.14)
(39, 0.15)
(41, 0.15)
(43, 0.15)
(45, 0.15)
(47, 0.16)
(49, 0.16)
(51, 0.17)
(53, 0.17)
(55, 0.17)
(57, 0.18)
(59, 0.18)
(61, 0.18)
(63, 0.19)
(65, 0.19)
(67, 0.19)
(69, 0.19)
(71, 0.19)
(73, 0.19)
(75, 0.2)
(77, 0.2)
(79, 0.2)
(81, 0.2)
(83, 0.2)
(85, 0.2)
(87, 0.2)
(89, 0.2)
(91, 0.21)
(93, 0.21)
(95, 0.21)
(97, 0.21)
(99, 0.21)
(101, 0.21)
(103, 0.21)
(105, 0.22)
(107, 0.22)
(109, 0.22)
(111, 0.22)
(113, 0.22)
(115, 0.22)
(117, 0.22)
(119, 0.22)
(121, 0.22)
(123, 0.22)
(125, 0.22)
(127, 0.22)
(129, 0.23)
(131, 0.23)
(133, 0.23)
(135, 0.23)
(137, 0.23)
(139, 0.23)
(141, 0.24)
(143, 0.24)
(145, 0.24)
(147, 0.24)
(149, 0.24)
};
\legend{Baseline HMM, HMM2Vec, Word2Vec}
\end{axis}
\end{tikzpicture}
\vglue-0.1in
\caption{$\kNN{k}$ results as a function of~$k$}\label{fig:kNN_overfit} 
\end{figure}

In Figure~\ref{fig:kNN_overfit}, we graph the accuracy of~$\kNN{k}$
as a function of~$k$ for the baseline HMM, HMM2Vec, and Word2Vec
feature sets. We see that all of these techniques perform more
poorly as~$k$ increases. In particular, for~$k\approx 75$, the performance
of each is poor in comparison to~$k\leq 3$, and this
effect is particularly pronounced in the case of the baseline HMM.
This provides strong evidence that small values of~$k$ in~$\kNN{k}$ 
results in overfitting for each feature set, and the overfitting is 
especially pronounced for the baseline HMM.

%%%% k-NN as a function of k

For a random forest, the overfitting that is inherent in decision trees
is mitigated by using more trees. In contrast, if the depth of the trees
in the random forest is too large, the effect is analogous to choosing~$k$ too
small in~$\kNN{k}$, and overfitting is likely to occur.

\begin{figure}[!htb]
\centering
    %\begin{tikzpicture}[scale=0.75]
\begin{tikzpicture}[scale=0.6]
\begin{axis}[no markers,smooth,
		   width=0.8\textwidth,
		   height=0.8\textwidth,
%		   /pgf/number format/1000 sep={},
%   	           symbolic x coords={1, 3, 5, 7, 9, 11, 13, 15, 17, 19, 21, 23, 25, 27, 29, 31, 33, 35, 37, 39, 41, 
%	           43, 45, 47, 49, 51, 53, 55, 57, 59, 61, 63, 65, 67, 69, 71, 73, 75, 77, 79, 81, 83, 85, 87, 89, 
%	           91, 93, 95, 97, 99, 101, 103, 105, 107, 109, 111, 113, 115, 117, 119, 121, 123, 125, 127, 
%	           129, 131, 133, 135, 137, 139, 141, 143, 145, 147, 149},
	 	   x tick label style={
   		 	/pgf/number format/.cd,
			/pgf/number format/1000 sep={},
   			fixed,
   			fixed zerofill,
    			precision=0
		   },
	 	   y tick label style={
    		 	/pgf/number format/.cd,
   			fixed,
   			fixed zerofill,
    			precision=2
		    },
                    xmin=0,xmax=30,
                    ymin=0.0,ymax=0.82,
                    legend pos=north east,
                    legend cell align={left},
                    xtick={2,4,6,8,10,12,14,16,18,20,22,24,26,28},
                    ytick={0.1,0.2,0.3,0.4,0.5,0.6,0.7,0.8},
                    xlabel={Depth of tree},
                    ylabel={Misclassification rate}] 
\addplot[color=red,thick] coordinates { % baseline HMM
(1, 0.73)
(2, 0.52)
(3, 0.45)
(4, 0.37)
(5, 0.32)
(6, 0.28)
(7, 0.18)
(8, 0.16)
(9, 0.13)
(10, 0.13)
(11, 0.11)
(12, 0.11)
(13, 0.1)
(14, 0.1)
(15, 0.1)
(16, 0.1)
(17, 0.09)
(18, 0.09)
(19, 0.09)
(20, 0.09)
(21, 0.09)
(22, 0.09)
(23, 0.09)
(24, 0.09)
(25, 0.09)
(26, 0.09)
(27, 0.09)
(28, 0.09)
(29, 0.09)
};
\addplot[color=blue,thick] coordinates { % HMM2Vec
(1, 0.4)
(2, 0.32)
(3, 0.24)
(4, 0.16)
(5, 0.13)
(6, 0.1)
(7, 0.09)
(8, 0.08)
(9, 0.08)
(10, 0.07)
(11, 0.07)
(12, 0.07)
(13, 0.07)
(14, 0.07)
(15, 0.06)
(16, 0.07)
(17, 0.06)
(18, 0.07)
(19, 0.07)
(20, 0.07)
(21, 0.06)
(22, 0.06)
(23, 0.07)
(24, 0.06)
(25, 0.06)
(26, 0.06)
(27, 0.07)
(28, 0.07)
(29, 0.06)
};
\addplot[color=black,thick] coordinates { % Word2Vec
(1, 0.62)
(2, 0.48)
(3, 0.38)
(4, 0.32)
(5, 0.25)
(6, 0.22)
(7, 0.18)
(8, 0.15)
(9, 0.13)
(10, 0.12)
(11, 0.1)
(12, 0.09)
(13, 0.08)
(14, 0.08)
(15, 0.07)
(16, 0.07)
(17, 0.06)
(18, 0.06)
(19, 0.07)
(20, 0.06)
(21, 0.06)
(22, 0.06)
(23, 0.06)
(24, 0.06)
(25, 0.06)
(26, 0.06)
(27, 0.06)
(28, 0.06)
(29, 0.06)
};
\legend{Baseline HMM, HMM2Vec, Word2Vec}
\end{axis}
\end{tikzpicture}
\vglue-0.1in
\caption{Random forest results as a function of tree depth}\label{fig:RF_overfit} 
\end{figure}
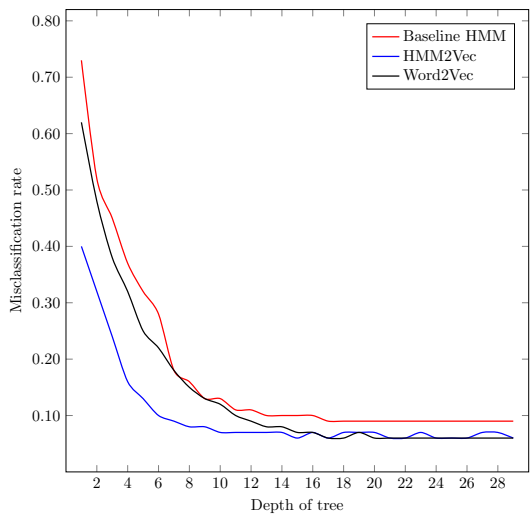

To explore overfitting in our RF experiments, in Figure~\ref{fig:RF_overfit},
we give the misclassification results for the baseline HMM, HMM2Vec, 
and Word2Vec features, as a function of the maximum depth of the trees.
In this case, Word2Vec performs best for smaller (maximum) 
depths, which indicates that the baseline HMM and HMM2Vec
features are more prone to overfitting.

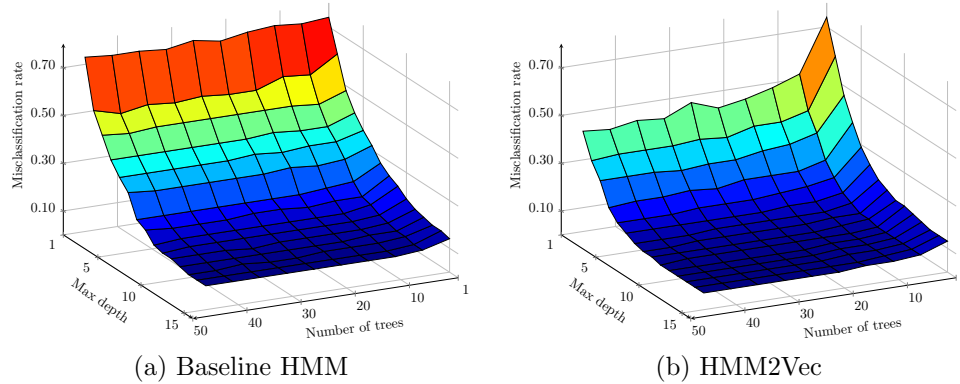
\begin{figure}[!htb]
\centering
\begin{tabular}{cc}
    %\begin{tikzpicture}[xscale=0.7,yscale=0.7]
\begin{tikzpicture}[scale=0.525]
%  \begin{axis} 
%   \addplot3[surf] coordinates
\begin{axis}[%
%view={45}{110},
view={64}{26},
%view={10}{26},
width=10cm,
height=8cm,
scale only axis,
xmin=1, xmax=16,
xtick={1,5,10,15},
xmajorgrids,
ymin=1, ymax=50,
ytick={1,10,20,30,40,50},
ymajorgrids,
y dir=reverse,
zmin=0, zmax=0.8,
ztick={0.1,0.3,0.5,0.7},
zmajorgrids,
axis lines=left,
grid=major,
xlabel=Max depth,
xlabel style={yshift=0.2cm,rotate=-35},
ylabel=Number of trees,
ylabel style={yshift=0.15cm,xshift=0.25cm,rotate=7},
zlabel=Misclassification rate,
zlabel style={xshift=0.1cm,yshift=-0.05cm},
x tick label style={
    		/pgf/number format/.cd,
		1000 sep={},
    		fixed,
    		fixed zerofill,
    		precision=0},
y tick label style={
    		/pgf/number format/.cd,
		1000 sep={},
    		fixed,
    		fixed zerofill,
    		precision=0},
z tick label style={
    		/pgf/number format/.cd,
		1000 sep={},
    		fixed,
    		fixed zerofill,
    		precision=2}]
%z filter/.code={\pgfmathparse{#1/100}\pgfmathresult}]
%
\addplot3[%
surf,
z buffer=sort,
colormap/jet,
shader=flat,
draw=black]
coordinates{

(1,1,0.74)
(1,6,0.73)
(1,11,0.74)
(1,16,0.73)
(1,21,0.71)
(1,26,0.73)
(1,31,0.71)
(1,36,0.72)
(1,41,0.72)
(1,46,0.73)

(2,1,0.61)
(2,6,0.54)
(2,11,0.55)
(2,16,0.51)
(2,21,0.51)
(2,26,0.52)
(2,31,0.51)
(2,36,0.52)
(2,41,0.5)
(2,46,0.53)

(3,1,0.48)
(3,6,0.44)
(3,11,0.45)
(3,16,0.45)
(3,21,0.45)
(3,26,0.45)
(3,31,0.45)
(3,36,0.45)
(3,41,0.44)
(3,46,0.45)

(4,1,0.41)
(4,6,0.37)
(4,11,0.37)
(4,16,0.38)
(4,21,0.37)
(4,26,0.37)
(4,31,0.37)
(4,36,0.37)
(4,41,0.37)
(4,46,0.37)

(5,1,0.36)
(5,6,0.33)
(5,11,0.33)
(5,16,0.32)
(5,21,0.33)
(5,26,0.32)
(5,31,0.32)
(5,36,0.32)
(5,41,0.32)
(5,46,0.32)

(6,1,0.32)
(6,6,0.28)
(6,11,0.27)
(6,16,0.28)
(6,21,0.27)
(6,26,0.27)
(6,31,0.27)
(6,36,0.27)
(6,41,0.27)
(6,46,0.28)

(7,1,0.27)
(7,6,0.21)
(7,11,0.2)
(7,16,0.19)
(7,21,0.19)
(7,26,0.19)
(7,31,0.19)
(7,36,0.19)
(7,41,0.19)
(7,46,0.19)

(8,1,0.24)
(8,6,0.17)
(8,11,0.17)
(8,16,0.16)
(8,21,0.17)
(8,26,0.17)
(8,31,0.16)
(8,36,0.16)
(8,41,0.16)
(8,46,0.17)

(9,1,0.22)
(9,6,0.16)
(9,11,0.14)
(9,16,0.14)
(9,21,0.14)
(9,26,0.14)
(9,31,0.14)
(9,36,0.14)
(9,41,0.14)
(9,46,0.13)

(10,1,0.19)
(10,6,0.14)
(10,11,0.13)
(10,16,0.13)
(10,21,0.13)
(10,26,0.13)
(10,31,0.13)
(10,36,0.13)
(10,41,0.13)
(10,46,0.12)

(11,1,0.17)
(11,6,0.13)
(11,11,0.12)
(11,16,0.12)
(11,21,0.12)
(11,26,0.12)
(11,31,0.11)
(11,36,0.12)
(11,41,0.12)
(11,46,0.12)

(12,1,0.16)
(12,6,0.12)
(12,11,0.12)
(12,16,0.11)
(12,21,0.11)
(12,26,0.11)
(12,31,0.11)
(12,36,0.11)
(12,41,0.11)
(12,46,0.11)

(13,1,0.15)
(13,6,0.12)
(13,11,0.11)
(13,16,0.11)
(13,21,0.11)
(13,26,0.1)
(13,31,0.1)
(13,36,0.11)
(13,41,0.1)
(13,46,0.1)

(14,1,0.15)
(14,6,0.11)
(14,11,0.1)
(14,16,0.1)
(14,21,0.1)
(14,26,0.1)
(14,31,0.1)
(14,36,0.1)
(14,41,0.1)
(14,46,0.1)

(15,1,0.14)
(15,6,0.11)
(15,11,0.1)
(15,16,0.1)
(15,21,0.1)
(15,26,0.1)
(15,31,0.1)
(15,36,0.1)
(15,41,0.1)
(15,46,0.1)

};
\end{axis}
\end{tikzpicture}
  
    & 
    %\begin{tikzpicture}[xscale=0.7,yscale=0.7]
\begin{tikzpicture}[scale=0.525]
%  \begin{axis} 
%   \addplot3[surf] coordinates
\begin{axis}[%
%view={45}{110},
view={64}{26},
%view={10}{26},
width=10cm,
height=8cm,
scale only axis,
xmin=1, xmax=16,
xtick={1,5,10,15},
xmajorgrids,
ymin=1, ymax=50,
ytick={1,10,20,30,40,50},
ymajorgrids,
y dir=reverse,
zmin=0, zmax=0.8,
ztick={0.1,0.3,0.5,0.7},
zmajorgrids,
axis lines=left,
grid=major,
xlabel=Max depth,
xlabel style={yshift=0.2cm,rotate=-35},
ylabel=Number of trees,
ylabel style={yshift=0.15cm,xshift=0.25cm,rotate=7},
zlabel=Misclassification rate,
zlabel style={xshift=0.1cm,yshift=-0.05cm},
x tick label style={
    		/pgf/number format/.cd,
		1000 sep={},
    		fixed,
    		fixed zerofill,
    		precision=0},
y tick label style={
    		/pgf/number format/.cd,
		1000 sep={},
    		fixed,
    		fixed zerofill,
    		precision=0},
z tick label style={
    		/pgf/number format/.cd,
		1000 sep={},
    		fixed,
    		fixed zerofill,
    		precision=2}]
%z filter/.code={\pgfmathparse{#1/100}\pgfmathresult}]
%
\addplot3[%
surf,
z buffer=sort,
colormap/jet,
shader=flat,
draw=black]
coordinates{

(1,1,0.74)
(1,6,0.52)
(1,11,0.48)
(1,16,0.45)
(1,21,0.43)
(1,26,0.47)
(1,31,0.42)
(1,36,0.43)
(1,41,0.41)
(1,46,0.42)

(2,1,0.6)
(2,6,0.39)
(2,11,0.36)
(2,16,0.36)
(2,21,0.33)
(2,26,0.34)
(2,31,0.33)
(2,36,0.33)
(2,41,0.33)
(2,46,0.32)

(3,1,0.46)
(3,6,0.32)
(3,11,0.29)
(3,16,0.28)
(3,21,0.26)
(3,26,0.26)
(3,31,0.25)
(3,36,0.26)
(3,41,0.26)
(3,46,0.26)

(4,1,0.37)
(4,6,0.23)
(4,11,0.2)
(4,16,0.2)
(4,21,0.2)
(4,26,0.18)
(4,31,0.19)
(4,36,0.18)
(4,41,0.18)
(4,46,0.17)

(5,1,0.29)
(5,6,0.18)
(5,11,0.16)
(5,16,0.15)
(5,21,0.14)
(5,26,0.14)
(5,31,0.14)
(5,36,0.13)
(5,41,0.13)
(5,46,0.13)

(6,1,0.24)
(6,6,0.14)
(6,11,0.12)
(6,16,0.12)
(6,21,0.11)
(6,26,0.11)
(6,31,0.11)
(6,36,0.11)
(6,41,0.11)
(6,46,0.11)

(7,1,0.2)
(7,6,0.12)
(7,11,0.1)
(7,16,0.1)
(7,21,0.1)
(7,26,0.1)
(7,31,0.1)
(7,36,0.09)
(7,41,0.09)
(7,46,0.09)

(8,1,0.18)
(8,6,0.1)
(8,11,0.1)
(8,16,0.09)
(8,21,0.09)
(8,26,0.09)
(8,31,0.09)
(8,36,0.09)
(8,41,0.08)
(8,46,0.08)

(9,1,0.16)
(9,6,0.1)
(9,11,0.09)
(9,16,0.08)
(9,21,0.08)
(9,26,0.08)
(9,31,0.08)
(9,36,0.08)
(9,41,0.08)
(9,46,0.08)

(10,1,0.16)
(10,6,0.09)
(10,11,0.08)
(10,16,0.08)
(10,21,0.08)
(10,26,0.08)
(10,31,0.08)
(10,36,0.08)
(10,41,0.08)
(10,46,0.07)

(11,1,0.15)
(11,6,0.09)
(11,11,0.08)
(11,16,0.08)
(11,21,0.07)
(11,26,0.07)
(11,31,0.07)
(11,36,0.07)
(11,41,0.07)
(11,46,0.07)

(12,1,0.14)
(12,6,0.09)
(12,11,0.08)
(12,16,0.08)
(12,21,0.07)
(12,26,0.07)
(12,31,0.07)
(12,36,0.07)
(12,41,0.07)
(12,46,0.07)

(13,1,0.13)
(13,6,0.09)
(13,11,0.08)
(13,16,0.08)
(13,21,0.07)
(13,26,0.07)
(13,31,0.07)
(13,36,0.07)
(13,41,0.07)
(13,46,0.07)

(14,1,0.13)
(14,6,0.09)
(14,11,0.08)
(14,16,0.07)
(14,21,0.07)
(14,26,0.07)
(14,31,0.07)
(14,36,0.07)
(14,41,0.07)
(14,46,0.07)

(15,1,0.13)
(15,6,0.09)
(15,11,0.08)
(15,16,0.08)
(15,21,0.07)
(15,26,0.07)
(15,31,0.07)
(15,36,0.07)
(15,41,0.07)
(15,46,0.07)

};
\end{axis}
\end{tikzpicture}
    \\
    (a) Baseline HMM
    & 
    (b) HMM2Vec
\end{tabular}
\caption{Random forest maximum depth vs number of trees}\label{fig:base_h2v_3d} 
\end{figure}

In Figures~\ref{fig:base_h2v_3d}~(a) and~(b), we give misclassification 
results as a function of both the maximum depth and the number of trees
for the baseline HMM and for HMM2Vec features, respectively.
From these results, we see that the baseline HMM
performs similarly as a function of the maximum depth, 
regardless of the number of trees. In contrast,
the HMM2Vec features yield consistently better
results than the baseline HMM (as a function of the 
maximum depth), except when the number of trees is very small.
This indicates that, with respect to the
maximum depth, overfitting is significantly worse for the
baseline HMM, since the overfitting cannot be overcome by increasing
the number of trees.

%\begin{figure}[!htb]
%\centering
%    \input figures/fig_3d_raw_smaller.tex
%\caption{Baseline HMM 3-d results}\label{fig:base_3d} 
%\end{figure}
%
%\begin{figure}[!htb]
%\centering
%    \input figures/fig_3d_h2v_smaller.tex
%\caption{HMM2Vec 3-d results}\label{fig:h2v_3d} 
%\end{figure}

%\begin{figure}[!htb]
%\centering
%    \input figures/fig_3d_rohit.tex
%\caption{Rohit's 3-d results}
%\end{figure}

From the discussion in this section, we see that all of our~$\kNN{k}$
experiments results suffer from some degree of overfitting, with this 
effect being most significant in the case of the baseline HMM.
For our RF results, overfitting is a relatively minor issue for the HMM2Vec 
and Word2Vec engineered features but, as with~$\kNN{k}$,
it is a significant problem for the baseline HMM. Consequently,
both the~$\kNN{k}$ and RF results we have reported
for the baseline HMM are overly optimistic, as these
represent cases where significant overfitting has occurred.

\subsection{Discussion}\label{sect:discuss}

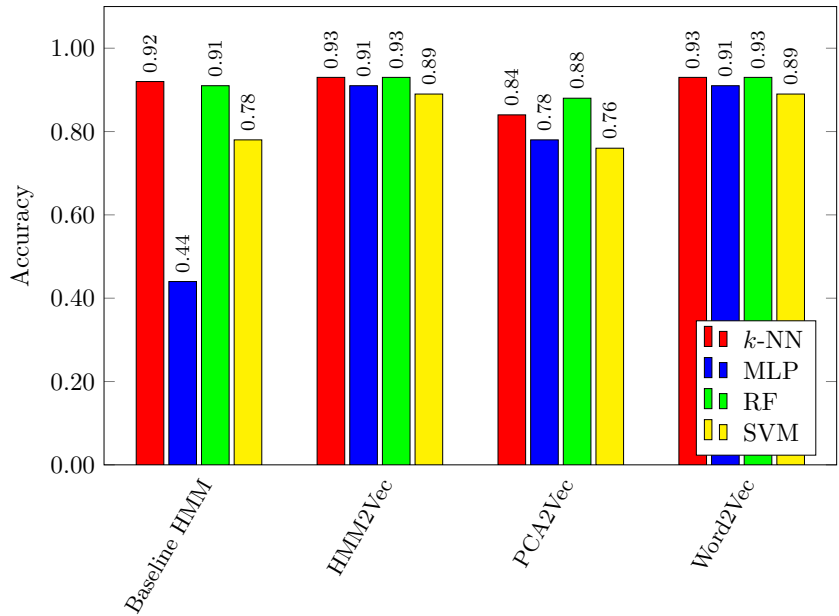
\begin{figure}[!htb]
\centering
    \begin{tikzpicture}[scale=0.95, every node/.style={scale=1.0}]
    \begin{axis}[
        width  = 0.8*\textwidth,
        height = 8cm,
        ymin=0.0,ymax=1.1,
        ytick={0.0,0.2,0.4,0.6,0.8,1.0},
        major x tick style = transparent,
        ybar=5*\pgflinewidth,
        bar width=11.0pt,
%        ymajorgrids = true,
        ylabel = {Accuracy},
        symbolic x coords={Baseline HMM,HMM2Vec,PCA2Vec,Word2Vec},
	y tick label style={
    		/pgf/number format/.cd,
   		fixed,
   		fixed zerofill,
    		precision=2},
%	yticklabel pos=right,
        xtick = data,
        x tick label style={
        		rotate=60,
		font=\small,
		anchor=north east,
		inner sep=0mm},
%		font=\small},
%        scaled y ticks = false,
	%%%%% numbers on bars and rotated
        nodes near coords,
        every node near coord/.append style={rotate=90, 
        								   anchor=west,
								   font=\footnotesize},
        %%%%%
%        enlarge x limits=0.03,
        enlarge x limits=0.175,
        legend cell align=left,
        legend style={
%                at={(1,1.05)},
%                anchor=south east,
%	        nodes={rotate=90},%%%%% rotate text in legend
%                at={(0.125,0)},
%                at={(0.125,0)},
%                at={(0.8775,0)},
                at={(0.89,0.02)},
                anchor=south,
                column sep=1ex
        }
    ]
\addplot[fill=red,opacity=1.00] %%%%% k-NN
coordinates {
(Baseline HMM,0.92)
(HMM2Vec,0.93)
(PCA2Vec,0.84)
(Word2Vec,0.93)
};
\addplot[fill=blue,opacity=1.00] %%%%% MLP
coordinates {
(Baseline HMM,0.44)
(HMM2Vec,0.91)
(PCA2Vec,0.78)
(Word2Vec,0.91)
};
\addplot[fill=green,opacity=1.00] %%%%% RF
coordinates {
(Baseline HMM,0.91)
(HMM2Vec,0.93)
(PCA2Vec,0.88)
(Word2Vec,0.93)
};
\addplot[fill=yellow,opacity=1.00] %%%%% SVM
coordinates {
(Baseline HMM,0.78)
(HMM2Vec,0.89)
(PCA2Vec,0.76)
(Word2Vec,0.89)
};
\legend{$\kNN{k}$,MLP,RF,SVM}
\end{axis}
\end{tikzpicture}
  \vglue-0.075in
\caption{Accuracies for combinations of features and classifiers}\label{fig:AFC} 
\end{figure}

Figure~\ref{fig:AFC} gives the overall 
accuracy for each of our multiclass experiments 
using~$\kNN{k}$, MLP, RF, and~SVM classifiers,
for our baseline HMM opcode experiments, 
and for each of the HMM2Vec, PCA2Vec, 
and Word2Vec engineered feature experiments.
In general, we expect~RF and~$\kNN{k}$ to perform somewhat similarly,
since both are neighborhood-based algorithms.
We also expect that in most cases, SVM and~MLP
will perform in a qualitatively similar manner to each other,
since these techniques are closely related.
We find that these expectations are generally met in our
experiments, which can be viewed as a confirmation
of the validity of the results.

From our~16 distinct experiments, we see that HMM2Vec and Word2Vec perform
best, with PCA2Vec lagging far behind. The  baseline HMM
does well with respect to the neighborhood-based classifiers, namely,
RF and~$\kNN{k}$. However, as discussed in the previous
section, these neighborhood based techniques
overfit the training data in the baseline HMM experiments. 
Neglecting these overfit results, we see that 
using the HMM2Vec and Word2Vec
engineered features with~SVM and~MLP classifiers,
give us the best results. Furthermore, these HMM2Vec and Word2Vec
results are substantially better than either of the reliable results 
obtained for the baseline HMM, that is, the baseline HMM
results using~SVM and~MLP classifiers.

\section{Conclusion and Future Work}\label{sect:conc}

In this paper, we have presented results for a number of experiments involving
word embedding techniques in malware classification.
We have applied machine learning techniques to raw features to
generate engineered features that are used for classification. 
Such a concept is not entirely unprecedented as, for example, 
PCA is often used to reduce the dimensionality of data
before applying other machine learning techniques. However, the authors 
are not aware of previous work involving the use word embedding techniques
in the same manner considered in this paper.

Our results show that word embedding techniques can be used to 
generate features that are more informative than the original data.
This process of distilling useful information from the data before 
classifying samples is potentially useful, not only in the field of malware
analysis, but also in other fields where learning plays a prominent role.

For future work, it would be interesting to consider other families and other types of
malware. It would also be worthwhile to consider more complex and higher dimensional
data---as with dimensionality-reduction techniques, such data would tend to offer 
more scope for improvement using the word embedding strategies
considered in this paper.

\bibliographystyle{plain}

\bibliography{references.bib,Stamp-Mark.bib}

\end{document}